\documentclass[]{aastex631}
\usepackage{multirow}
\usepackage{gensymb}
\usepackage{xcolor}
\usepackage{amsmath}
\usepackage[bottom]{footmisc}
\usepackage{tikz}
\usetikzlibrary{arrows.meta, decorations.markings, angles, quotes}
\usepackage{float}
\usepackage{xcolor}

\newcommand{\tb}{\textcolor{black}}
\begin{document}
\title{Tidal disruptions of close white dwarf binaries by intermediate mass black holes}
\correspondingauthor{Aryabrat Mahapatra}
\email{aryabratm23@iitk.ac.in}

\author{Aryabrat Mahapatra}
\affiliation{Department of Physics, Indian Institute of Technology Kanpur,
	Kanpur 208016, India}

\author{Adarsh Pandey}
\affiliation{Department of Physics, Indian Institute of Technology Kanpur,
	Kanpur 208016, India}

\author{Pritam Banerjee}
\affiliation{Department of Physics, Indian Institute of Technology Kharagpur,
	Kharagpur  721302, India}

\author{Tapobrata Sarkar}
\affiliation{Department of Physics, Indian Institute of Technology Kanpur,
	Kanpur 208016, India}

\begin{abstract}
We perform a suite of numerical simulations of tidal disruption events, using smoothed particle hydrodynamics,
for a close binary system consisting of two low-mass white dwarfs, and an intermediate mass non-spinning black hole.
The binary components are considered to be detached and on the same plane with the black hole. 
Our results quantify how the outcomes of these events depend crucially on the positional configuration of the binary components 
at the orbital pericenter, and we also show 
how distinctive behaviour for non-identical mass binaries arise, as compared to identical ones. 
We highlight these differences on observables such as mass fallback rates, kick velocities and gravitational waves, and
also compute clump formation time within the stellar debris. 
In our setup, prograde binary motion, where the angular momentum of the binary is in the same direction as that of 
the center of mass motion around the black hole, is qualitatively similar to multiple events of single star tidal disruptions. 
However, we argue that interactions between stellar 
debris in the corresponding retrograde scenarios result in different and distinct outcomes. Our results should serve as
indicative benchmarks in the observational aspects of tidal interactions between close white dwarf binaries and intermediate
mass black holes. 
\end{abstract}
	
\section{Introduction}
\label{sec1}

Tidal interactions of binary stellar systems with black holes (BHs) are important and interesting in the context of 
astrophysical consequences of General Relativity (GR). In the point particle approximation, this is the
classical three-body problem modified by GR effects, whose stability and perturbative dynamics has been studied for over 
three centuries, the oldest three-body problem being the Sun-Earth-Moon configuration. The literature on the subject is vast, and we
only point to a modern exposition to the basic concepts that can be found in the book by \cite{VK} and the recent review by \cite{MQ}. 
There are fascinating features in such systems, for example the Kozai-Lidov oscillations for inclined binary orbits 
(see \cite{Naoz} for a recent review). Now, when a stellar binary comes close enough to a BH, there are non-trivial effects of
the finite size of the stars, as tidal effects are large enough to deform them. In this situation, one or both stars of
the system may be disrupted, either fully or partially, and the physics of the resulting situation is rich and offers an
ideal laboratory to study both the binary as well as the BH. In this paper, we study the interactions between a 
close white dwarf (WD) binary and an intermediate mass black hole (IMBH), both of which are interesting to study
in an astrophysical scenario, as we discuss below. 

Recall that BHs were predicted by Einstein's GR, more than a hundred years ago, and
are by now known to be inhabiting the centres of most galaxies. These are some of the most interesting objects in astrophysical 
studies of strong gravity. Till the end of the last century, it was commonly believed that BHs can be broadly categorised into two types,
depending on their masses, namely the stellar mass ($\sim 3 - 20 M_{\odot}$) and the supermassive ($\sim 10^6 - 10^{10} M_{\odot}$) ones
\citep{Coleman}. More recently, growing evidence for the existence of intermediate mass black holes (IMBHs) with masses $\sim 10^2 - 10^{5} M_{\odot}$ 
(see \cite{Green} for a review) has resulted in considerable efforts to understand these. In fact IMBHs are 
by now commonly known as the `missing link' in the theory of BH formation and evolution, and are believed to exist in
dwarf galaxies and globular clusters, and are possible seeds of supermassive black holes (SMBHs), see e.g., \cite{Volonteri}. 

Compared to the large amount of literature that exists on SMBHs and stellar mass BHs, observational evidences for IMBHs are relatively fewer, \citep{IMBH0, 
IMBH1, IMBH2, IMBH3, IMBH4}. An important phenomenon that can serve as a distinctive
feature of IMBHs are tidal disruption events (TDEs) \citep{Holoien} involving BHs and stars, since observables associated to these often 
show strong dependence on BH mass. In a few recent previous works \citep{Garain1, Garain2, Garain3}, we have addressed some issues related to TDEs involving
IMBHs (both of the Schwarzschild and Kerr types) and WDs, focussing in particular on partial disruption events (when the
star is not fully disrupted and retains a core in the disruption process) in eccentric, inclined orbits, and have also studied these with the
inclusion of stellar spin. The mass-radius relation for WDs indicate that these be tidally disrupted by IMBHs, 
and are captured as a whole by SMBHs, (see the chapter of Maguire et. al. in \cite{Jonker}). TDEs involving WDs and IMBHs can thus
be an ideal laboratory to glean further insight into the distinction between IMBHs and SMBHs. 
The situation becomes more interesting in TDEs involving IMBHs and WD binaries, the issue that we study in this paper for non-spinning, 
Schwarzschild IMBHs. In particular, we will consider close WD binaries, which were predicted more than four decades ago. In particular,
we consider an initially detached (i.e., non-accreting) WD binary (see \cite{Korol} for a recent survey) which wanders close to an IMBH
to undergo a TDE. Note that $\sim 25$ percent of 
all WDs are in binary or multiple-star systems \citep{Toonen}. Although this is lower compared to the fact that $\sim 50$ percent of all main 
sequence solar-like stars are in binary/multiple-star formations, the fraction of WD binaries is large enough to warrant a detailed 
analysis on their tidal disruption scenarios in the background of IMBHs. some relevant observational aspects of WD binaries were reported in \cite{Kilic1},
\cite{Roelofs}, \cite{Hermes}, \cite{Kilic2}, \cite{Brown} \cite{Ren}, \cite{Amaral}. 

Before we discuss TDEs involving stellar binaries, it would be prudent to briefly review some known facts about TDEs from 
solitary stars by BHs, in order to set the stage for the discussions in the rest of the paper. 
Recall that a star is tidally disrupted by a BH when its self gravity is overcome by the gravitational effects of the BH. 
To obtain an approximate Newtonian estimate, recall that for a star of mass $M_{\star}$ and radius $R_{\star}$ 
the tidal radius $r_t\sim R_{\star}(M_{\rm BH}/M_{\star})^{1/3}$,  with $M_{\rm BH}$ being the mass of the BH, see \cite{Hills} and related 
early works by \cite{FrankRees}, \cite{Lacy1}, \cite{CarterLumineta}, \cite{CarterLuminetb}. An order of magnitude estimate for obtaining
the strength of a TDE is given by the impact factor $\beta = r_t/r_p$, where $r_p$ is the pericenter distance of the star. Namely, small (large) values
of $\beta$ signify weak (strong) tidal interactions. 
In realistic situations, a star can be considerably deformed before actually being disrupted. Nonetheless, the above is a reasonable
estimate to indicate a TDE. Broadly speaking, TDEs can be separated into two distinct classes. In the first, a star is fully disrupted, wherein 
the entire stellar content is reduced to a stream of debris \citep{Kochanek1994, Guillochon2014}. In the second scenario, the disruption is partial, 
so that a remnant core forms, along with stellar debris \citep{Manukian2013, Gafton2015, Banerjee2023}. Upon full tidal disruption, it is well known that
roughly half of the debris is bound to the BH and falls into it with time. Partial tidal disruption effects might be more intricate, since the
self-bound core exerts gravitational force on the debris, and in extreme cases might cause a re-collapse of stellar debris to form
a circumstellar disc around the core, see \cite{Nixon2021}. Note that partial tidal disruption can in principle occur multiple times \citep{MacLeod2013}, 
as initial conditions might allow the remnant core to come back to a new pericenter position. 
For both full and partial disruptions, in-falling stellar debris dissipates energy and results in accretion flow, and can cause a 
luminous event \citep{Evans1989, Hayasaki2013, Hayasaki2016, Bonnerot2016, Liptai2019, Clerici2020}, which is a significant
astrophysical phenomenon. Not surprisingly therefore, TDEs have been extensively investigated
over more than four decades now, and continue to be an active research area of research. For a sampling of the more recent literature,
see \cite{Lodato, Guillochon2013, Rosswog_TR, Coughlin2015,  Law2017, Chen2018, Law2019, Gafton2019, Golightly2019a, Golightly2019b, 
Kagaya2019, Sacchi2019, Darbha2019, Milesetal, Park2020, Ryu2020a, Ryu2020b, Wang2021, 
Cufari2022}. A comprehensive recent review of various features of TDEs can be found in \cite{Jonker}. 

We note further that inclusion of stellar spin is a natural addition to the physics of TDEs. Indeed, spinning stars are perhaps the most
natural objects to study TDEs in the background of BHs. However, numerical studies of TDEs involving such stars is a 
fairly recent addition to the literature, see \cite{Golightly2019b, Kagaya2019, Sacchi2019}. 
Since it is known that tidal torques greatly spin up a star, stellar spin was possibly thought to be of less importance, 
as mentioned by Rossi et. al. in page 16 of article no. 40 of \cite{Jonker}. Contrary to this belief however, 
the works of \cite{Golightly2019b, Kagaya2019, Sacchi2019} indicate interesting effects due to stellar spin, such as steeper late time 
fallback rates \citep{Golightly2019b}, or failed disruption events \citep{Sacchi2019}. A comprehensive analysis of 
spinning stars undergoing TDEs in the background of Kerr IMBHs has recently been performed in \cite{Garain2} where the effects
of ``coupling'' between BH and stellar spins was elucidated upon. 

Having recalled a few facts about TDEs involving solitary stars, we now come to the issue of TDEs involving stellar binaries. As we have
mentioned, a natural question to ask is how TDEs affect binary stars. 
In a series of recent papers \citep{Maeda1, Maeda2, Maeda3} have studied the dynamics of a binary star system in the background of
a SMBH in this context. Their method consists of setting up a locally inertial frame along the geodesic trajectory of the center of mass of the system 
using the Fermi-Walker transport formalism, and solving the resulting equations of motion in this frame. \cite{Manzaneda} has 
studied the role of relativistic effects in the process of separation of the two stars in a binary system near a SBH due to tidal effects, considering
this as a three-body problem. In the same spirit, 
\cite{YuLai1} have performed a comprehensive analysis of binary stars that approach SMBHs, using numerical three-body scattering. 

It is then natural to study TDEs in a scenario in which the hydrodynamics of the stars become relevant. This is important and interesting,
as the deformation and disruption of stars may crucially affect the physics of the three-body problem compared to the situation in which 
these are considered in a point particle approximation. Analytical methods become less relevant here, and one has to resort to 
numerical simulations. Indeed, in a followup
work, \cite{YuLai2} extended their previous results by considering hydrodynamics of the component stars (considered as polytropes), 
and studying various possibilities of stellar collision due to the resulting gravitational forces in a SMBH background, using smoothed 
particle hydrodynamics (SPH) (a related older work appears in \cite{Antonini}; to our knowledge, some of the first SPH simulations of binary 
stellar systems with polytropic equations of state were reported in \cite{RasioShapiro1, RasioShapiro2}). 

%%%%%%%%%%%%%%%%%%%%%%%%%%
\begin{figure}[H]
	\epsscale{0.7}
	\plotone{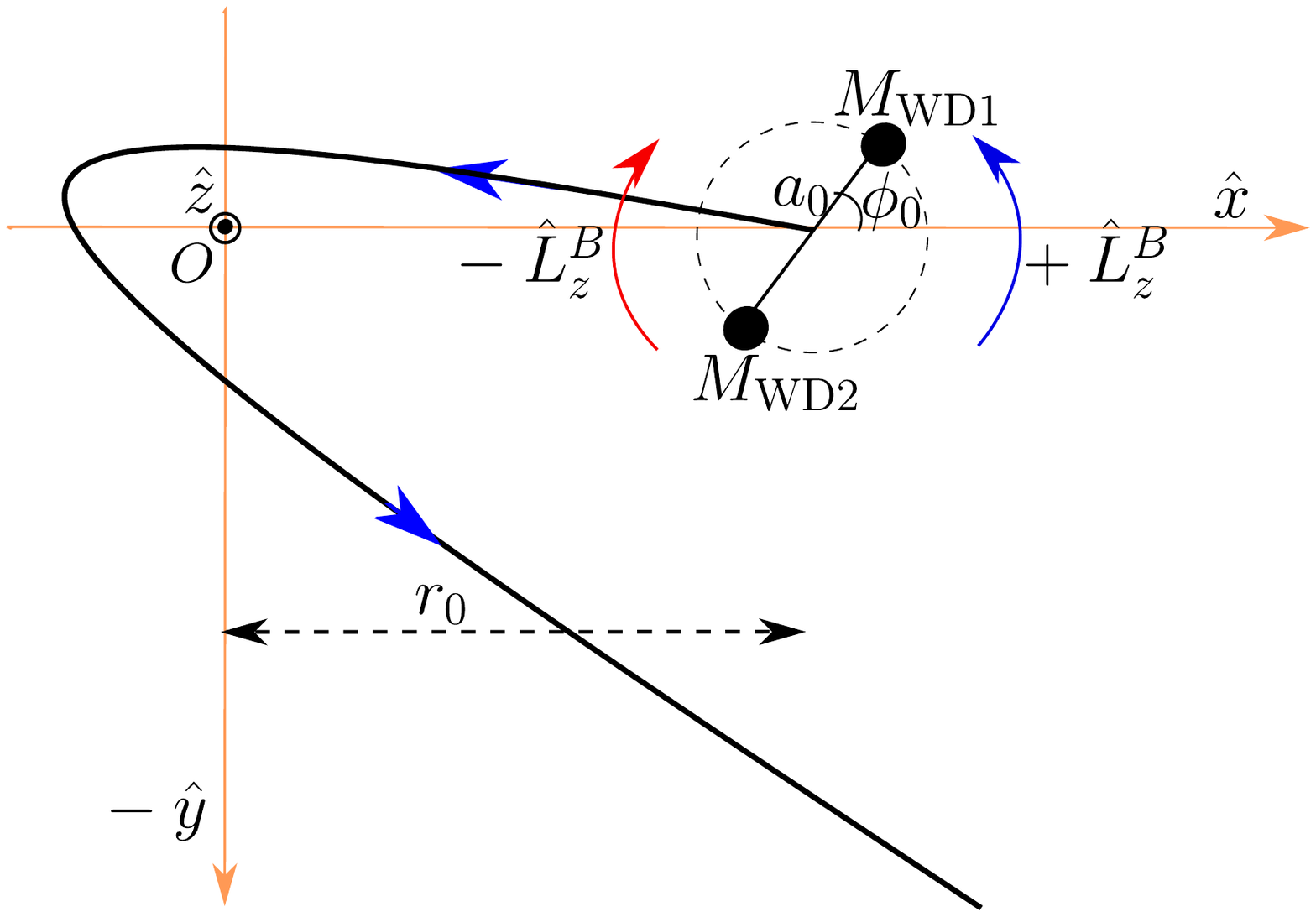}
	\caption{{\small Qualitative depiction of a binary system on a parabolic orbit around a BH located at the origin $O$. The motion lies in the $x–y$ plane, with the $z$-direction pointing out of the page. The binary centre of mass is initially placed at a distance $r_0$ along the $x$-axis. The initial binary separation is $a_0$ with phase $\phi_0$. The binary’s internal orbital motion is prograde or retrograde depending on whether its initial angular momentum about the centre of mass is aligned with ($+\hat{L}^B_z$) or opposite to ($-\hat{L}^B_z$) the orbital angular momentum.}}
	\label{img}
\end{figure}
%%%%%%%%%%%%%%%%%%%%%%%%%%

In this paper, we use SPH to perform numerical analyses of TDEs involving WD binaries
that are in parabolic orbits in a Schwarzschild IMBH background. The binary components are considered to be in the equatorial plane without
loss of generality, and the situation is qualitatively depicted in Figure \ref{img}. The numerical algorithm used in this study has been developed
by us \citep{Banerjee2023, Garain2023a, Garain1, Garain2} and extensively tested against many available results that often use the popular publicly available
code PHANTOM \citep{2018PASA} which inspired an initial version of our code. With our choice of stellar and binary parameters (mentioned in sequel), 
the pericenter distance of the center of mass of the system is $r_p \sim 25 r_g$, with $r_g = G M_{\rm BH} / c^2$ denoting 
the gravitational radius of the BH, so that the stellar hydrodynamics can be
assumed to be non-relativistic \citep{Stone2019}. Here, $G$ is Newton's gravitational constant and $c$ is the speed of light. 
Our analysis here, that uses a hybrid formalism that integrates Schwarzschild geodesics along with a Newtonian treatment of self-gravity of the
fluid, is then expected to be sufficiently robust.

There are a few key differences between the methodology 
considered here, as compared to the one in \cite{YuLai2}. First, as we have already mentioned, we will consider here close WD binaries in the background
of Schwarzschild IMBHs,  whose center of mass evolves in a parabolic trajectory. Close binaries were first considered by \cite{Hills1} to study 
TDEs from SMBHs, resulting in possible hypervelocity stars (HVSs). 
Secondly, we follow \cite{Sari} and study the evolution of our binary systems with different initial phases (the angle $\phi_0$ in
Figure \ref{img}). We find here that there is significant dependence of our results on this phase, which is natural, as the 
stellar positional configuration near 
periapsis is expected to be a strong function of the initial configuration of the system. Thirdly, we consider both pro and retrograde orbits, 
with the former (latter) denoting the cases when the angular momentum of the binary about its center of mass
is aligned towards (opposite) to the angular momentum of the center of mass about the BH, the latter always being taken along
the positive $z$ direction.

There are a few immediate consequences that result when we consider numerical simulations of binaries as compared to solitary stars. 
First, we note that as compared to solitary stars, the tidal radius for a binary is defined by \citep{Sari}
\begin{eqnarray}\label{r_t}
r_t^b  = R_{\rm sep} \left(\frac{M_{\rm BH}}{M}\right)^{1/3},
\end{eqnarray} 
where we consider
$R_{\rm sep}=a_0/2$ and $a_0$ is the initial separation between the two stars in the binary and $M$ is its total mass, so that the``impact factor'' is
$\beta^b =  r_t^b/r_p$, where $r_p$ is the pericenter distance of the center of mass of the binary. The greater is the impact factor, the deeper is 
the encounter, but it is not guaranteed that both stars will face complete disruption, unlike a TDE involving a solitary star. In fact, as we will elaborate 
below, there can be a variety of scenarios (that can be inferred from \cite{Sari} as well) -- and these result in different behaviour of stellar debris. 
Secondly, the fact that the two stars can be fully and/or partially disrupted, might give rise to fall-back rates that show distinct behaviour 
from those in single-star TDEs. Finally, when one star is fully disrupted and the other only partially so, the gravitational effects of the remaining
core of the latter is expected to play a significant role in the future evolution of the system, given that we are sufficiently away from the IMBH. 
We further note that there can be two distinct types of WD binaries that can in principle undergo TDEs with IMBHs, namely ones with equal/unequal
individual masses. We have performed a suite numerical simulations including both scenarios, and find that the results can be substantially different in the
two cases. 

In the next sections, we elaborate upon these comments above. In section \ref{sec2}, we present the methodology used in this paper. 
Section \ref{sec3} discusses our main results and we conclude this paper with section \ref{sec4} that contains a summary of our main 
findings and some related discussions. Throughout this paper, Greek indices will run from $0$ to $4$ and the Latin ones from
$1$ to $3$, whenever these indicate coordinates. 

\subsection{\textsc{Physical context of the setup}}

\tb{Since our focus is on tidal encounters involving IMBHs, we consider globular clusters as the natural astrophysical environment for such events. Globular clusters are widely regarded as promising sites for hosting IMBHs due to their high stellar densities, which facilitate runaway stellar collisions and repeated mergers of stellar-mass BHs. Several observational studies have suggested possible evidence for IMBHs in globular clusters \citep{IMBH0, 2008ApJ...676.1008N, 2002ApJ...578L..41G}.}

\tb{To probe these systems through TDEs, WDs provide ideal stellar probes, as discussed earlier. In order to explore additional variability in observable properties of TDEs, such as kick velocities and mass fallback rates, we consider tidal encounters involving WD binaries. This scenario is also astrophysically plausible because dynamical interactions in dense cluster environments efficiently form and harden compact binaries, leading to a significant population of close WD binaries \citep{shara, Ivanova}. For example, the population synthesis study by  \cite{Ivanova2} has shown that clusters with masses of $2\times 10^5~M_{\odot}$ can produce $\sim10-20$ double white dwarf (DWD) binaries per Gyr through dynamical interactions, implying the presence of numerous close WD binaries in dense cluster environments.}

\tb{According to the Heggie–Hills law, hard (close) binaries tend to become harder through dynamical encounters, whereas soft (wide) binaries are preferentially disrupted by interactions with other stars \citep{Heggie}. As a result, surviving binaries in globular clusters are expected to be predominantly close systems. These close binaries further evolve toward nearly circular orbits due to tidal dissipation and gravitational-wave emission, both of which efficiently damp orbital eccentricity \citep{Zahn, Hut, Peters}. Dynamical encounters within the cluster can also contribute to binary hardening, producing compact binaries that subsequently circularize. Observational population studies provide further support this behavior. For instance, binaries with orbital periods $\lesssim 5$–$10$ days are observed to exhibit nearly circular orbits due to efficient tidal circularization \citep{Meibom, Kirk}. Even more compact systems, such as ultracompact DWD binaries, have been discovered with orbital periods as short as $\sim5$–$10$ minutes, demonstrating the existence of extremely tight and nearly circular systems driven by gravitational-wave orbital decay \citep{Brown2016, Hermes}. Large surveys of binary stars also show that short-period systems typically exhibit very small eccentricities ($e \lesssim 0.1$), consistent with strong orbital circularization in compact binaries \citep{Raghavan, Price-Whelan, Moe}.}

\tb{In clusters hosting an IMBH, close binaries can also interact with the central BH through dynamical scattering processes. Mass segregation and two-body relaxation drive compact binaries toward the cluster core, where the stellar density is highest and interactions with the IMBH become more likely. Numerical studies of star clusters containing IMBHs suggest that stellar tidal disruption rates can reach $\sim10^{-7}$–$10^{-6},\mathrm{yr^{-1}}$ per cluster, depending on cluster properties and stellar populations \citep{Baumgardt2004}. Given that population synthesis models predict the presence of multiple compact WD binaries in typical globular clusters \citep{Ivanova2}, these systems can occasionally be scattered onto loss-cone orbits that bring them within the tidal radius of the IMBH. Such encounters can lead to binary tidal separation or tidal disruption of one or both WDs, providing a plausible pathway for DWD–IMBH tidal disruption events in dense stellar environments.}

\section{Methodology}
\label{sec2}

In this section, we present our numerical methodology for simulating the tidal disruption of WD binaries by an IMBH. For a detailed description of the numerical methods employed
here, we refer the reader to \cite{Banerjee2023} and avoid repeating the details here for brevity. 
Our simulation of the fluid star uses the SPH technique, which discretizes the fluid star into a set of particles, each possessing density, position, velocity, as
well as thermodynamic properties. To efficiently compute fluid properties and forces on each particle, we implement a binary tree with a tree opening angle parameter set at $\theta_{\rm MAC} = 0.5$. 
Standard SPH artificial viscosity parameters, $\alpha^{\rm {AV}} = 1.0$ and $\beta^{\rm {AV}} = 2.0$ \citep{Hayasaki2013}, are adopted, while the Balsara switch is used to reduce the viscosity in shear flows \citep{Balsara1995}. 
The SPH fluid equations are integrated using the leapfrog integrator, ensuring the exact conservation of energy and angular momentum throughout the simulation. 
In modeling the external gravitational influence on each particle originating due to the Schwarzschild BH, we adopt the same approach as detailed in \cite{Banerjee2023}. In this work, we employ a global time step for the evolution of the system. Here, the numerical simulations are carried over through the following three stages:

\begin{itemize}
	\item In the first stage, individual WDs are constructed in isolation. 
	SPH particles are initially arranged in a close-packed spherical configuration, which is then stretched to reproduce the desired stellar density profile. 
	This target profile, representing a carbon-oxygen (C-O) WD in hydrostatic equilibrium, is obtained following the methodology outlined in \cite{Garain2023}, where the electron degeneracy pressure balances the star's self-gravity. The stretched SPH configuration is then evolved in isolation (i.e., without the presence of the BH or binary companion) until transient density fluctuations subside. 
	The final relaxed model is accepted once the SPH density profile closely matches the desired stellar structure.  The pressure $P$ 
	contributed by degenerate electrons and the density $\rho$ arising from carbon nuclei within the WD are given by
	\tb{\begin{equation}
		P  = K_{P} \Bigl[x_f (1+x_f^2)^{1/2}(2x_f^2/3-1)+ \log_{e}\bigl[x_f+(1+x_f^2)^{1/2}\bigr] \Bigr]~, ~~\rho  =  K_{\rho} x_f^3
	\end{equation}}
	Here, $K_{P} = 1.4218\times10^{24}/(8\pi^2)$  $\rm N~m^{-2}$, $K_{\rho} = 1.9479 \times 10^9$ $\rm kg~m^{-3}$ are constants, and \tb{$x_f$} denotes the dimensionless Fermi momentum, often referred to as the “relativity parameter”. This parameter \tb{$x_f$} relates the pressure and density, effectively defining the equation of state employed in our analysis. \tb{For a shocked WD, the associated temperature and the thermal energy generated by shock heating can be roughly estimated using the Virial theorem \citep{Rosswog2009}:
	\begin{eqnarray}
	T &\sim& 5 \times 10^8 \left(\frac{M_{\rm{WD}}}{0.5 M_{\odot}} \right) \left(\frac{R_{\rm{WD}}}{0.0141R_{\odot}} \right)^{-1} ~\rm{K}\\
	E_{\rm{therm}} &\sim& 7 \times 10^{49}  \left(\frac{M_{\rm{WD}}}{0.5 M_{\odot}} \right)^2 \left(\frac{R_{\rm{WD}}}{0.0141R_{\odot}} \right)^{-1} ~\rm{erg}
	\end{eqnarray}}
	
%	
%	\tb{The total thermal energy of the WD is then associated with its total fermi energy (excluding rest mass energy), $u=\epsilon_k(x_f)$. From Virial equilibrium conditions, we can approximate the related temperature of the relaxed WDs
%	\begin{eqnarray}
%		T \approx 5 \times 10^8 \left(\frac{M_{WD}}{0.5 M_{\odot}} \right) \left(\frac{R_{WD}}{0.0141R_{\odot}} \right)^{-1} ~\rm{K}
%	\end{eqnarray}}
	
	\item In the second stage, two such equilibrated WDs are placed at a specified separation and assigned orbital velocities corresponding to a circular binary orbit. The dynamical stability of the identical binary system is verified by evolving the binary for up to 15 orbital cycles, as shown in Figure~\ref{xtyt}, similar to \cite{2018PASA}. This ensures the absence of significant structural deformation or orbital drift. A similar analysis holds for non-identical 
	binaries, which we do not show here. 
	
	%%%%%%%%%%%%%%%%%%%%%%%%%%%%%%%%%%%%%%%%%%%%%%%%
	\begin{figure}[H]
		\epsscale{1.1}
		\plottwo{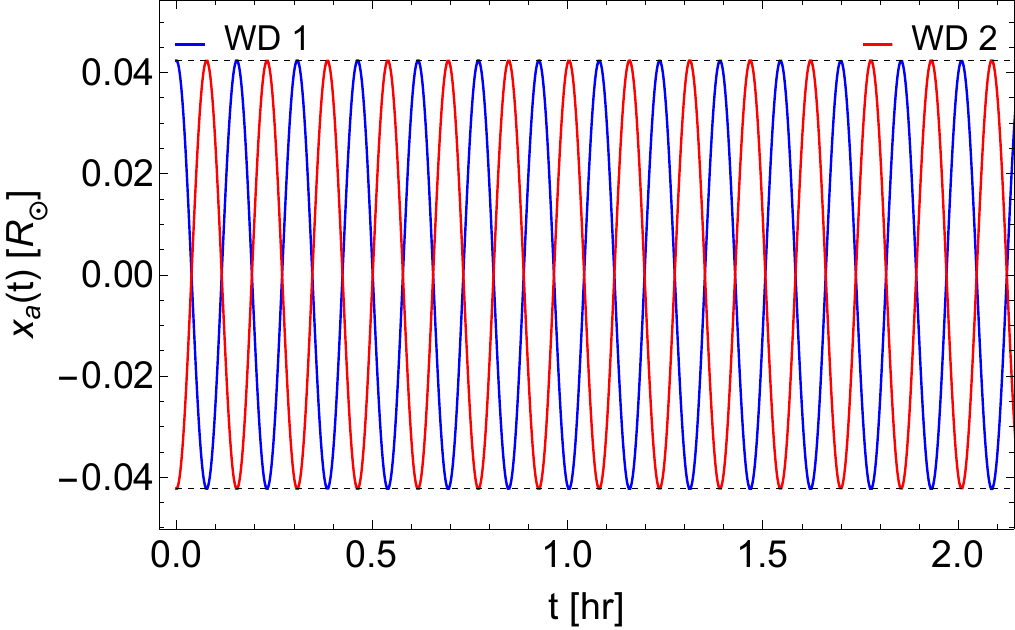}{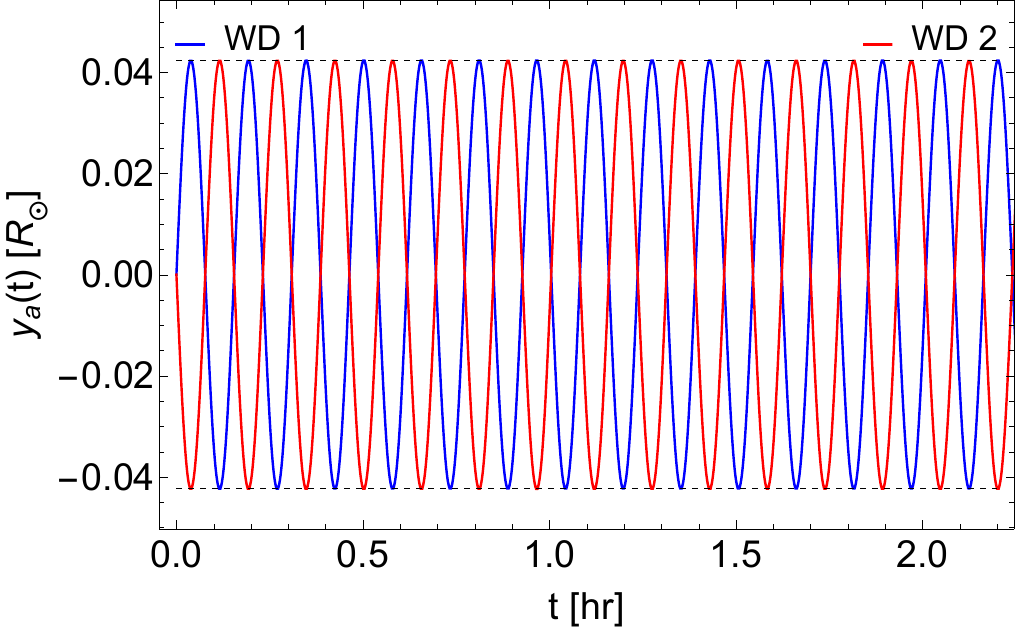}
		\caption{{\small Centre of mass coordinates, $x$ and $y$ of both WDs plotted against $t$ to check the equilibrium of the identical binary system. \textbf{Left Panel: }$x_1(t)$  \& $x_2(t)$ oscillates within binary separation $a_0$ of both WDs. \textbf{Right Panel: }$y_1(t)$  \& $y_2(t)$ oscillates within binary separation $a_0$ of both WDs.}}
		\label{xtyt}
	\end{figure}
	%%%%%%%%%%%%%%%%%%%%%%%%%%%%%%%%%%%%%%%%%%%%%%%%
	
	\item Finally, in the third stage, the stable WD binary is introduced into the gravitational potential of a Schwarzschild BH to investigate the dynamics of tidal disruption.
	To simulate TDEs with the center of mass of the binary in a parabolic orbit, we adopt a hybrid formalism, in which we integrate
	the Schwarzschild geodesics while implementing a Newtonian form of the fluid self-gravity. The geodesic equation of GR written in terms of 
	proper time $\tau$ reads \citep{Misner1973}
	\begin{equation}
		\frac{d^2 x^{\mu}}{d \tau^2} + \Gamma_{\gamma\delta}^{\mu}\frac{d x^{\gamma}}{d \tau}\frac{d x^{\delta}}{d \tau} = 0~,~~
		\Gamma_{\gamma\delta}^{\mu} = \frac{1}{2}g^{\mu\sigma}\left(\partial_{\delta} g_{\gamma\sigma} + \partial_{\gamma} g_{\delta\sigma} - \partial_{\sigma} g_{\gamma\delta}\right)~.
		\label{Eq.geodesic1}
	\end{equation}
	\tb{Here we consider the Schwarzschild spacetime metric $g_{\mu\nu}$. The corresponding Christoffel symbols are denoted by $\Gamma^{\mu}_{\gamma\delta}$.} A standard computation can be used to write the spatial acceleration in terms of the coordinate time $t$ as \citep{Tejeda2017}
	\begin{equation}
		\frac{d^2 x^i}{d t^2}  = -\left(g^{i\sigma}-\dot{x}^i g^{0\sigma}\right)\Bigg( \frac{\partial g_{\gamma\sigma}}{\partial x^{\delta}} -\frac{1}{2}\frac{\partial g_{\gamma\delta}}{\partial x^{\sigma}} \Bigg)\dot{x}^{\gamma}\dot{x}^{\delta}~,
		\label{Eq.geodesic}
	\end{equation}
	where an overdot indicates a derivative with respect to the coordinate time $t$.
	Newtonian self gravity is then implemented locally by adding the term $\sim \nabla\Phi$ with $\Phi$ being the Newtonian
	potential following the Poisson's equation $\nabla^2\Phi=4\pi G\rho$, with $\rho$ being the local density. The contribution of an artificial viscosity is also added, following \cite{2018PASA}. 
	Here, we consider the BH at the origin and the motion of the binary components is restricted to the equatorial plane $\theta = \pi/2$ without loss of generality. 
	Our simulations use Cartesian 
	coordinates, and conversion from the usual spherical coordinates involving the radius $r$ and the azimuthal angle $\varphi$ 
	is implemented following \cite{Rosswog_TR} as 
	\begin{equation}
		r = \sqrt{x^2 + y^2 + z^2}~,~
		r\dot{r} = x \dot{x} + y \dot{y} + z\dot{z}~,~
		r^4 \dot{\varphi}^2 = (x \dot{y} - y \dot{x})^2 +  (x \dot{z} - z \dot{x})^2 +  (z \dot{y} - y \dot{z})^2.   
	\end{equation}
	\end{itemize}

%The stability of our identical binary configuration is validated in Figure~\ref{xtyt}, which shows the evolution of center of mass coordinates 
%$x_1$, $y_1$ and $x_2$, $y_2$ exhibiting stable oscillations,
%i.e., negligible dissipation or drift around their respective mean values and within their amplitude bounds. A similar analysis holds for non-identical 
%binaries, which we do not show here. 

\subsection{\textsc{Parameters}}\label{sec:param}

In our study, we consider an IMBH treated as a Schwarzschild BH situated at the origin. For an identical WD binary, we construct the relaxed WDs, 
each of which has a mass $M_{\rm WD} = 0.5\, M_\odot$ and a radius $R_{\rm WD} = 0.0141\, R_\odot$, as follows 
from the WD mass-radius relation. For non-identical WDs, we use $M_{\rm WD1}=0.25\, M_\odot$, $M_{\rm WD2}=0.5\, M_\odot$ and their corresponding radii are $R_{\rm WD1}=0.019\, R_\odot$, $R_{\rm WD2}=0.0141\, R_\odot$, respectively. Here, we label the WD that is initially on the right of 
the center of mass of the binary as WD1 and the one on the left as WD2 (see Figure \ref{img}). The initial separation of the identical binary is set
to be $6R_{\rm WD} = 0.0846 \, R_\odot$, while that for the non-identical one is initially fixed at $ 0.152 \, R_\odot$, which is $8R_{\rm WD1}$. 
For identical binaries, we take the BH mass $M_{\rm BH} = 8 \times 10^3\, M_\odot$, while for non-identical binaries, 
$M_{\rm BH} = 2 \times 10^4\, M_\odot$. These choices ensure that  $r_p \sim 25r_g$ in all cases that we consider. Further, 
the impact parameter is fixed to $\beta ^b = 2$, so that meaningful comparison can be made between different cases. 
With these choice of parameters, the initial time period of the binary is $\sim 4.1$ minutes for
the identical and $\sim 11.6$ minutes for the non-identical cases, respectively. We note here that as of now, the minimum observed time period of
the WD binary HM Cancri is $\sim 5.4$ minutes \citep{Munday}. 

\tb{Within this tidal interaction regime, we find two typical outcomes for the interaction between the binary system and the BH. In the first case, the binary undergoes an exchange-type scattering interaction with the BH (Hills mechanism). In the second case, the binary remains gravitationally bound and survives the tidal interaction with the BH. We find that these two outcomes can occur for a wide range of initial inclinations between the binary orbital angular momentum (about its center of mass) and the orbital angular momentum of the binary center of mass around the BH. In the present study, we focus on two coplanar configurations that capture these interaction scenarios, prograde and retrograde orientations of the binary motion relative to its orbital motion around the BH. Interestingly, we also observe complete orbital flips in our hierarchical triple configuration, a behavior that has been reported to occur in coplanar configurations in previous studies \citep{Will2017}. Observational studies of hierarchical stellar systems further suggest that many triples evolve toward low mutual inclinations over time, leading to approximately coplanar configurations \citep{Tokovinin}. TDEs involving a generic non-planar configuration (and with the inclusion of BH spin) have been discussed recently by us \citep{MahapatraLatest}.}

%In future, we plan to extend this analysis to non-coplanar configurations and include a spinning BH, which would allow us to investigate the coupling between the binary 
% angular momentum, the orbital angular momentum around the BH, and the BH spin. Such configurations represent a more general dynamical scenario.}

The center of mass of the relaxed WD binary is placed at an initial distance of $r_0 = 10 \, r_t^b$ from the BH. With
$\beta ^b = 2$, the center of mass of the WD binary has pericenter distance $\sim 25 \, r_g$. Finally, we are left with initializing the velocities of the individual WDs. Each WD’s initial velocity components consists of two parts: the circular orbital velocity due to the binary’s motion around its centre of mass, and the velocity of the centre of mass of the binary system as it moves in the gravitational field of the BH. The latter is computed using the constants of motion in the Schwarzschild BH spacetime.

We employ a total of \(6 \times 10^5\) particles, i.e., \(3 \times 10^5\) particles per star to simulate relaxed WD binaries in the case of identical components. For non-identical binaries, we use \(4 \times 10^5\) and \(2 \times 10^5\) particles for the more massive and less massive star, respectively, to ensure equal-mass SPH particles. \tb{In both cases, simulations with a total of \(2 \times 10^5\) particles show converged results, ensuring that the numerical results are independent of SPH particle resolution. We further verify convergence by performing a higher-resolution run with \(10^6\) SPH particles for one configuration (see appendix \ref{appendix}).}

\subsection{\textsc{Remarks}}
\tb{ \noindent{\emph {Tidal compression: }} As our aim is to study the evolution of WDs in the background of IMBHs, the WDs can experience extreme tidal compression at the pericentre. In SPH, these effects can be captured through the evolution of the WD's thermal energy via the tidal heating rate, $\dot{E}_{\rm heat}\propto\rm{d}u/\rm{d}t$. For a compact object like a WD, this can significantly affect the stellar dynamics due to the intense tidal forces exerted by the BH. In many instances, extreme tidal heating can trigger a thermonuclear runaway, often referred to as a `{\it tidal novae}' \citep{Vick2017, Fuller, Xuan}. }

\tb{To estimate these effects in our current setup, we evaluate the characteristic timescales at which they become relevant. Based on the present parameters (Section~\ref{sec:param}), the orbital dynamical timescale of the tidal encounter can be approximated by the WD dynamical timescale,
$\tau_{\rm dyn}=\sqrt{R_{\rm WD}^3/G M_{\rm WD}}$.
For the selected WD configurations, the dynamical times are $\tau_{\rm dyn}\sim3.8~{\rm s}$ and $8.3~{\rm s}$, with their corresponding effective single-star impact factors $\beta\sim0.84$ and $0.77$, respectively.	
As we consider C-O type WDs, the initial nuclear burning timescale can be estimated from the carbon ignition rate \citep{Woosely2004}
\begin{eqnarray}
	\tau_b = \frac{C_p T}{\dot{S}_{\rm nuc}}\approx 15 \left(\frac{7}{T_8}\right)^{22} \left(\frac{2}{\rho_9}\right)^{3.3} \rm{s},
\end{eqnarray}
where $C_p$ is the specific heat at constant pressure, $\dot{S}{\rm nuc}$ is the energy generation rate from the carbon fusion reaction, and $T_8$, $\rho_9$ are defined as $T/10^8~{\rm K}$ and $\rho/10^9~{\rm g~cm^{-3}}$ respectively. Furthermore, this nuclear fusion timescale can be significantly affected as the WD experiences tidal compression in the BH background. Its variation can be predicted from the TDE parameters, particularly through the impact factor $\beta$. \cite{CarterLuminetb} provides an analytical treatment of tidal compression that leads to scaling relations in $\beta$ for the maximum values of temperature, $T_m$, and density, $\rho_m$. 
When the stellar object passes through pericentre, the maximum values of these quantities can be approximated in terms of their initial values as
\begin{eqnarray}
	\rho_m \approx \beta^3 \rho, \quad T_m \approx \beta^2 T.
\end{eqnarray}}
\tb{With these scaling relations and for the values of $\beta\sim0.84,~0.77$, we estimate the burning times to be $\tau_b\sim10^{21}$ s and $\tau_b\sim10^{35}$ s, respectively. 
Therefore, in the present dynamical regime, $\tau_b \gg \tau_{\rm dyn}$ and the contribution from nuclear burning is negligible. If we move to the deep encounter regime where $\beta\sim2$ or larger, higher peak temperatures are expected in the tidally compressed WDs. In that case, the burning time may reduce to $\tau_b\sim1$ s (or when $\tau_b \ll \tau_{\rm dyn}$), allowing appreciable nuclear fusion and potentially triggering runaway nuclear detonation. 
Consequently, the tidal disruption outcome of the WD may be significantly altered, as demonstrated in \cite{Rosswog2009,Rosswog2008}. 
For instance, a recent study by \cite{Vynatheya} shows that close encounters (higher $\beta$) can produce TDE scenarios accompanied by runaway nuclear burning, where a large fraction of the disrupted material becomes unbound from the BH. 
In contrast, wider encounters (lower $\beta$) lead to the standard TDE outcome. Hence, we can safely neglect the nuclear effects of tidal compression as we restrict our analysis to the present parameter regime.}\\

\noindent{\emph {Mass-fallback rate calculations: }}When the WD binary approaches the BH, the bound WD system can tidally break apart, with some portion of the WD’s material being disrupted. After the interaction, once the disrupted WD binary recedes from the BH and reaches a significant distance, we expand the accretion radius to $3 r_t^b \simeq 150 r_g$. 
We calculate the rate of fallback of debris when it reaches $3r_t^b$. 
To calculate the peak fallback rate directly from our simulations, we follow the methods outlined in \cite{Coughlin2015}, \cite{Golightly2019a}, \cite{Milesetal}, \cite{Garain2023}. 

However, since most of our simulations include hydrodynamical interactions between core(s) and the surrounding debris, we chose not to evolve the system using sink particles. As a result, the simulations become computationally expensive, making it difficult to track the late-time behavior of the fallback rate using the aforementioned method. Therefore, we compute the fallback rates at the latest instant when the distribution of debris differential mass with respect to specific energy, ${\rm d}M/{\rm d}\epsilon$, can be considered ‘frozen-in’ in our simulations. At this instant, we use the specific energies, $\epsilon$, and the specific orbital angular momenta, $l$, of all available bound debris elements to calculate their respective half-orbital periods (as defined in \cite{Rosswog_TR}), by numerically integrating the corresponding elliptic integrals. We then calculate the distribution of $\dot{M} \equiv {\rm d}M/{\rm d}t$ as a function of these orbital return times, yielding the desired fallback rate profile.

Recall that following the partial disruption of a solitary WD, it may retain a remnant mass due to its own self-gravity, forming a self-bound core. In the binary case, there can be scenarios where the self-bound cores merge, or where bound cores accumulate additional disrupted material from the surroundings, resulting in a mutual self-bound core. To identify these core particles, we employ an energy-based iterative method as discussed in more details in section~\ref{core_mass}.

\subsection{\textsc{Comparative Dynamics in the Three-Body Problem}}\label{point_particle}
In this section, we present a brief comparison of the dynamics of a three-body system involving our SPH modeled binary WD configuration and a binary point particle system. 
We emphasize a few key aspects that are qualitatively different from other works, e.g., that of \cite{Sari}, \cite{Sari2}, \cite{Manzaneda} highlighting the effects
of the finite size of the stars, which were not considered in these works.
The finite size of stars becomes significant when a close binary system evolves in the gravitational field of a BH. 
Furthermore, a proper relativistic treatment of the BH background is necessary to accurately capture encounters in which the binary approaches the IMBH closely. This is particularly important given our chosen configuration of the IMBH and WD binary.
Before addressing these effects, we first quantify the amount of fluctuations associated with our binary system in two different cases: in the absence of a BH background and at a large distance from the BH. 
We also compare these fluctuations in our SPH-modeled identical WD binary system with those of a binary system composed of identical point particles. Here, we keep the initial conditions to be fixed by considering only $\phi_0 = 0$.

In order to study the binary system treated as point-like particles, we can solve a binary system of bound test particles evolving around a BH of mass $M$ at a large distance, using the standard Newtonian equations of motion:
\begin{eqnarray}\label{3bodyNewt}
\ddot{\mathbf{x}}_a = - \frac{G M}{|\mathbf{x}_a|^3} \mathbf{x}_a - \frac{G m_b}{|\mathbf{x}_a - \mathbf{x}_b|^3} (\mathbf{x}_a - \mathbf{x}_b),
\end{eqnarray}
where $a,b$ refers to star 1 or star 2, $\mathbf{x}$ is the position vector and $m$ is the mass. Following the analysis of \cite{Sari}, the differential equation for the binary separation $r(t) = |\mathbf{x}_1 - \mathbf{x}_2|$ can be solved under the condition 
that the binary system orbits the BH on a parabolic trajectory.

Near the BH, the equations of motion take the form:
\begin{eqnarray}\label{hrna}
\ddot{x}^i_a = - \left( g^{i\lambda} - g^{0\lambda} \dot{x}^i \right)_a \left[ \left( \frac{\partial g_{\mu\lambda}}{\partial x^\nu} - \frac{1}{2} \frac{\partial g_{\mu\nu}}{\partial x^{\lambda}} \right) \dot{x}^{\mu} \dot{x}^{\nu} \right]_a - \frac{G m_b}{|x^i_a - x^i_b|^3} (x^i_a - x^i_b)~.
\end{eqnarray}
Here, the first term describes the motion of test-particle in curved spacetime and the second term is the Newtonian gravitational interaction.
%%%%%%%%%%%%%%%%%%%%%%%%%%%%%%%%%%%%%%%%%%%%%%%%
\begin{figure}[H]
	\epsscale{1.1}
    \plottwo{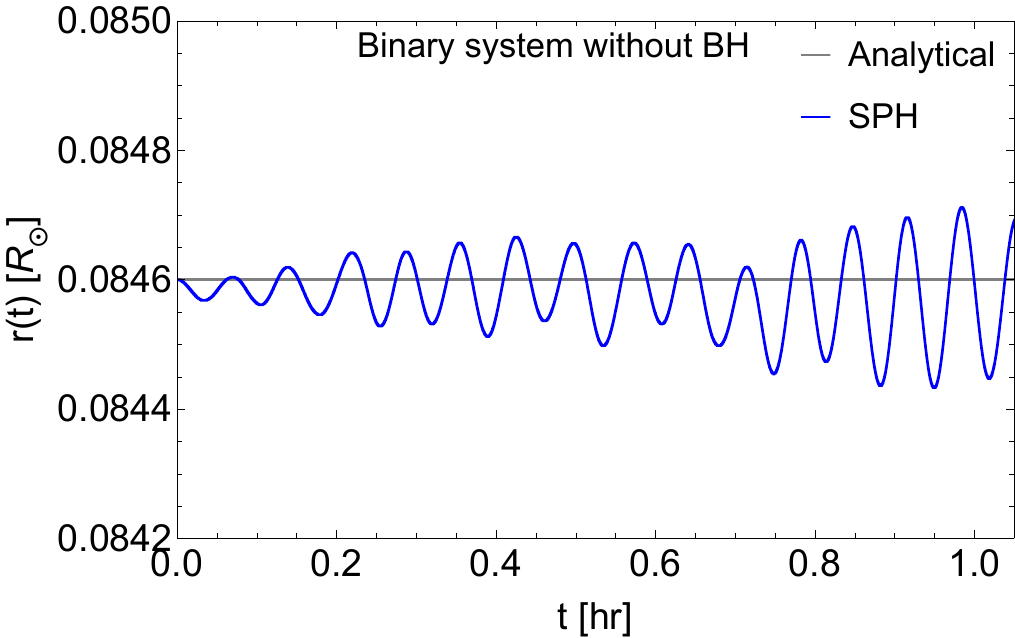}{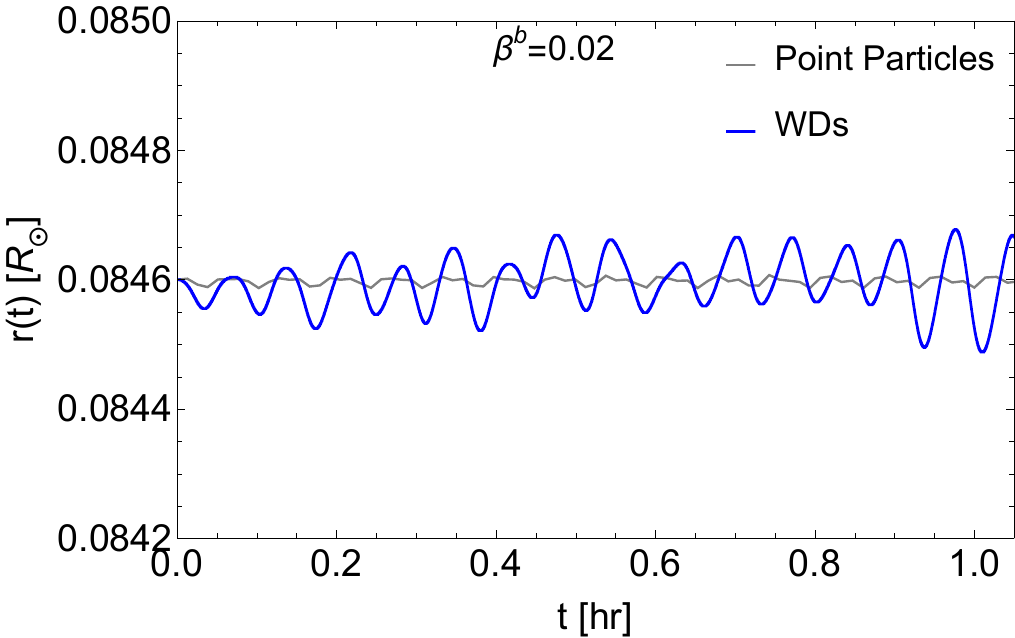}
    \caption{{\small Separation $r(t) = |\mathbf{x}_1 - \mathbf{x}_2|$ plot to check the stability of binary system. \textbf{Left Panel:} shows the change in binary separation for an isolated oscillating binary system (without any BH background). The blue line represents the SPH evolution of fluctuations in the binary system WDs. The gray line indicates the analytical value of initial separation of binary. \textbf{Right Panel: } shows the binary separation for an oscillating binary system, moving in a parabolic orbit around BH at a large distance. The blue line indicates the SPH simulation of WDs and the gray is for the binary test particles separation.}}
    \label{sep_nbh}
\end{figure}
%%%%%%%%%%%%%%%%%%%%%%%%%%%%%%%%%%%%%%%%%%%%%%%%
In Figure \ref{sep_nbh} Left Panel, we show the fluctuation in binary separation observed in SPH numerical simulations over approximately fifteen orbital cycles of the binary system in the absence of a BH background. The initial separation of the binary is set to $r(t_0) = a_0 = 6 R_{\text{WD}}$, which corresponds to an analytical value of $0.0846 R_{\odot}$. By setting the BH mass to zero in equation \ref{3bodyNewt}, one can show this analytical separation for the binary system of point particles. The SPH numerical result show that the separation fluctuates around this value, with a maximum relative deviation remaining within $1\%$.
In the Right Panel of this figure, we show a similar scenario but with the binary system placed in the presence of a BH background at a large distance. Here, the 
impact parameter is taken as $\beta^b = 0.02$, corresponding to a pericenter distance of approximately $2500\, r_g$. In this regime, the tidal effect of the BH on the length scale of binary separation is negligible. As a result, the relative fluctuations in the SPH simulations again remain within $1\%$, and are even smaller for the point-particle case, which is solved using equation~\ref{3bodyNewt} and the method described in \cite{Sari}.

The stars used in these results are made of $10^5$ SPH particles each or a total of $2 \times 10^5$ particles for the binary system. Increasing this particle resolution, these numerical fluctuations go down further. Henceforth in our simulations, we will use the higher particle resolutions detailed before.

\begin{figure}[h]
	\epsscale{1.1}
	\plottwo{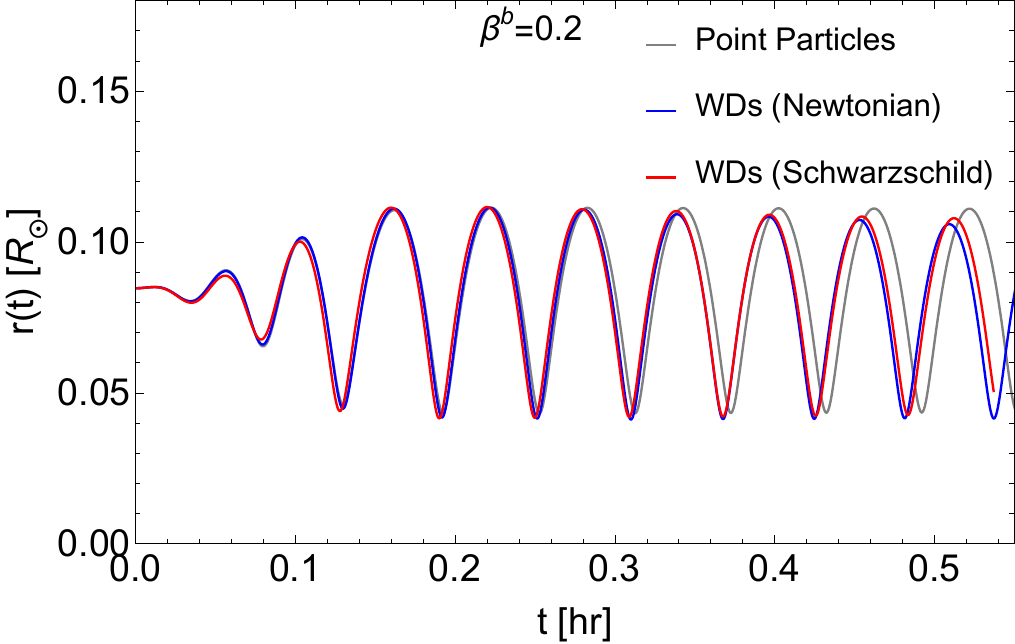}{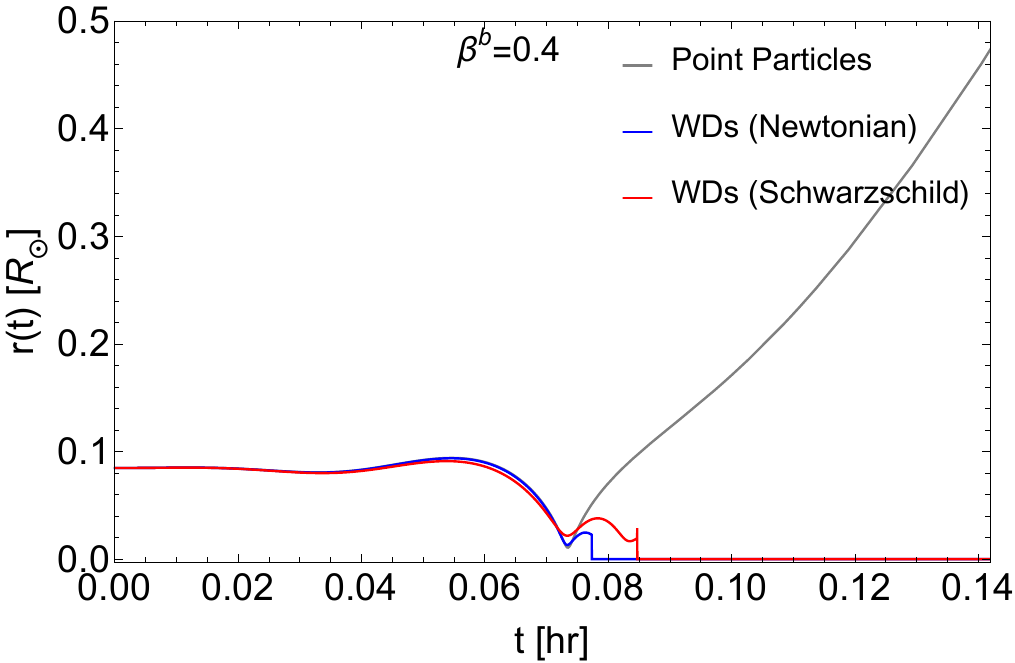}
	\caption{{\small Separation $r(t) = |\mathbf{x}_1 - \mathbf{x}_2|$ plot for values of $\beta^b < 0.5$ for initial phase, $\phi_0$. The gray line represents the binary test particle separation and blue and red line denotes the SPH WD binary for Newtonian and Schwarzschild case respectively. \textbf{Left Panel: }shows binary separation for $\beta^b = 0.2$ corresponding to a pericentre distance of $r_p \sim 250 r_g$.  \textbf{Right Panel: }shows binary separation for $\beta^b = 0.4$ corresponding to a pericentre distance of $r_p \sim 125 r_g$. }}
	\label{sep_b0p2}
\end{figure}

According to \cite{Manzaneda}, when a WD binary system encounters a SMBH, tidal separation of the binary is expected for encounters with impact factors $\beta^b \gtrsim 1$. Similarly, in our case of a WD binary interacting with an IMBH, we find tidal separation to occur for $\beta^b \gtrsim 0.2$, according to the value of $r_t^b$ calculated from equation \ref{r_t}. 
In the Left Panel of Figure \ref{sep_b0p2}, the binary separation is shown for $\beta^b = 0.2$, which corresponds to a pericentre distance 
$r_p \sim 250 r_g$. 
The separation exhibits oscillatory behavior in all three scenarios. However, the SPH modeled WDs, both in the Newtonian and Schwarzschild cases, begin to deviate from the test-particle separation later in the evolution. 
This deviation arises from the finite size of the stars: they experience mutual tidal interactions of the other companion, particularly when the separation reaches a minimum and the stars are at the
closest distance from each other. 
These tidal effects cause deviations that become more pronounced at subsequent minima in the separation. 
This process continues, leading to increasingly significant differences between the SPH and test-particle trajectories. In the Right Panel of Figure \ref{sep_b0p2}, 
we present another case, this time with $\beta^b = 0.4$, corresponding to a pericentre distance, $r_p \sim 125 r_g$. 
Initially the separation matches well for both stars and particles. But, after achieving the closest separation, the point particles fly apart to infinity, which is not the same situation for SPH WDs. This is 
because at that closest separation, tidal interactions draw them together to result into a binary merger. This causes the binary separation to approach zero,
as depicted in this figure. 

%%%%%%%%%%%%%%%%%%%%%%%%%%%%%%%%%%%%%%%%%%%%%%%%
%%%%%%%%%%%%%%%%%%%%%%%%%%%%%%%%%%%%%%%%%%%%%%%%

\begin{figure}[htpb]
	\epsscale{1.1}
    \plottwo{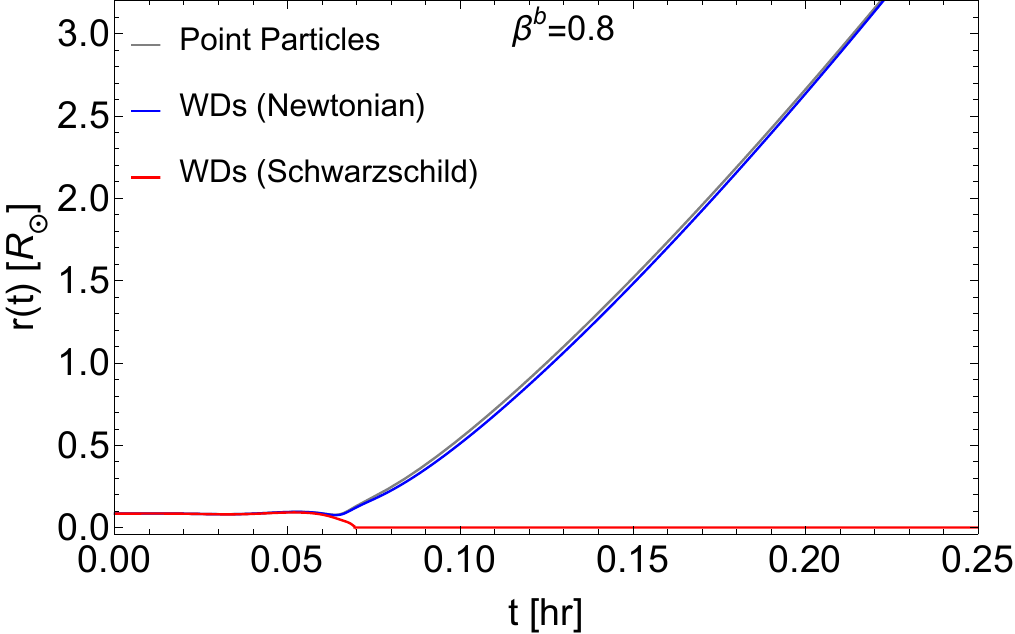}{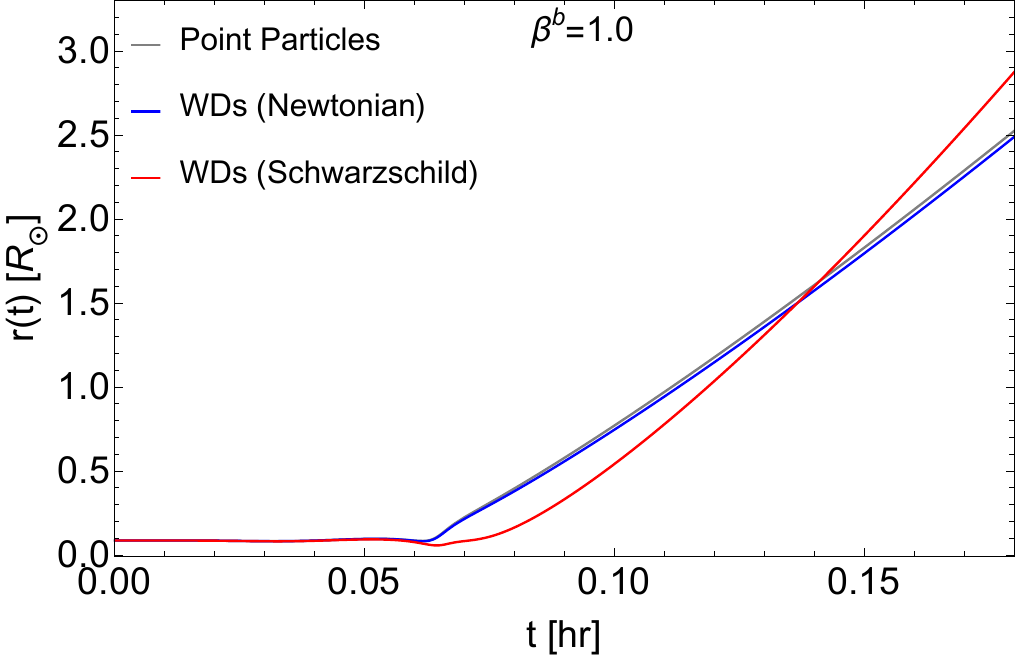}
     \caption{{\small Separation $r(t) = |\mathbf{x}_1 - \mathbf{x}_2|$ plot for values of $\beta^b > 0.5$ for initial phase, $\phi_0$. The gray line represents the binary test particle 
     separation and blue and red line denotes the SPH modelled WD binary for Newtonian and Schwarzschild case respectively. The gray and blue lines are almost visually overlapping. \textbf{Left Panel: }shows 
     binary separation for $\beta^b = 0.8$ corresponding to a pericentre distance of $r_p \sim 62 r_g$.  \textbf{Right Panel: }shows binary 
     separation for $\beta^b = 1.0$ corresponding to a pericentre distance of $r_p \sim 50 r_g$. }}
    \label{sep_b0p8}
\end{figure}

Up to this point, we observe that the behavior exhibited by the SPH binary is largely similar in both Newtonian and Schwarzschild backgrounds. 
%This suggests that it is generally safe to work within either gravitational background, as long as we remain in the regime where $\beta^b \lesssim 0.5$, or equivalently, when the pericenter distance is greater than approximately $100 r_g$. 
\tb{This suggests that, if we fix the present parameter domain, i.e., for the adopted BH mass, binary separation, and the masses and radii of the WDs, it is generally safe to work within either gravitational background, provided that $\beta^b \lesssim 0.5$, corresponding to pericenter distances larger than $\approx100,r_g$.}
However, at higher values of $\beta^b$, such as in Figure \ref{sep_b0p8}, the SPH binary's behavior in the Schwarzschild case deviates significantly.
%%%%%%%%%%%%%%%%%%%%%%%%%%%%%%%%%%%%%%%%%%%%%%%%
%%%%%%%%%%%%%%%%%%%%%%%%%%%%%%%%%%%%%%%%%%%%%%%%

In the Left Panel of Figure \ref{sep_b0p8}, for $\beta^b=0.8$, we see that the Newtonian SPH and point particle separations evolve similarly, with the binary separation increasing over time. In contrast, the SPH binary in the Schwarzschild background undergoes a merger situation, causing its separation to drop to zero. 
The Right Panel of this figure further illustrates this distinction: the steepness in the increase of binary separation for the Schwarzschild SPH binary contrasts significantly with the Newtonian SPH and point particle cases. Thus, it becomes essential to not ignore the relativistic effect of background geometry. 

To illustrate this further, we compare other dynamical features of binary system due to the BH backgrounds. 
%This will help us  provide a better 
%framework to understand various underlying tidal effects that will arise in our subsequent analysis when we consider stellar hydrodynamics.
We treat the binary system to be a pair of test particles evolving around a non-spinning BH at the regime of our interest, i.e., $r_p \sim 25 r_g$. 
Now, we will make use of equation \ref{hrna}, along with Schwarzschild BH metric components to evolve the binary test particles in the relativistic background.

%%%%%%%%%%%%%%%%%%%%%%%%%%%%%%%%%%%%%%%%%%%%%%%%
\begin{figure}[H]
	\centering
	\begin{minipage}{0.475\textwidth}
		\includegraphics[width=\textwidth]{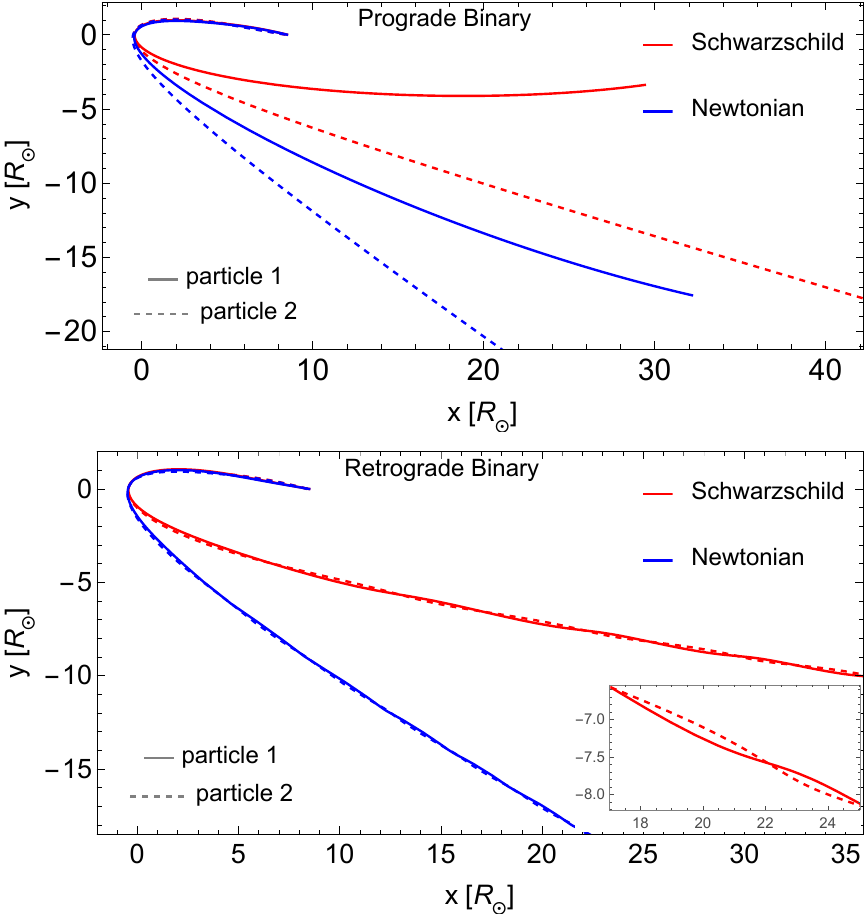}
	\end{minipage}
	\begin{minipage}{0.5\textwidth}
		\includegraphics[width=\textwidth]{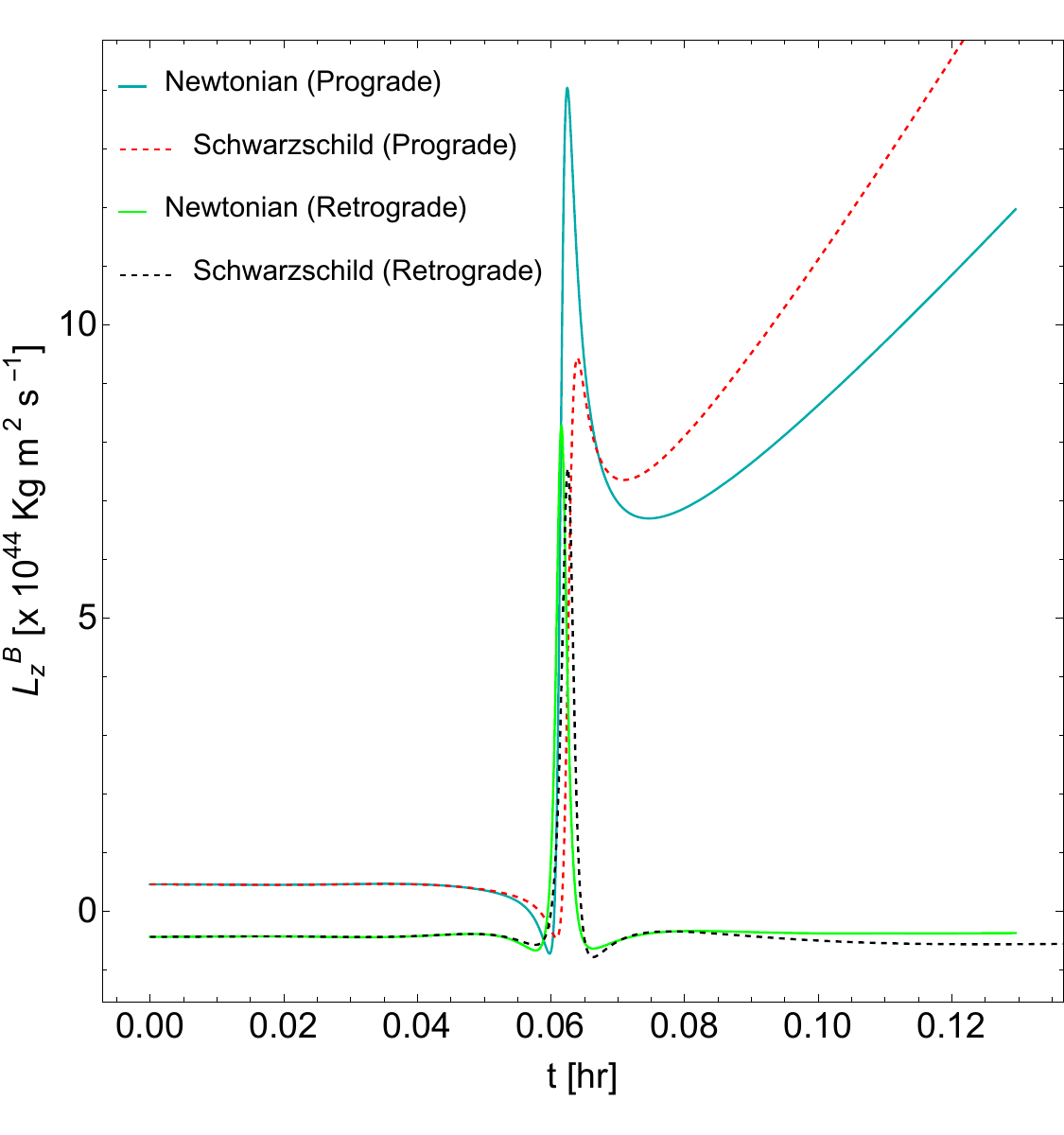}
	\end{minipage}
	\caption{{\small Binary test particle around the IMBH with $\beta^b = 2.0$ for both BH backgrounds for initial phase, $\phi_0=0^\circ$. We show the individual test particle trajectory of binary in the left panel. \textbf{Top Left Panel: }prograde binary where binary's intrinsic orbital angular momentum is along $+z$ axis. \textbf{Bottom Left Panel:} retrograde binary, where binary's intrinsic orbital angular momentum is along $-z$ axis. A zoomed-in view is shown in the bottom corner, illustrating the revolution of the binary particles around each other. \textbf{Right Panel: }shows the evolution of binary's intrinsic orbital angular momentum}}
	\label{trajlz}
\end{figure}
%%%%%%%%%%%%%%%%%%%%%%%%%%%%%%%%%%%%%%%%%%%%%%%%

In the Top Left Panel of Figure \ref{trajlz}, we show the individual test particle trajectories of the binary system for prograde motion. 
Here, the binary system undergoes tidal break-up, i.e., disintegrates from a bound system to two individual particles evolving independently around the 
BH. Typically, in case of actual binary stars, one star is tidally captured by the BH, while the other star is ejected, potentially as a HVS. 
In the Schwarzschild background, the captured star settles into an elliptical orbit with low eccentricity, whereas in the Newtonian case, it follows a highly eccentric orbit. The ejected star in both backgrounds continues along a parabolic trajectory.
This picture changes in retrograde motion, as shown in the Bottom Left Panel of Figure \ref{trajlz}. After passing the pericentre, the binary system does not become unbound. 
We observe that the binary test particle trajectories revolve around each other, indicating that the particles continue to orbit each other with varying binary orbits. This is reflective of the known fact that binaries in retrograde motion are more stable \citep{Sari}. One interesting aspect of these binaries is that, as the particles revolve around each other, their orbital energies oscillate, causing them to become periodically bound and unbound. We will explore this phenomenon further in the future scope of our work.

To complement this picture, we also show the evolution of the binary’s angular momentum before and after the pericenter passage in the Right Panel 
of Figure \ref{trajlz}. In the retrograde case, we observe a sudden increase in angular momentum at the pericenter, effectively causing the binary to become temporarily prograde. As the system moves away from the BH, it gradually restores its original (retrograde) angular momentum state. On the other hand, in the prograde motion, the angular momentum increases further at periastron, resulting in a high centrifugal force that facilitates the separation of the stars. 
\tb{The increasing trend in the angular momentum after the pericentre passage is a consequence of the Hills mechanism \cite{Hills}. In this process, one star becomes bound to the BH while the other is ejected, leading to a continuous increase in the magnitude of the binary’s angular momentum as the separation between the two stars grows progressively larger.}

\section{Results}
\label{sec3}

Now we present our main results. There are some abbreviations that we use in the rest of the paper for brevity. IP and IR refer to the pro and
retrograde systems of identical WD binaries in the sense discussed above, while NIP and NIR refer to these for non-identical 
binaries.\footnote{For unequal-mass cases, we focus only on NIR-type binaries. Since the WDs evolve independently after pericentre passage and binary tidal interactions are almost absent, so only one type of prograde binary case is sufficient to capture the relevant dynamics.}
Our analysis highlights the distinct outcomes observed during the closest approach of the binary’s orbital motion around the BH, which includes stellar disruption, binary tidal breakup, and binary tidal interactions. 
We examine the fate of the disrupted WD material, such as self-disrupting cores, mergers of disrupted cores and companion debris, or the mutual aggregation of tidal debris into multiple small core-like fragments. 
Next, we study observables pertaining to the unique features of tidal disruption in binary WDs, as compared to solitary WDs, such as the accretion rates of tidal debris. 
We also identify signatures of HVSs resulting from both tidal breakup and disruption of each individual WD, using kick velocity estimates. 
Finally, we present results on the generation of gravitational wave (GW) amplitudes from the TDEs of WD binaries.

%%%%%%%%%%%%%%%%%%%%%%%%%%%%%%%%%%%%%%%%%%%%%%%%
\begin{figure}[H]
	\epsscale{1.1}
	\plottwo{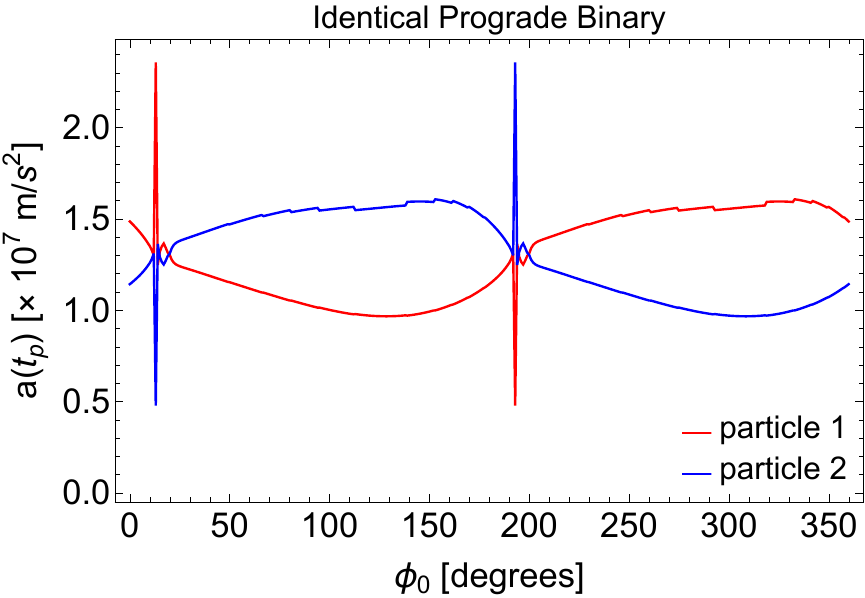}{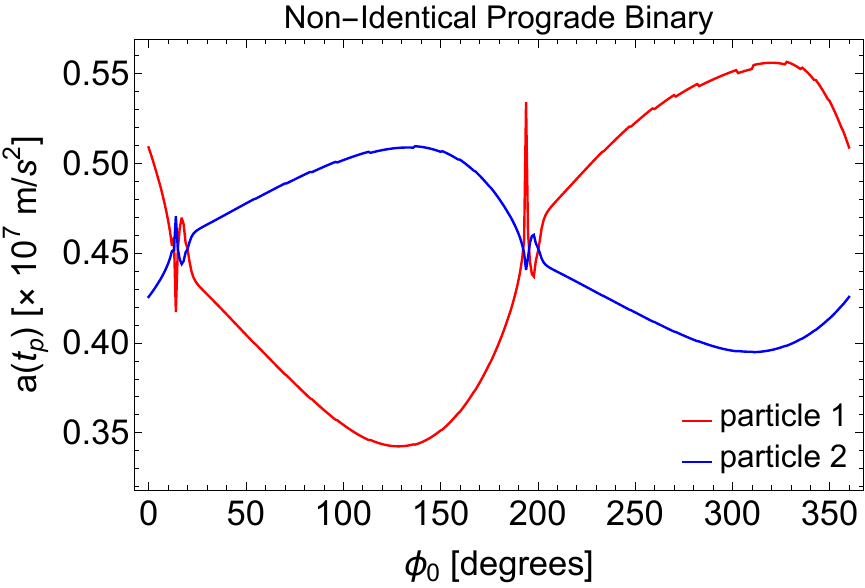}
	\caption{{\small Total magnitude of acceleration experienced by individual WDs at pericentre time $t_p$ for all different $\phi_0$}}
	\label{acc}
\end{figure}
%%%%%%%%%%%%%%%%%%%%%%%%%%%%%%%%%%%%%%%%%%%%%%%%

We mainly focus on the outcomes that arise specifically from the coupling between the binary’s intrinsic orbital angular momentum and its orbital angular momentum around the BH. 
As illustrated in Figure \ref{trajlz}, from the dynamical standpoint of point-like particle approximations, different coupling scenarios can lead to unique outcomes: the binary may break apart, allowing each WD’s disruption to evolve independently, or the disruption dynamics of one WD may be significantly influenced by tidal interactions with its companion. 
In what follows, we will demonstrate the unique signatures associated with prograde and retrograde motion across different choices of the initial phase angle of binary, $\phi_0$. 

The orientation of binary and the relative positions of WDs at periastron are dependent on the initial phase $\phi_0$ (refer to Figure \ref{img}).  
It can also be noted that for a given $\phi_0$, different $r_0$ or $\beta^b$ can still lead to different orientations of the binary at the same pericentre. Our objective is to investigate how different orientations of our binary configuration at the same periastron location, $r_p \sim 25\, r_g$, give rise to diverse tidal interaction outcomes for a fixed parabolic orbit. In our simulations, the only way that we can control these orientations is through parameter $\phi_0$. 
Hence, to understand the binary system’s dynamics at periastron for different values of $\phi_0$, we compute the magnitude of the total acceleration experienced by each WD due to both the BH and its companion.
To study how the acceleration evolves with different initial phase angles, we evaluate the acceleration using equation~\ref{hrna}, treating the binary of WDs as test particles.

In Figure \ref{acc}, for both the IP and NIP binaries, we observe two distinct phase angles, approximately $13^\circ$ and $193^\circ$, at which the acceleration of one particle sharply increases while that of the other sharply decreases. 
Between these two phases, the overall trend in acceleration remains consistent: one of the particles consistently experiences a higher acceleration than its companion. 
This behavior correlates with the behavior found in our SPH simulations, where WD2, which becomes bound in an elliptical orbit, experiences stronger acceleration compared to WD1, which is ejected on a nearly parabolic trajectory. 
As a result, in the IP binary case, WD2 is fully disrupted, while WD1 experiences only partial disruption.

Interestingly, this behavior reverses across either of the two identified phase angles, indicating a switch in the dynamical roles of the two particles. 
This switch is symmetric in the IP binary but not in the NIP binary. We can refer to these critical angles as transition or offset phases, as they mark the boundary between two qualitatively different dynamical outcomes. 
At these phases, the abrupt rise or fall in acceleration primarily results from the binary system’s self-gravity, as the separation between the two particles becomes very small. In the SPH simulation, we find that each WD experiences an overwhelmingly strong tidal interaction from its companion, ultimately leading to a binary merger.
Among all phase configurations of the prograde binary, this transition phase results in the most impactful binary interaction.

Unfortunately, this type of analysis is difficult to perform for retrograde tidal encounters. As previously mentioned, in the retrograde case, there is a high likelihood of the binary being disrupted and, at the same time it can exhibit various kinds of interactions with its companion. We will substantiate this in the following sections.

\subsection{\textsc{Self-bound core mass}}\label{core_mass}
As the WD binary encounters the IMBH, familiar partial disruption scenarios will result. Moreover, along with it, we also uncover  scenarios from our simulations, especially identified in retrograde motions. These include scenarios such as the collision between the disrupted bound core of one WD and the tidal debris of the other, resulting in mass gain by the core. 
In other instances, two distinct bound cores, or a bound core and an intact WD may approach each other and exchange mass, or lead to accretion onto the compact remnant. All of these cases involve one or two self-gravitating bound cores. 

The bound mass of each core can be estimated using the energy-based iterative method described in \cite{Guillochon2013}. While this method is primarily developed for single-star disruptions, it remains applicable to our binary system with appropriate consideration of certain subtleties. To assess whether a particle is gravitationally bound to a particular core, we calculated the specific enthalpy $h_{i,a}$ used in \cite{2021MNRAS.501.1621L} defined for each particle $i$ belonging to WD cores labeled $a$, given by 
\begin{eqnarray}
h_{i,a} = \frac{1}{2}(v_i-v_{{\mathrm{peak}},a})^2+\Phi_{i,a}+u_i+\frac{P_i}{\rho_i}
\end{eqnarray}
In order for a particle to be considered bound to a particular WD core, the following conditions must be satisfied:
\begin{itemize}
	\item If $h_{i,a} < 0$ and $h_{i,b} > 0$~: the particle is bound to WD core $a$.
	\item If $h_{i,a} < 0$, $h_{i,b} < 0$, and $h_{i,a} < h_{i,b}$~: the particle is bound to WD core $a$.
	\item If $h_{i,a} > 0$ and $h_{i,b} > 0$~: the particle is unbound from both WD cores.
\end{itemize}

Here, $\Phi_{i,1}$ and $\Phi_{i,2}$ represent the potential energies of the $i$-th particle due to core 1 and core 2, respectively. Initially, for each core, the particle with the peak density is identified, and its velocity ($v_{{\mathrm{peak}},a}$) is set to the center of mass velocity of the corresponding bound core. 

As the bound particles are identified based on the conditions described above, the velocity $v_{{\mathrm{peak}},a}$ for each core is updated to the center-of-mass velocity of the corresponding bound core:
\begin{eqnarray}
	\mathbf{v}_{{\rm core},a} = \frac{\sum_{i,a} \mathbf{v}_{i} m_{i}}{\sum_{i,a} m_{i}}~,
\end{eqnarray}
where the summation is over all $i$ particles bound to the core $a$. 
Similarly, we can estimate the position 
of the centre of mass, $\mathbf{x}_{{\rm core},a}$. Using this, we can find other properties of the bound core such as specific energy, $\epsilon_{{\rm core},a}$ and specific orbital angular momentum $l_{{\rm core},a}$.

In instances where a disrupted core $a$ gains an $i$-th particle from the debris of its companion core $b$, such particles originating from core $b$ can accrete onto core $a$ over time, leading to a gradual mass gain by core $a$. Furthermore, if the two cores collide and merge, the entire mass of one core is transferred to the other. In some scenarios, a core may be disrupted by its companion, and its disrupted material may subsequently accrete onto the surviving core.
%%%%%%%%%%%%%%%%%%%%%%%%%%%%%%%%%%%%%%%%%%%%%%%%
\begin{figure}[H]
	\epsscale{1.1}
	\plottwo{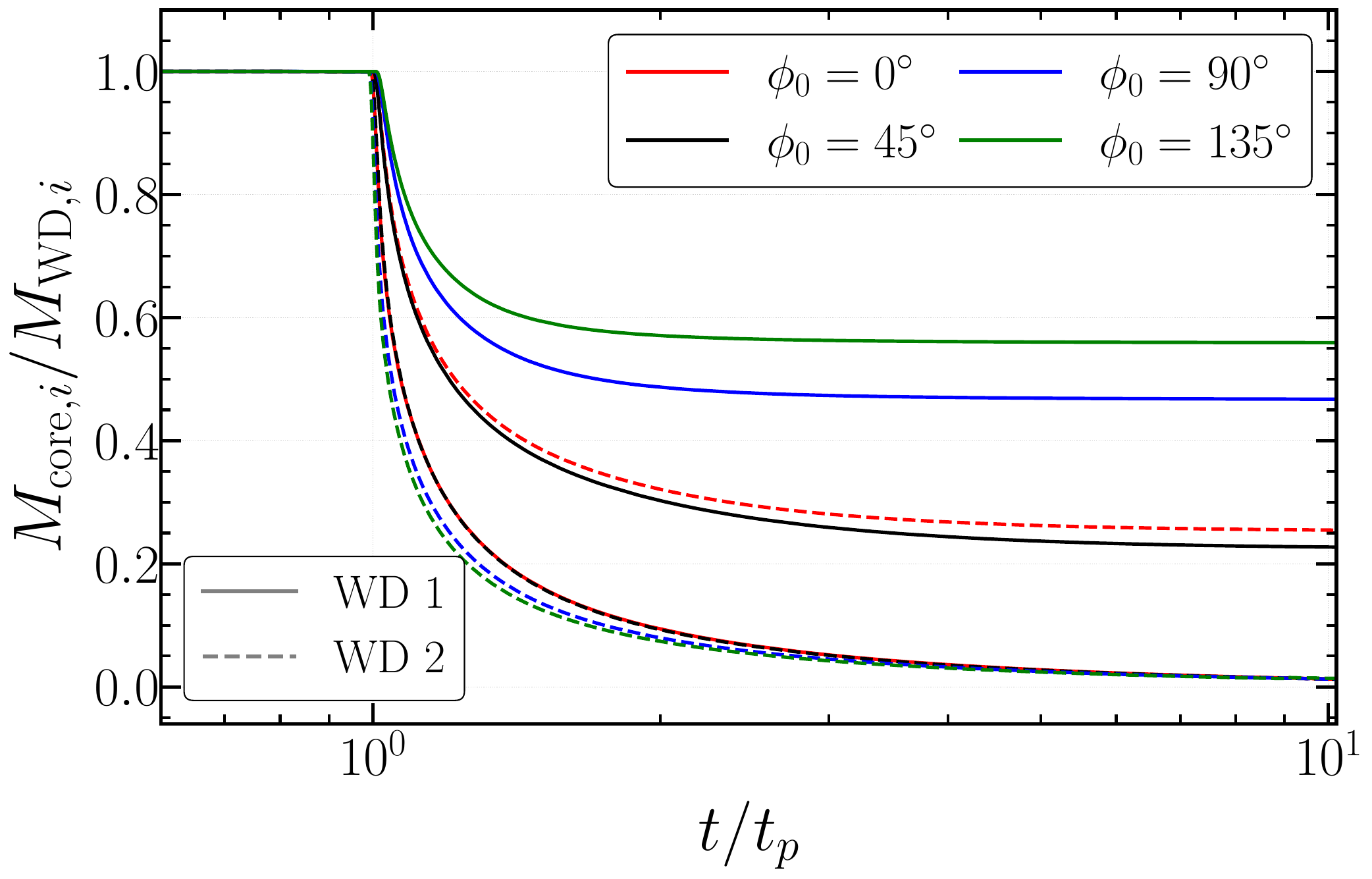}{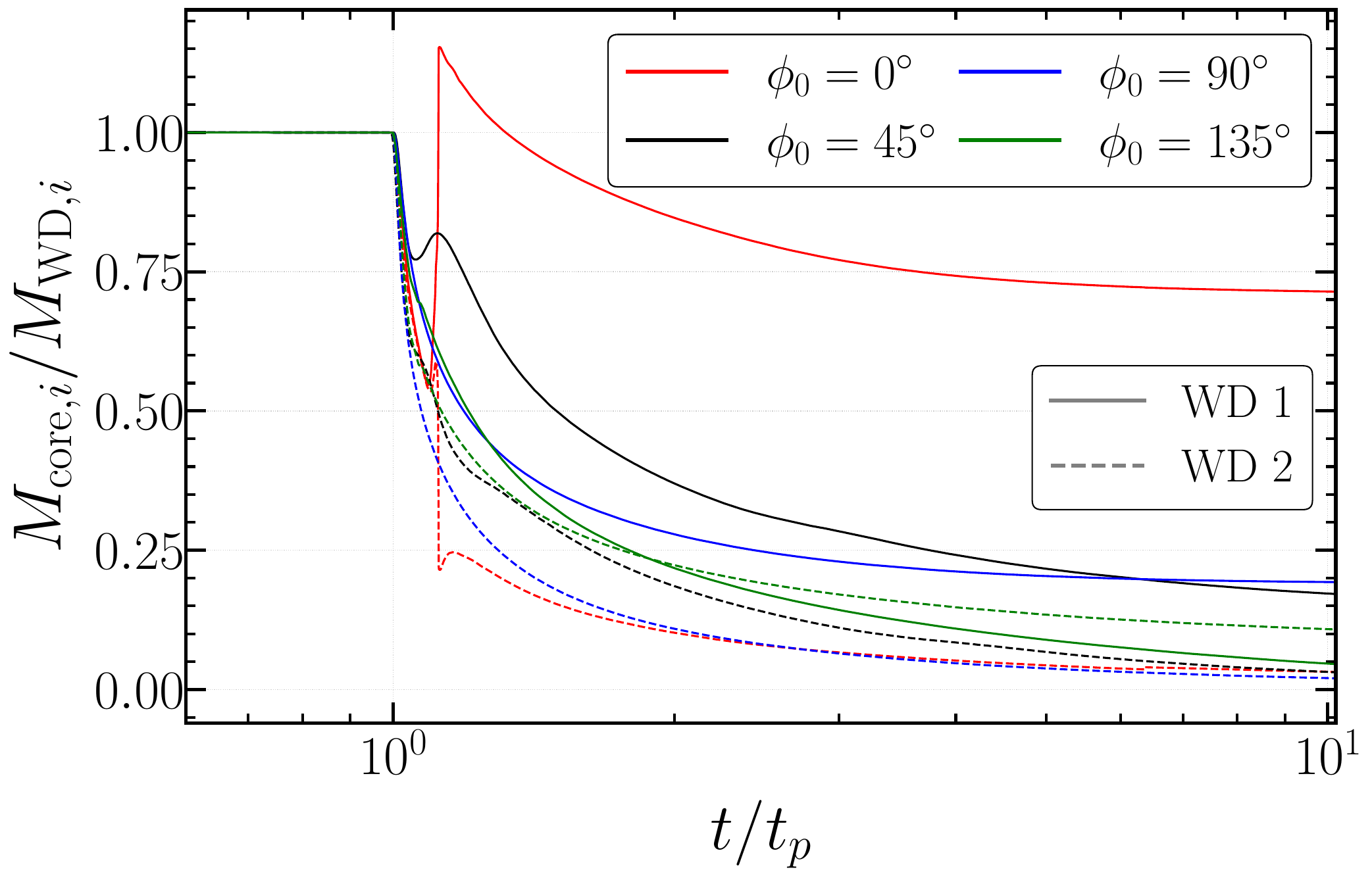}
	\caption{{\small The bound core mass fraction \( M_{{\rm core},i} / M_{{\rm WD},i} \) of each WD is shown as a function of \( t / t_p \), where \( i = 1, 2 \) and \( t_p \approx 0.062\,\mathrm{hr} \) represents the time of pericenter passage. \textbf{Left Panel:} In Identical Prograde (IP) binary, the tidal breakup of the binary leads to independent stellar disruptions of WDs. The WDs that become tidally captured in bound orbits undergo full disruption. \textbf{Right Panel:} In Identical Retrograde (IR) binary, the binary system survives. As a result, the disrupting WD cores dynamically gain or lose mass through interaction with the disrupted material from their respective companions.}}
	\label{mcore_IRIP}
\end{figure}
%%%%%%%%%%%%%%%%%%%%%%%%%%%%%%%%%%%%%%%%%%%%%%%%

%%%%%%%%%%%%%%%%%%%%%%%%%%%%%%%%%%%%%%%%%%%%%%%%
\begin{figure}[H]
	\epsscale{0.6}
	\plotone{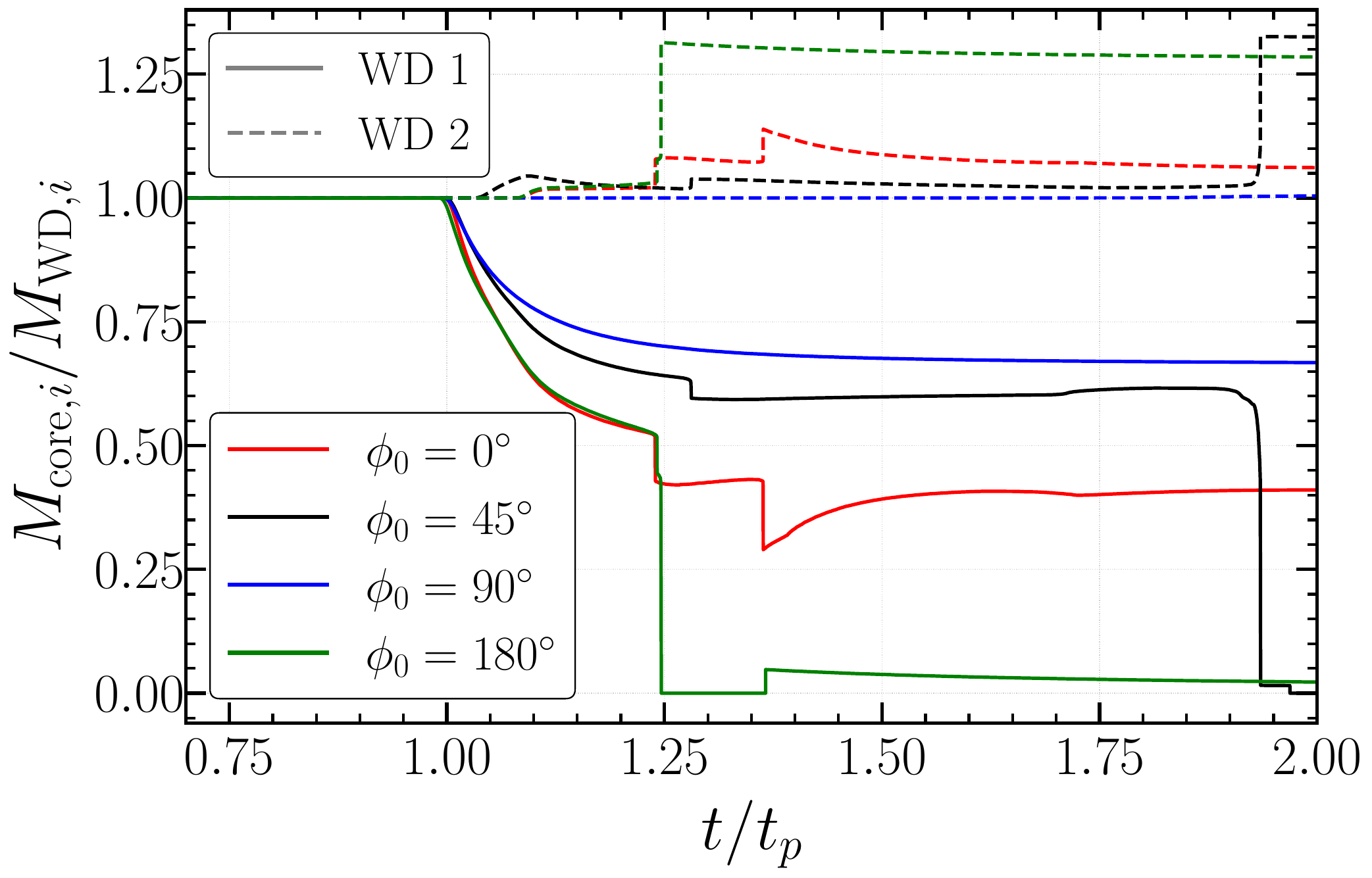}
	\caption{{\small The bound core mass fraction \( M_{{\rm core},i} / M_{{\rm WD},i} \) of each WD in the Non Identical Retrograde (NIR) binary is shown as a function of \( t / t_p \), where \( i = 1, 2 \) and \( t_p \approx 0.17\,\mathrm{hr} \) represents the time of pericenter passage. The disrupting core of WD1 becomes bound to its undisrupted companion, WD2.}}
	\label{mcorenIR} 
\end{figure}
%%%%%%%%%%%%%%%%%%%%%%%%%%%%%%%%%%%%%%%%%%%%%%%%

In the Left Panel of Figure~\ref{mcore_IRIP}, we show the core mass fractions of individual WDs in the disruption of an IP binary. The behavior resembles that of isolated stellar disruptions, as the WDs detach from the binary and evolve independently. 
WDs captured in elliptical orbits undergo full disruption, while the ejected ones experience partial disruptions. Except for $\phi_0 = 0^\circ$, WD2 is typically captured and fully disrupted, while WD1 is ejected and partially disrupted. At $\phi_0 = 0^\circ$, this trend reverses: WD1 is fully disrupted, and WD2 is ejected. 
This reversal is explained by the Left Panel of Figure~\ref{acc}, where WD1 experiences greater acceleration at periastron, where it remains bound for longer period and the behavior shifts to WD2 beyond the transition phase $\phi_0 = 13^\circ$.

%%%%%%%%%%%%%%%%%%%%%%%%%%%%%%%%%%%%%%%%%%%%%%%%%
\begin{figure}[H]
	\epsscale{1.15}
	\plottwo{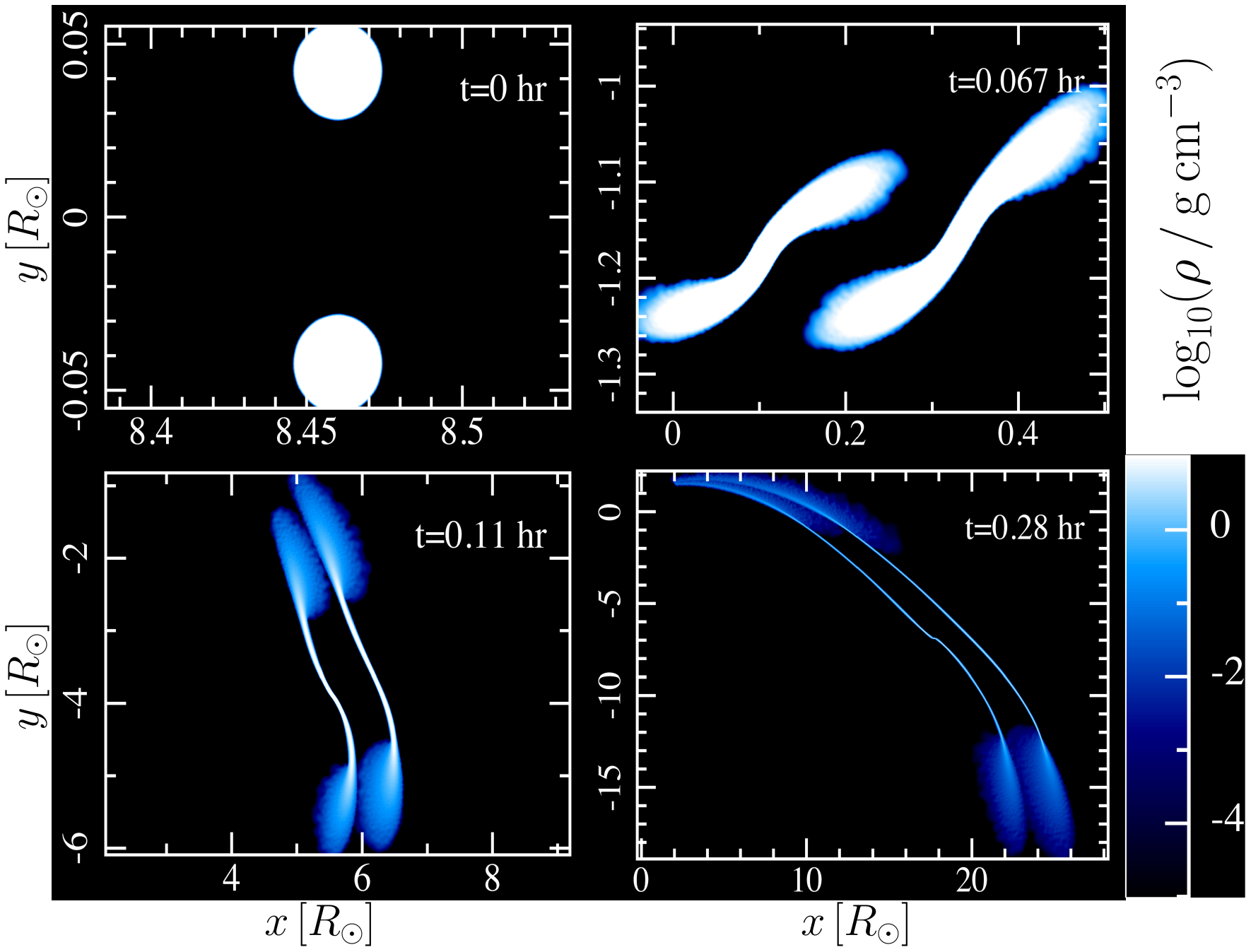}{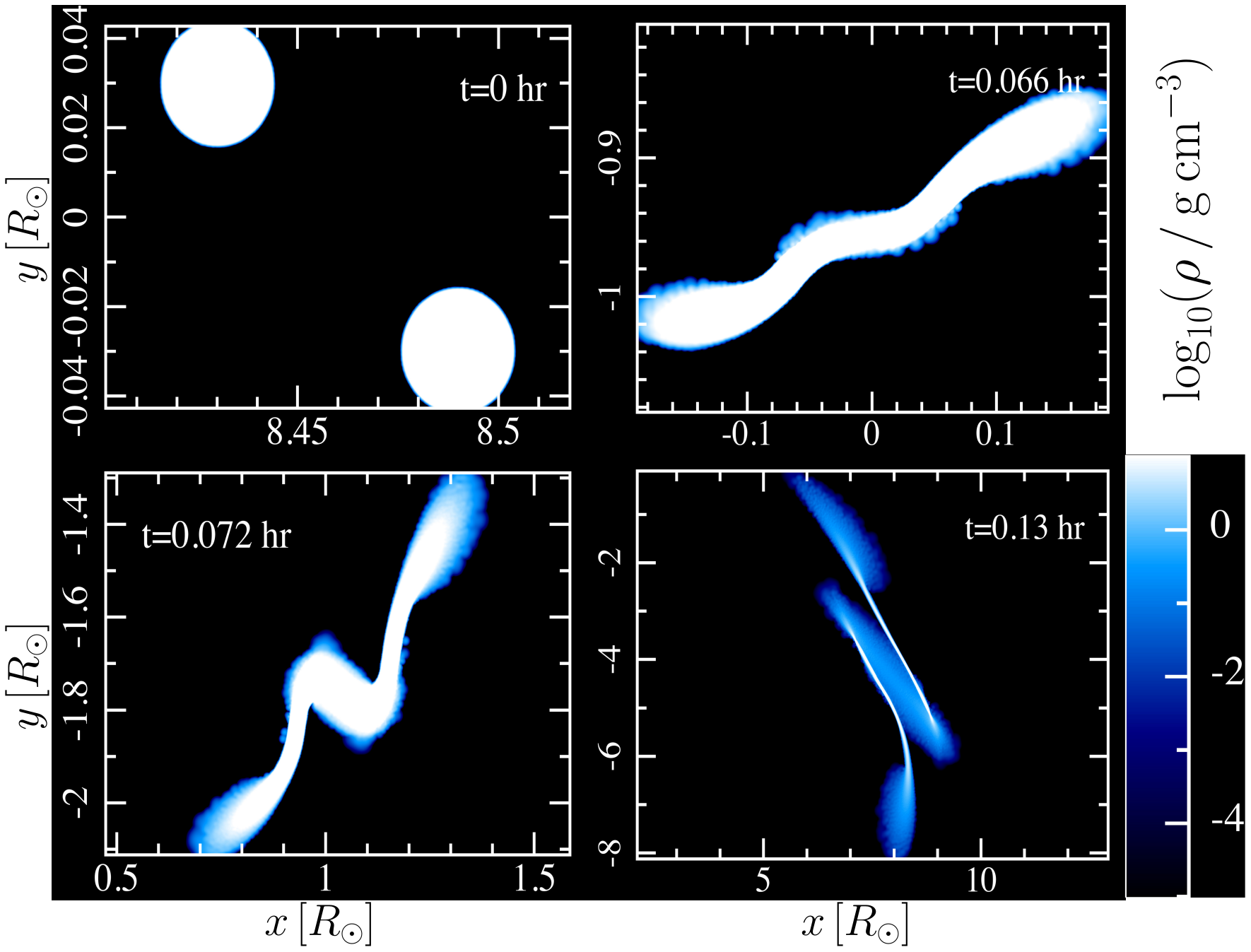}
	\caption{{\small Snapshots for the two different Identical Retrograde binaries (at times mentioned in the figures). The visualizations were produced with the \textsc{SPLASH} software \citep{splash}. \textbf{Left Panel:}  $\phi_0 = 90^{\circ}$ \textbf{Right Panel: } $\phi_0 = 135^{\circ}$}}
	\label{snapIR}
\end{figure}

%%%%%%%%%%%%%%%%%%%%%%%%%%%%%%%%%%%%%%%%%%%%%%%%
%%%%%%%%%%%%%%%%%%%%%%%%%%%%%%%%%%%%%%%%%%%%%%%%
\begin{figure}[H]
	\epsscale{1.15}
	\plottwo{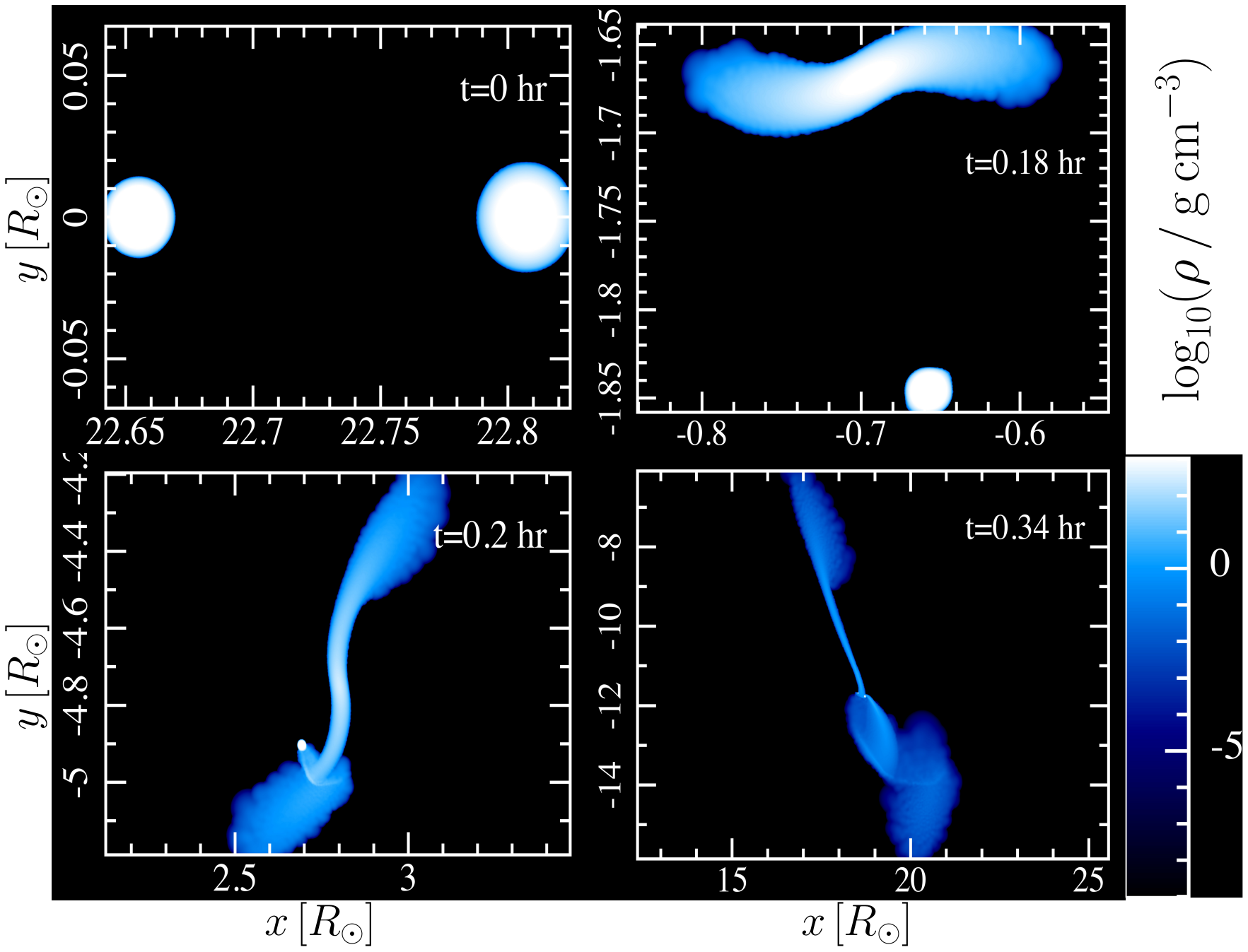}{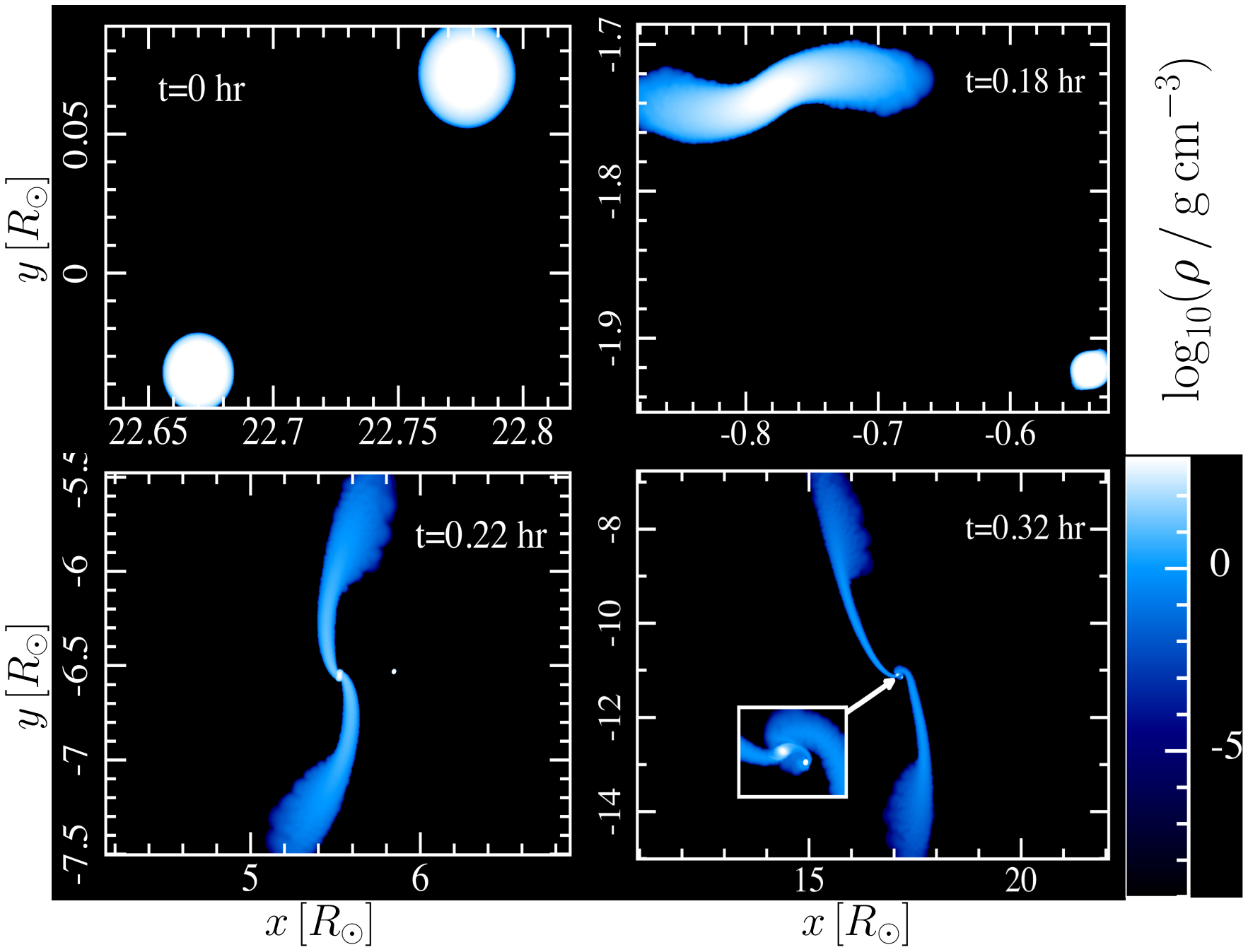}
	\caption{{\small Snapshots for the two different Non Identical retrograde binaries (at times mentioned in the figures). The visualizations were produced with the \textsc{SPLASH} software \citep{splash}.  \textbf{Left Panel:}  $\phi_0 = 0^{\circ}$ \textbf{Right Panel: } $\phi_0 = 45^{\circ}$ \tb{(Inset displays the zoomed region where the undisrupted WD is interacting with the core of the disrupted WD).} }}
	\label{snapNIR}
\end{figure}
%%%%%%%%%%%%%%%%%%%%%%%%%%%%%%%%%%%%%%%%%%%%%%%%%

Next, in the Right Panel of Figure~\ref{mcore_IRIP}, we present the core mass fractions for the IR binary. We observe both gradual and sudden increase in core mass fractions shortly after disruption, resulting from interactions between the disrupted stellar material, as anticipated. For example, at $\phi_0 = 0^\circ$, an immediate collision between the two disrupting cores causes the core of WD1 to gain more mass than its initial value. In contrast, for phases like $\phi_0 = 90^\circ$ and $135^\circ$, where no immediate core interaction occurs (see snapshot sequence in Figure~\ref{snapIR}), the core mass fractions evolve regularly, following typical disruption behavior.

More dramatic interactions occur in the NIR binary configuration\footnote{Unlike the IR binary, the NIR binary lacks inherent symmetry; therefore, for $\phi_0 \ge 180^\circ$, the outcomes are expected to differ. However, since we did not observe any qualitatively distinct outcomes, except for some quantitative differences, so we varied $\phi_0$ within $0^\circ$ to $180^\circ$ for NIR binaries too.}, as reflected in the bound core fraction shown in Figure~\ref{mcorenIR}. In this setup, the more massive WD (WD2, with $M_{\rm WD2} = 0.5, M_{\odot}$) remains undisrupted, while the less massive WD (WD1, with $M_{\rm WD1} = 0.25, M_{\odot}$) undergoes partial disruption. However, due to the inherent effects of retrograde motion, WD2 consistently gains mass—either from the debris or the core of the partially disrupted WD1. For most values of $\phi_0$. For $\phi_0 = 0^\circ$ and $180^\circ$, WD2 initially accretes mass from one of the disrupted tails of WD1 (see Left Panel of Figure~\ref{snapNIR}), and subsequently orbits around the core of WD1. Notably, we have found, at $\phi_0 = 180^\circ$, WD2 disrupts the remaining core of WD1, ultimately accreting nearly all of its material. In contrast, for $\phi_0 = 45^\circ$ and $90^\circ$, the two WDs evolve independently. However, in the case of $\phi_0 = 45^\circ$ (see the Right Panel of Figure \ref{snapNIR}), WD2 is later drawn toward the core of WD1 and follows a similar final outcome as observed for $\phi_0 = 180^\circ$.

\subsection{\textsc{Fragmentations of debris}}
\begin{figure}[H]
	\epsscale{1.1}
	\plottwo{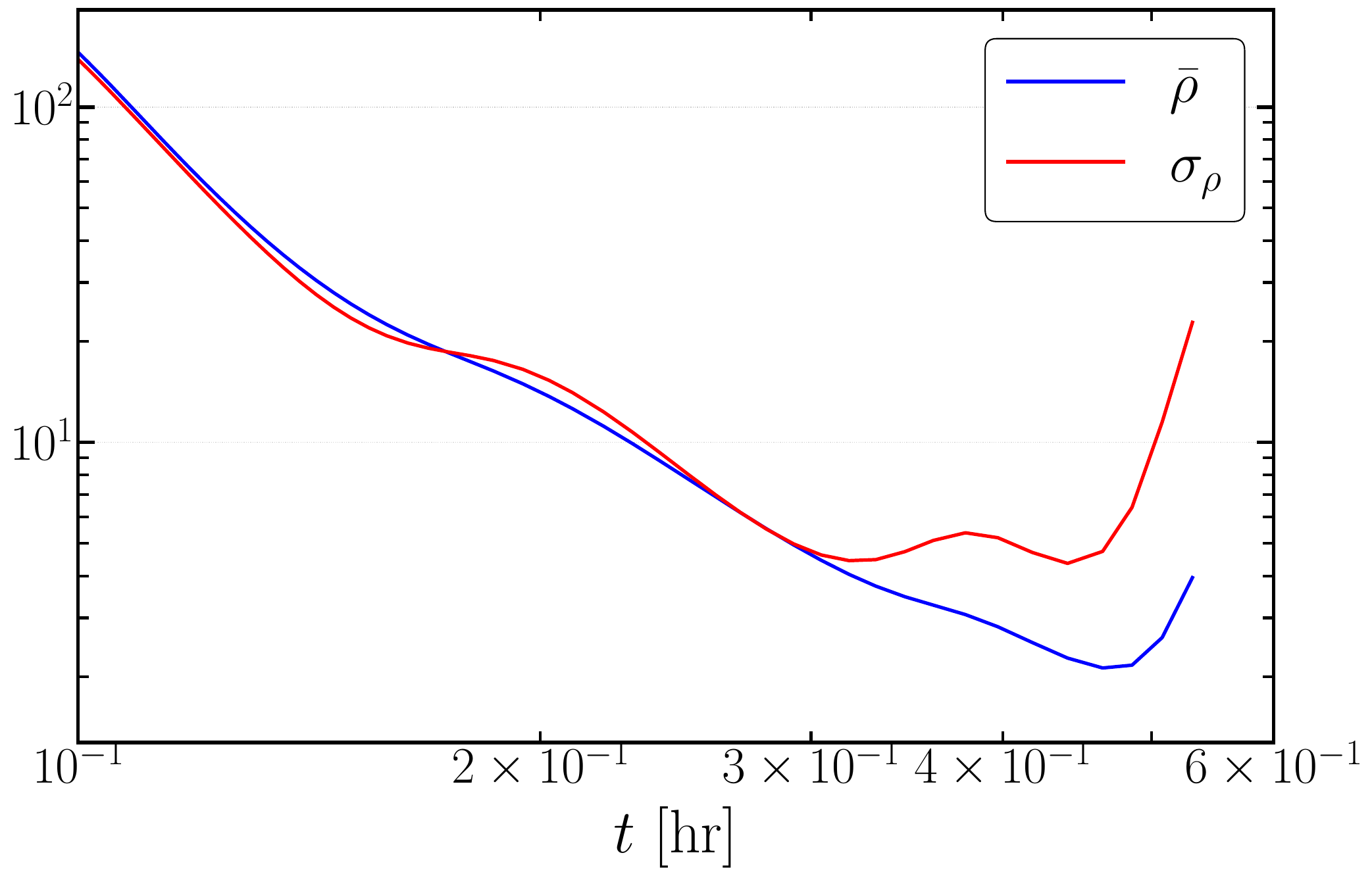}{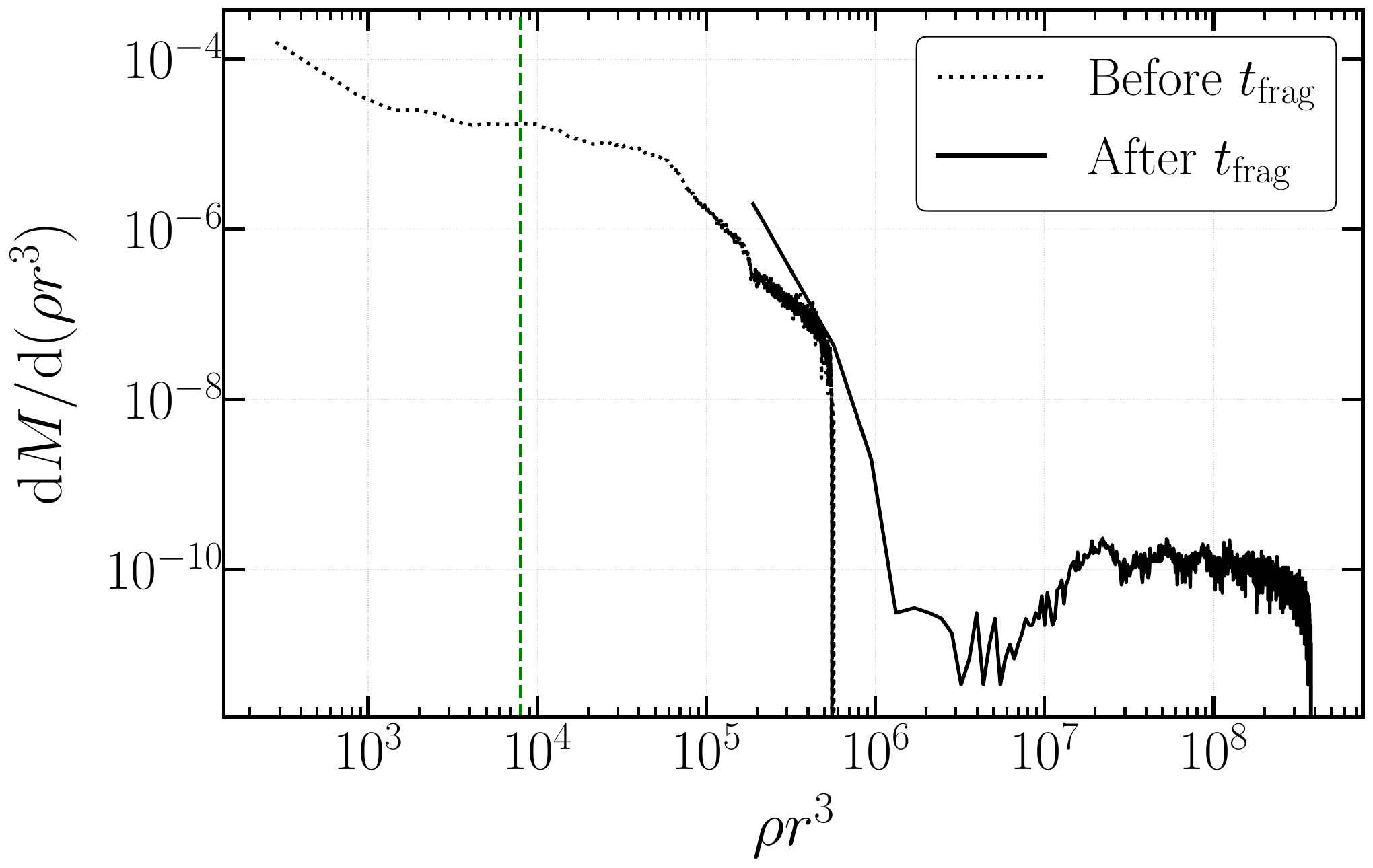}
	\caption{{\small Clump formation in Identical Retrograde (IR) binary for initial phase $\phi_0 = 135^{\circ}$. \textbf{Left Panel:} illustrates the variation of mean density, $\bar{\rho}$ \& standard deviation, $\sigma_{\rho}$ over time post disruption. The fragmentation time is obtained at the turning point of $\sigma_{\rho}$ at $t_{\rm frag}\sim0.46 ~\rm hr$ . \textbf{Right Panel:} gives the $\mathrm{d}M / \mathrm{d}(\rho r^3)$ distribution of $\rho r^3$ at two instants: before $t_{\rm frag}$, $t_{\rm before} \sim 0.4~t_{\rm frag}$ and after $t_{\rm frag}$, $t_{\rm after} \sim 2 ~t_{\rm frag}$ .}}
\label{clump}
\end{figure}

At the aftermath of binary TDE, the stellar debris stream becomes gravitationally unstable, fragmenting into small, spherical clumps. 
Initially, the mean density of the debris stream and its fluctuations fall as a power law with time, and at some later point of time, 
the fluctuations reach a minimum, and then increases and takes values greater than the mean density, indicating the formation
of clumps inside the debris \citep{Sacchi 2020}.  
Here, using the method outlined in \cite{Sacchi 2020}, we computed the mean density and its fluctuations over several post-disruption timesteps to estimate the fragmentation time. In the Left Panel of Figure~\ref{clump}, density fluctuations show a turning point  at $t_{\rm frag} \sim 0.46~{\rm hr}$, marking the onset of fragmentation. Note that this $t_{\rm frag}$ is much earlier than that reported for solar-like stars disrupted by SMBHs (which is in years). This
is due to the fact that for IMBHs, the self gravity of the debris can be comparable to the gravity due to the BH at smaller distances, i.e, at shorter
time scales. We also compared it to some typical single WD-IMBH cases, where fragmentation was further delayed, likely due to the limited amount of debris available in 
those scenarios.

%%%%%%%%%%%%%%%%%%%%%%%%%%%%%%%%%%%%%%%%%%%%%%%%

%%%%%%%%%%%%%%%%%%%%%%%%%%%%%%%%%%%%%%%%%%%%%%%%

%\begin{figure}[H]
%	\centering
%	\begin{minipage}[b]{0.4\textwidth}
%		\includegraphics[width=1.2\textwidth]{Plots/kick_vel_nlz_n.pdf}
%	\end{minipage}
%	\begin{minipage}[b]{0.4\textwidth}
%		\includegraphics[width=1.2\textwidth]{Plots/dmdrhor3_.pdf}
%	\end{minipage}
%	\caption{{\small Clump formation in Identical Retrograde (IR) binary for initial phase $\phi_0 = 135^{\circ}$. \textbf{Left Panel:} illustrates the variation of mean density, $\bar{\rho}$ \& standard deviation, $\sigma_{\rho}$ over time post disruption. The fragmentation time is obtained at the turning point of $\sigma_{\rho}$ at $t_{\rm frag}\sim0.46 ~\rm hr$ . \textbf{Right Panel:} gives the $\mathrm{d}M / \mathrm{d}(\rho r^3)$ distribution of $\rho r^3$ at two instants: before $t_{\rm frag}$, $t_{\rm before} \sim 0.4~t_{\rm frag}$ and after $t_{\rm frag}$, $t_{\rm after} \sim 2 ~t_{\rm frag}$ .}}
%	\label{clump}
%\end{figure}
%%%%%%%%%%%%%%%%%%%%%%%%%%%%%%%%%%%%%%%%%%%%%%%%

\begin{figure}[H]
	\centering
	% Left figure
	\begin{minipage}[t]{0.32\textwidth}
		\centering
		\includegraphics[width=\textwidth]{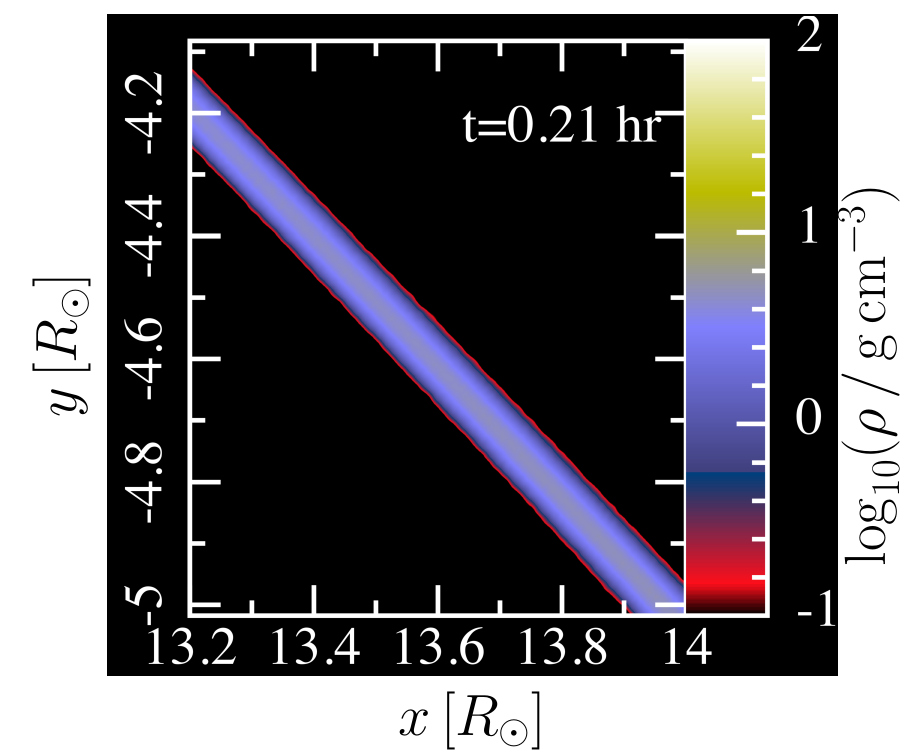}
	\end{minipage}%
	\hspace{0.005\textwidth} % spacing
	% Center figure
	\begin{minipage}[t]{0.32\textwidth}
		\centering
		\includegraphics[width=\textwidth]{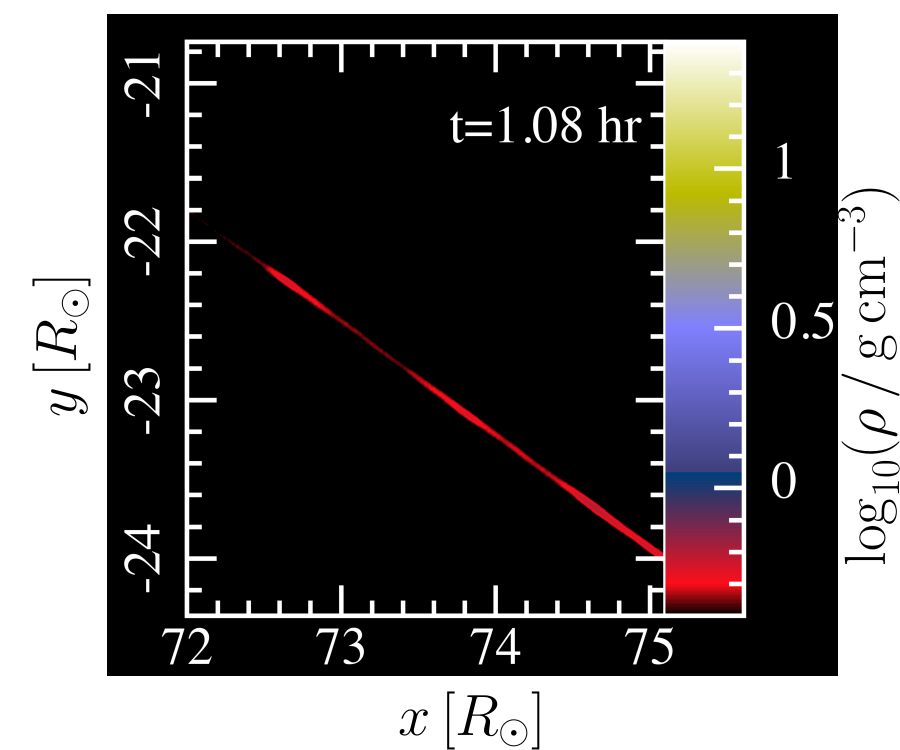}
	\end{minipage}%
	\hspace{0.005\textwidth} % spacing
	% Right figure
	\begin{minipage}[t]{0.32\textwidth}
		\centering
		\includegraphics[width=\textwidth]{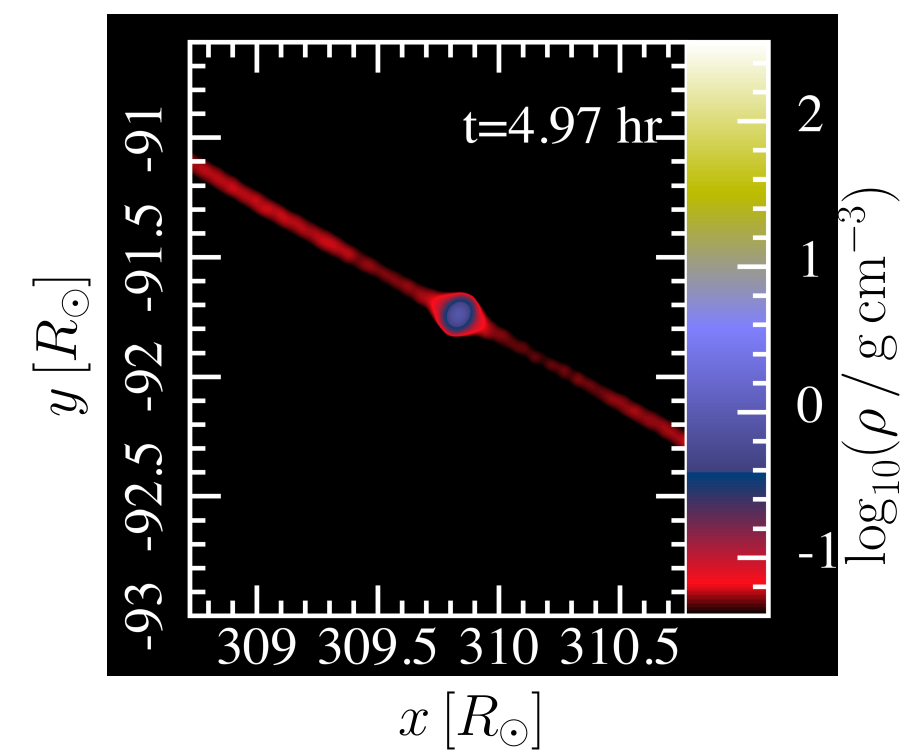}
	\end{minipage}
	\caption{{\small Snapshots of the debris at three distinct stages of the Identical Retrograde (IR) binary corresponding to $\phi_0=135^\circ$: \textbf{Left:} before fragmentation, at $t_{\rm before} \sim 0.4~t_{\rm frag}$, \textbf{Center:} after fragmentation, $t_{\rm after} \sim 2~t_{\rm frag}$, and \textbf{Right:} late after fragmentation, at $t_{\rm late} \sim 11~t_{\rm frag}$}. This figure was produced
	using the \textsc{SPLASH} software \citep{splash}.}
	\label{clump135nlz}
\end{figure}

As previously reported by \cite{Coughlin2015}, an instability criterion for a TDE debris stream is defined using a critical density, $\rho_h \sim M_h / r^3$, where $r$ is the radial distance of a debris fluid element from a BH of mass $M_h$. The stream becomes gravitationally unstable when its local density $\rho$ exceeds this critical value, i.e., $\rho > \rho_h$. Physically, this implies that the rate of self-gravitational aggregation of materials exceeds the rate at which the material is being pulled apart by tidal forces. To analyze $t_{\rm frag}$ using this criterion, we compute the mass distribution $\mathrm{d}M / \mathrm{d}(\rho r^3)$ as a function of $\rho r^3$ at two instants of time, one before fragmentation ($t_{\rm before} \sim 0.4\,t_{\rm frag}$), and one after fragmentation ($t_{\rm after} \sim 2 \, t_{\rm frag}$). We show this in the Right Panel of Figure \ref{clump}. 
As seen from this figure, at $t_{\rm after}$, we find that the entire stream satisfies the instability condition, as $\rho r^3$ exceeds the BH mass ($M_h = 8000~M_\odot$), denoted by the vertical dotted line. We provide visuals of these instants from our simulation snapshots in Figure \ref{clump135nlz}.

%%%%%%%%%%%%%%%%%%%%%%%%%%%%%%%%%%%%%%%%%%%%%%%%
%%%%%%%%%%%%%%%%%%%%%%%%%%%%%%%%%%%%%%%%%%%%%%%%

\subsection{\textsc{Kick velocity}}
%%%%%%%%%%%%%%%%%%%%%%%%%%%%%%%%%%%%%%%%%%%%%%%%
The ejected WD and/or the partially disrupted WD core can suddenly attain extremely high specific energy, becoming a HVS \citep{Hills}. This is characterized by the kick velocity, defined as
$v_{{\rm kick},a} = \sqrt{\epsilon_{{\rm core},a} - \epsilon_{{\rm initial},a}}$,
where $\epsilon_{{\rm core},a}$ is computed from the final bound core of each WD (as described in Section~\ref{core_mass}), and $\epsilon_{{\rm initial},a}$ is its initial value. We observe significantly higher $v_{\rm kick}$ values compared to the earlier study of solitary WD and IMBH \citep{Garain3}, with gains differing by nearly an order of magnitude, showing relative increase of up to $\sim 90 \%$ in the binary WD case.

\begin{figure}[H]
	\epsscale{1.1}
	\plottwo{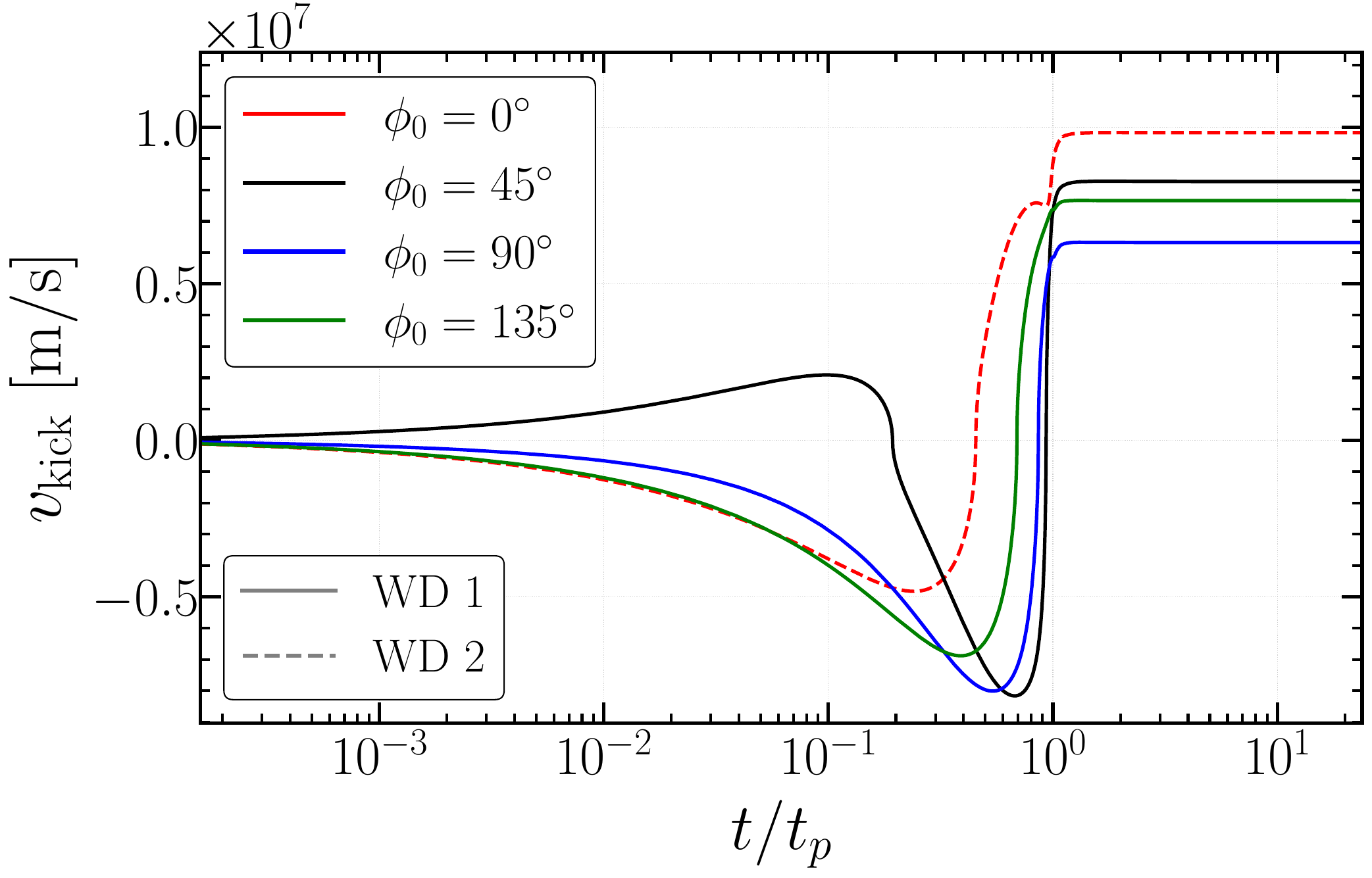}{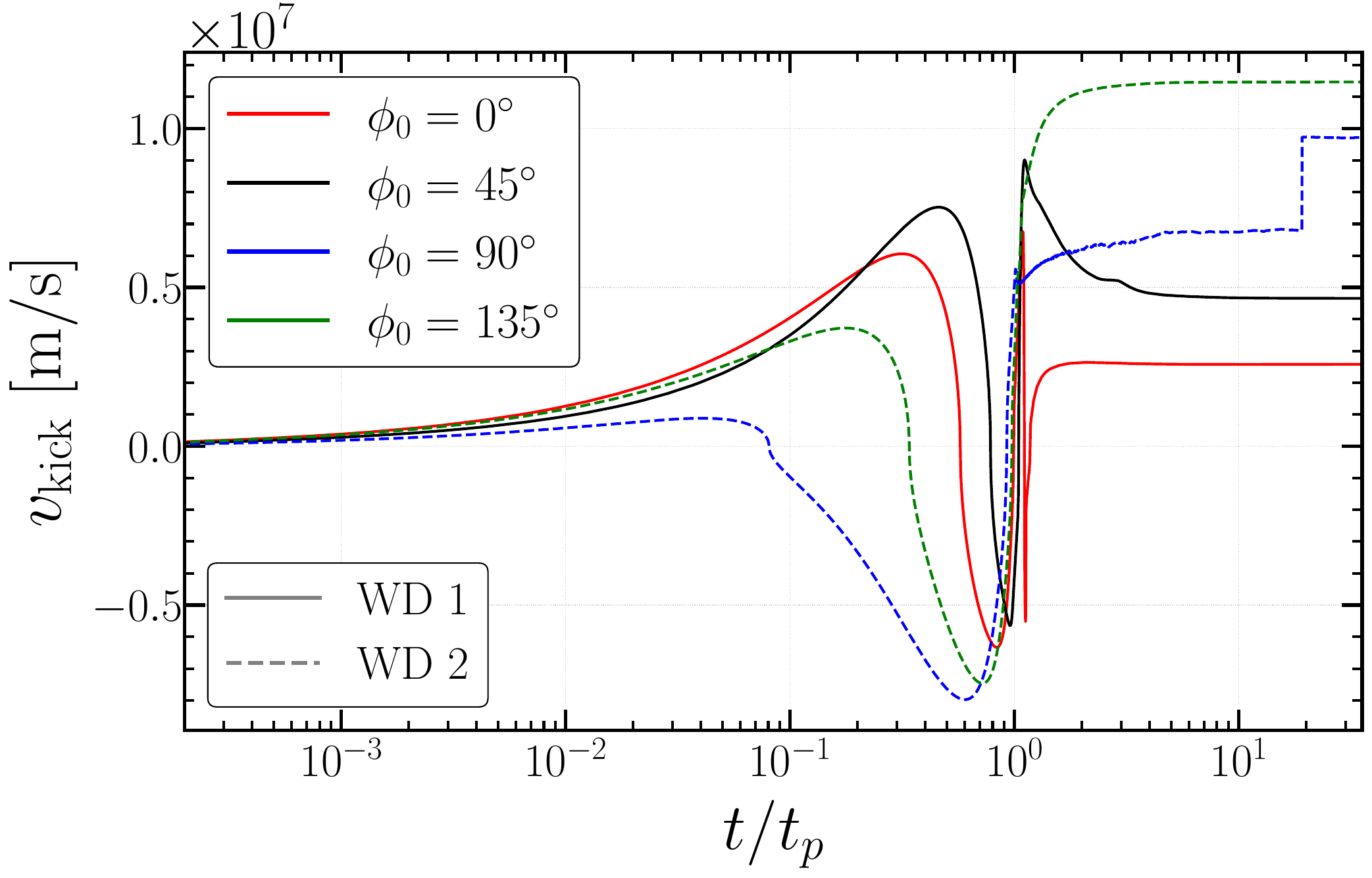}
	\caption{{\small Kick velocities for identical WD are shown as functions of $t/t_p$ where \( t_p \approx 0.062\,\mathrm{hr} \) represents the time of pericenter passage. \textbf{Left Panel:} In the Identical Prograde (IP) binary, the kick velocities originate from both the tidal break-up and the partial disruption of the non-zero self-bound ejected core.\textbf{ Right Panel:} In the Identical Retrograde (IR) binary, the kick velocities emerge from the partial disruption of the non-zero self-remnant core.}}
	\label{vIRIP}
\end{figure}

In the IP binary (see Left Panel of Figure~\ref{vIRIP}), we observe a combined effect, particularly for WD2 at $\phi_0 = 0^\circ$. Here, 
an initial increase in $v_{\rm kick}$ results from the tidal breakup. This is followed by a slightly delayed second rise due to partial disruption. In the IR binary case, the behavior is somewhat different due to multiple dynamical obstructions during the interaction (see, e.g., WD2 at $\phi_0 = 90^\circ$ in the Right Panel of Figure~\ref{vIRIP}). In certain cases, $v_{\rm kick}$ exceeds even that of the IP binary, for instance, WD2 at $\phi_0 = 135^\circ$ shows a significantly higher gain, likely due to more extensive disruption than in the IP binary.
%%%%%%%%%%%%%%%%%%%%%%%%%%%%%%%%%%%%%%%%%%%%%%%%
%%%%%%%%%%%%%%%%%%%%%%%%%%%%%%%%%%%%%%%%%%%%%%%%
\subsection{\textsc{Fallback rates}}
In this section, we highlight the unique features of the mass fallback rates for the IP and IR binaries. As mentioned in section \ref{sec2}, we calculate the fallback rates from the `frozen-in' estimation of specific energies, $\epsilon$ and specific angular momenta, $l$ of the bound debris. We have verified the early-time behavior of these, by comparing them with results from numerical differentiation. The two methods show consistent physical features and initial slopes of the fallback curves for each value of $\phi_0$. The primary differences lie in the peak fallback rate, $\dot{M}_{\rm peak}$, and the corresponding peak time, $t_{\rm peak}$, which differ by no more than approximately $20\%$.

\begin{figure}[H]
	\epsscale{1.1}
	\plottwo{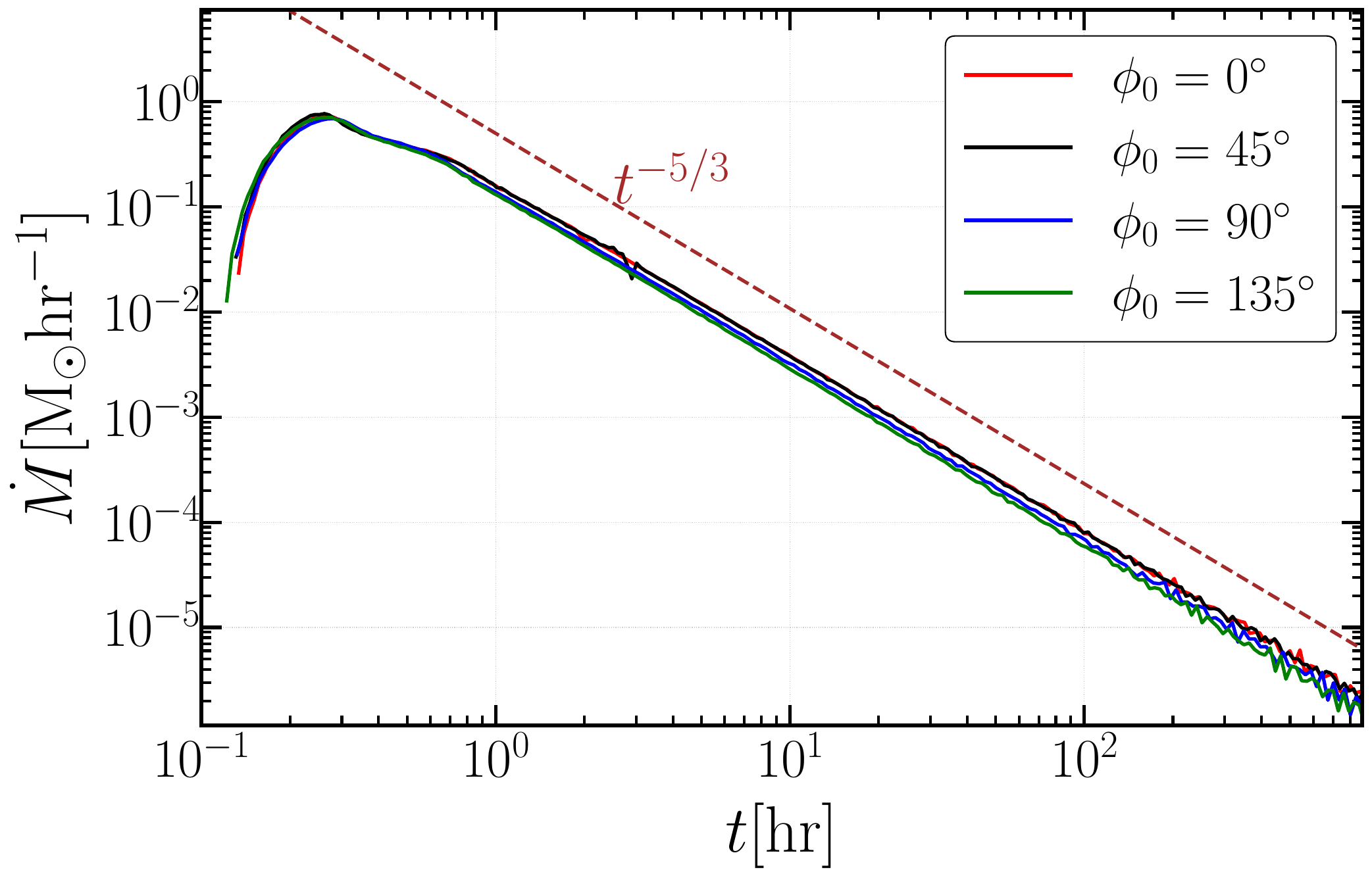}{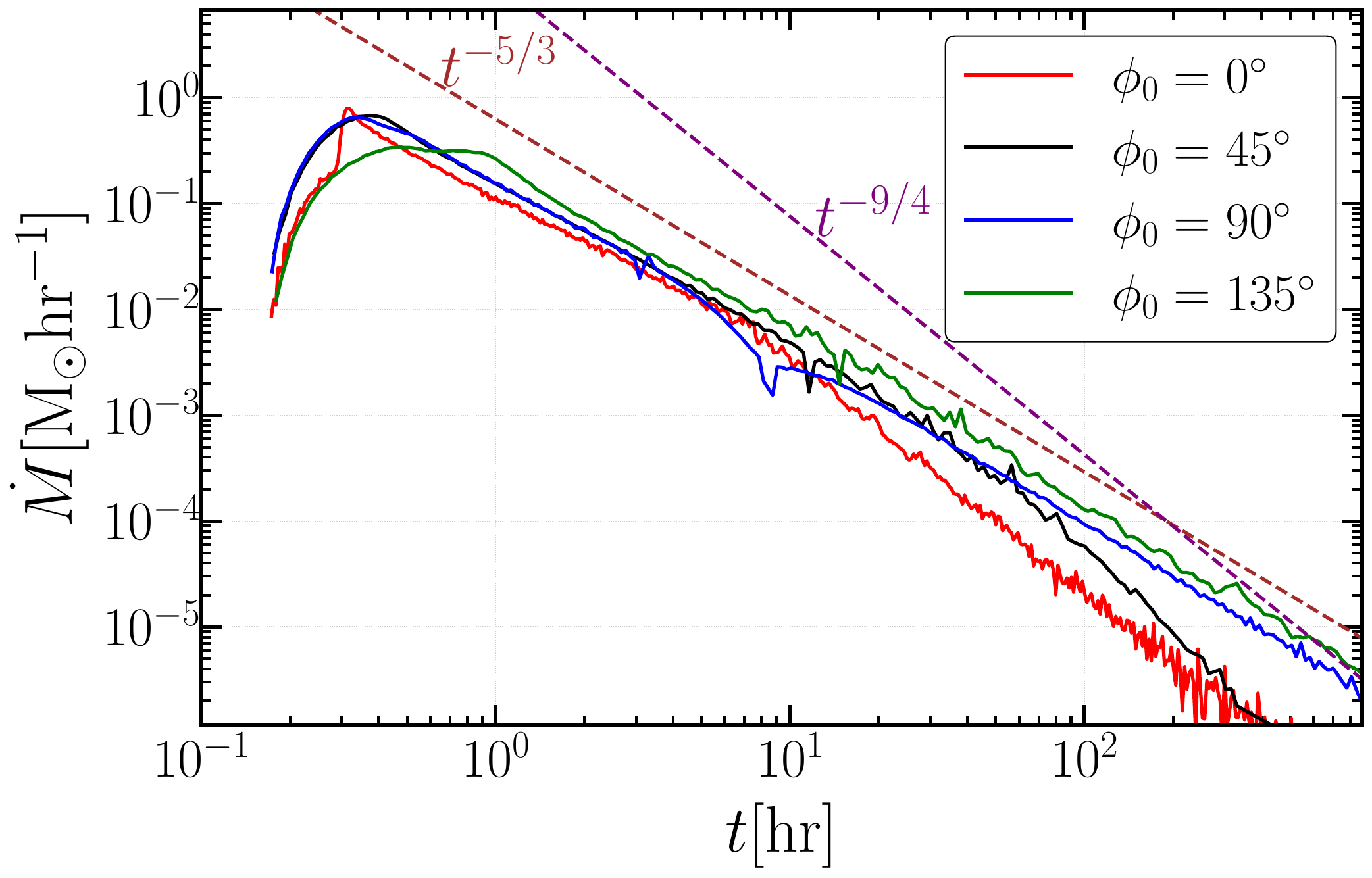}
	\caption{{\small Fallback rates are shown as functions of time \( t \). \textbf{Left Panel:} In the Identical Prograde (IP) binary, the $\dot{M}$ curve displays a characteristic double peak and asymptotically follows a $t^{-5/3}$ power-law. \textbf{Right Panel:} In the Identical Retrograde (IR) binary, the $\dot{M}$ curve has no definitive double-peaked feature and asymptotically deviates from the $t^{-5/3}$ power-law.}}
\label{dmdtIPIR}
\end{figure}

In the Left Panel of Figure~\ref{dmdtIPIR}, the fallback curve exhibits a signature unique to binary systems: a double-peaked structure \citep{2015ApJ...805L...4M}, but its late-time slope matches quite well with the known standard power-law slope, $t^{-5/3}$ \citep{Rees, Phinney}. This arises because the material from the bound tail of the fully disrupted WD (captured in an elliptical orbit) begins accreting earlier than that from the ejected WD (which is partially disrupted). As a result, there is a time delay between the fallback peaks of the two WDs, and at the end, we get the effective late-time slope. This behavior remains 
almost the same across all values of $\phi_0$ presented in the figure. The effective late-time slope of $t^{-5/3}$ is followed 
inspite of the compact stellar structure of the WD and in the presence of a self-bound WD core. This power law is well known in a qualitatively 
different situation that was studied by \cite{CoughlinNixon, Lodato} for the disruption of a solar mass star by a SMBH.
%%%%%%%%%%%%%%%%%%%%%%%%%%%%%%%%%%%%%%%%%%%%%%%%
%%%%%%%%%%%%%%%%%%%%%%%%%%%%%%%%%%%%%%%%%%%%%%%%

%%%%%%%%%%%%%%%%%%%%%%%%%%%%%%%%%%%%%%%%%%%%%%%%
\begin{figure}[H]
	\epsscale{1.1}
	\plottwo{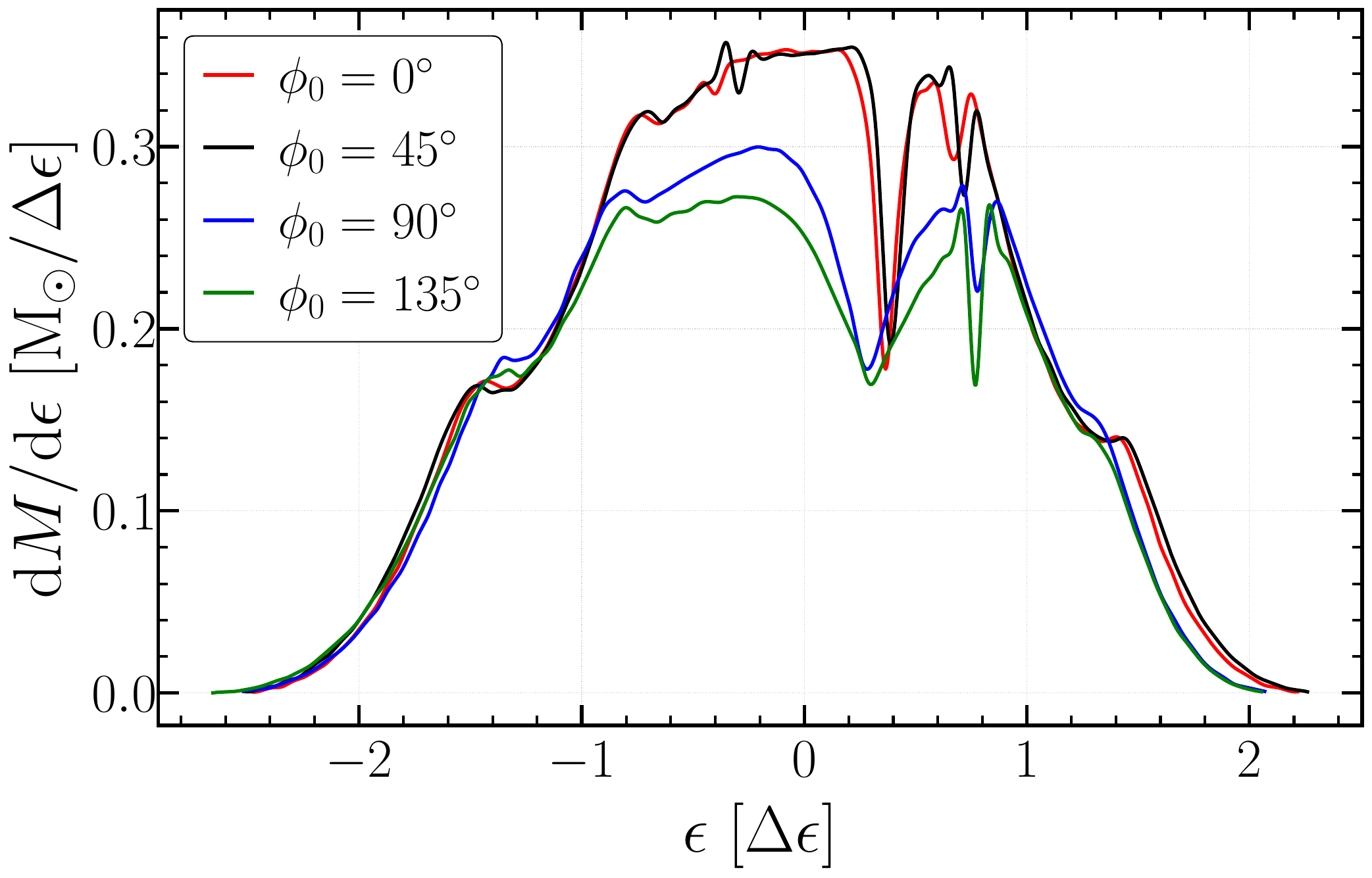}{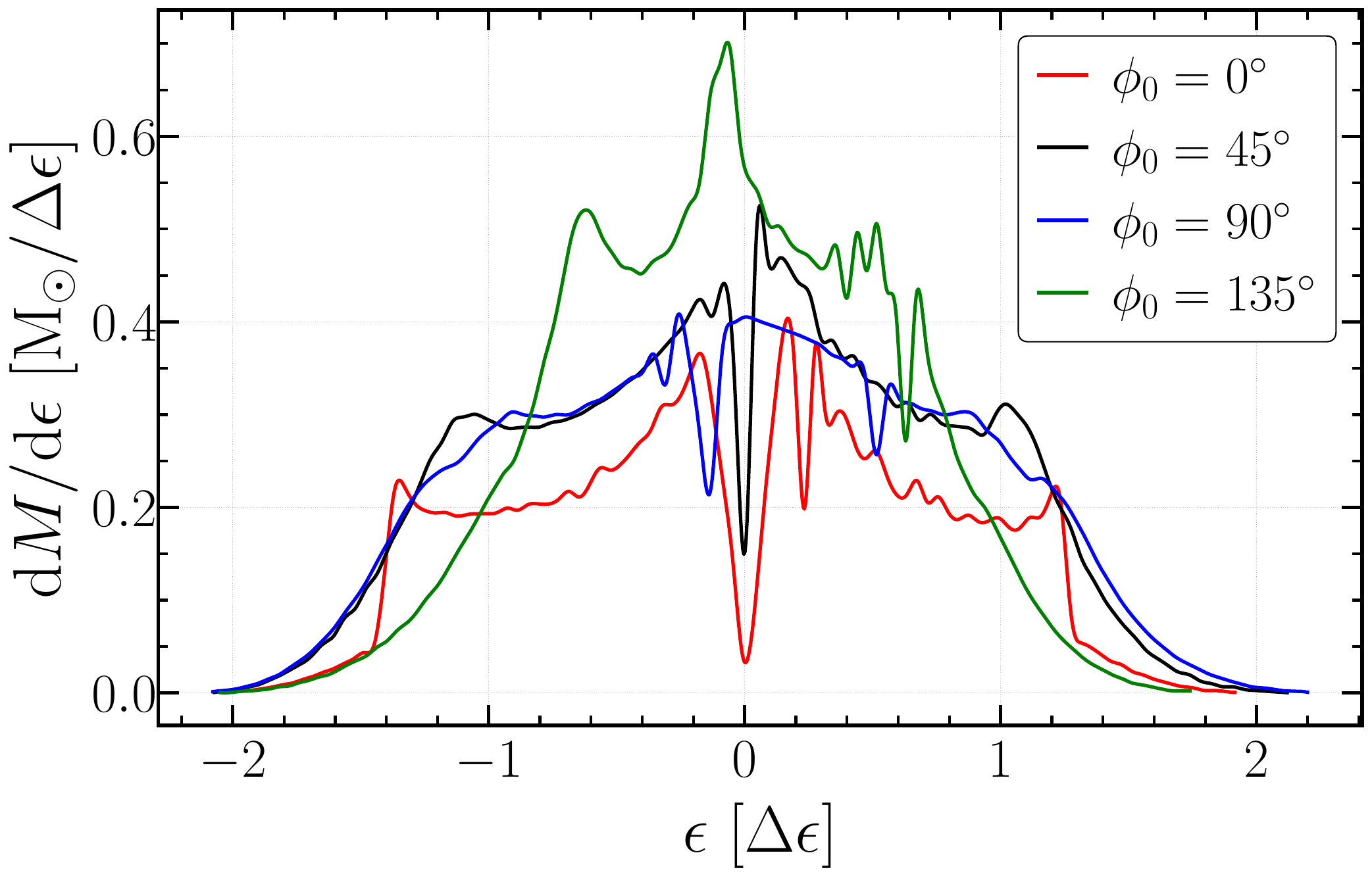}
	\caption{{\small  The variation in the total debris differential mass distribution, ${\rm dM/d}\epsilon$ with respect to the specific energy. The specific energies are scaled using the characteristic energy spread in the debris, $\Delta \epsilon = G M_{\rm BH} R_{\rm bin}/{r_t^{b}}^2$, where $R_{\rm bin} = R_{\rm sep} + R_{\rm WD}$. \textbf{Left Panel:} Identical Prograde (IP) binary. \textbf{Right Panel:} Identical Retrograde (IR) binary.}}
	\label{dmdeIPIR}
\end{figure}
%%%%%%%%%%%%%%%%%%%%%%%%%%%%%%%%%%%%%%%%%%%%%%%%

In contrast, these features are not always present in the IR binary case. Since the binary system remains bound in retrograde motion, material from both WDs tends to accrete around the same time, producing fallback curves that mimic those from a solitary WD disruption (see Right Panel of Figure~\ref{dmdtIPIR}, e.g., for $\phi_0 = 45^\circ$ and $90^\circ$). However, for phases such as $\phi_0 = 0^\circ$ and $135^\circ$, the fallback curves show different behaviour, as influenced by the strong interactions inherent to retrograde disruption dynamics. These accretion rates are influenced due to events such as dragging away of material from one WD's disrupted tail by the other (as shown in the Right Panel of Figure~\ref{snapIR}), or collisions between debris streams, leading to the accumulation and simultaneous accretion of a large amount of material. In the IR binary, the late-time slope interestingly shows a change from the initial $t^{-5/3}$ to the $t^{-9/4}$ power law behaviour 
for $\phi_0=0^\circ$.

To capture the subtle features of these fallback curves, we present the debris differential mass distribution with respect to specific energy, ${\rm dM/d}\epsilon$ in Figure \ref{dmdeIPIR}. The left panel shows the ${\rm dM}/{\rm d}\epsilon$ distribution for the IP binary.
The most bound debris elements yield nearly same distributions across different $\phi_0$, leading to similar early-time behavior of $\dot{M}$, including the first peak. The least bound distribution has more weightage for $\phi_0 = 0^\circ$ and $\phi_0 = 45^\circ$, which corresponds to higher $\dot{M}$ from the second peak onward throughout the late-time tail.
In the right panel, the ${\rm d}M/{\rm d}\epsilon$ distribution for the IR binary lacks a smooth profile, due to inherent interactions of IR binary. 
However, here, the behavior of $\dot{M}$ is affected by the presence of loosely bounded cores ($\epsilon\lesssim0$). For instance, in the $\phi_0 = 0^\circ$ and $45^\circ$ cases, the presence of a large core (also seen in the right panel of Figure \ref{mcore_IRIP}) significantly steepens the late-time slope (it is $t^{-9/4}$ for the case $\phi_0=0^\circ$). Similarly, the core in the $\phi_0 = 90^\circ$ case introduces a small dip in the $\dot{M}$ curve, though the overall slope at late times remains largely unaffected. Moreover, the sharp rise in the $\dot{M}_{\rm peak}$ value for $\phi_0 = 0^\circ$ is also suggested by its ${\rm dM}/{\rm d}\epsilon$ distribution. The amount of debris with the most bound specific energies is relatively small, which leads to a gradual initial rise in $\dot{M}$. However, the presence of a large amount of bound debris clustered around $\epsilon \gtrsim -1.5$ causes a sudden and sharp increase in $\dot{M}$, resulting in the observed $\dot{M}_{\rm peak}$ in this case.

Finally, we observe that the peak fallback rates, $\dot{M}_{\rm peak}$, and its time of arrival (at time $t_{\rm peak}$) remain relatively same across different phases in the IP binary, while they show greater variability in the IR binary over the phases presented. Moreover, these obtained values of $\dot{M}_{\rm peak}$ are higher compared to those in cases of TDEs involving solitary WDs of similar mass by an IMBH.

\subsection{\textsc{Gravitational wave emissions}}

%TDEs involving WDs and BHs are potential sources of gravitational waves, particularly in the form of short, intense bursts.  The characteristic strain amplitude \(h\) and duration \(t\) of the emission are linked through the GW frequency \(f\), with \(t \sim 1/f\).  
%As the WD approaches the pericenter of its orbit, it experiences extreme tidal forces from the BH, leading to its partial or complete disruption. This violent interaction produces a sharp spike in GW emission, with the peak amplitude typically occurring near the point of closest approach. Following the pericenter passage, the stellar material rapidly spreads out, diminishing the coherence of the mass quadrupole and thereby quenching further strong GW emission. As a result, the signal resembles a transient burst rather than a continuous waveform. The characteristic strain amplitude \(h\) and duration \(t\) of the emission are linked through the GW frequency \(f\), with \(t \sim 1/f\). 
%These features make TDEs promising targets for future space-based GW detectors operating in the millihertz to hertz frequency band.

\begin{figure}[H]
	\epsscale{1.15}
\plottwo{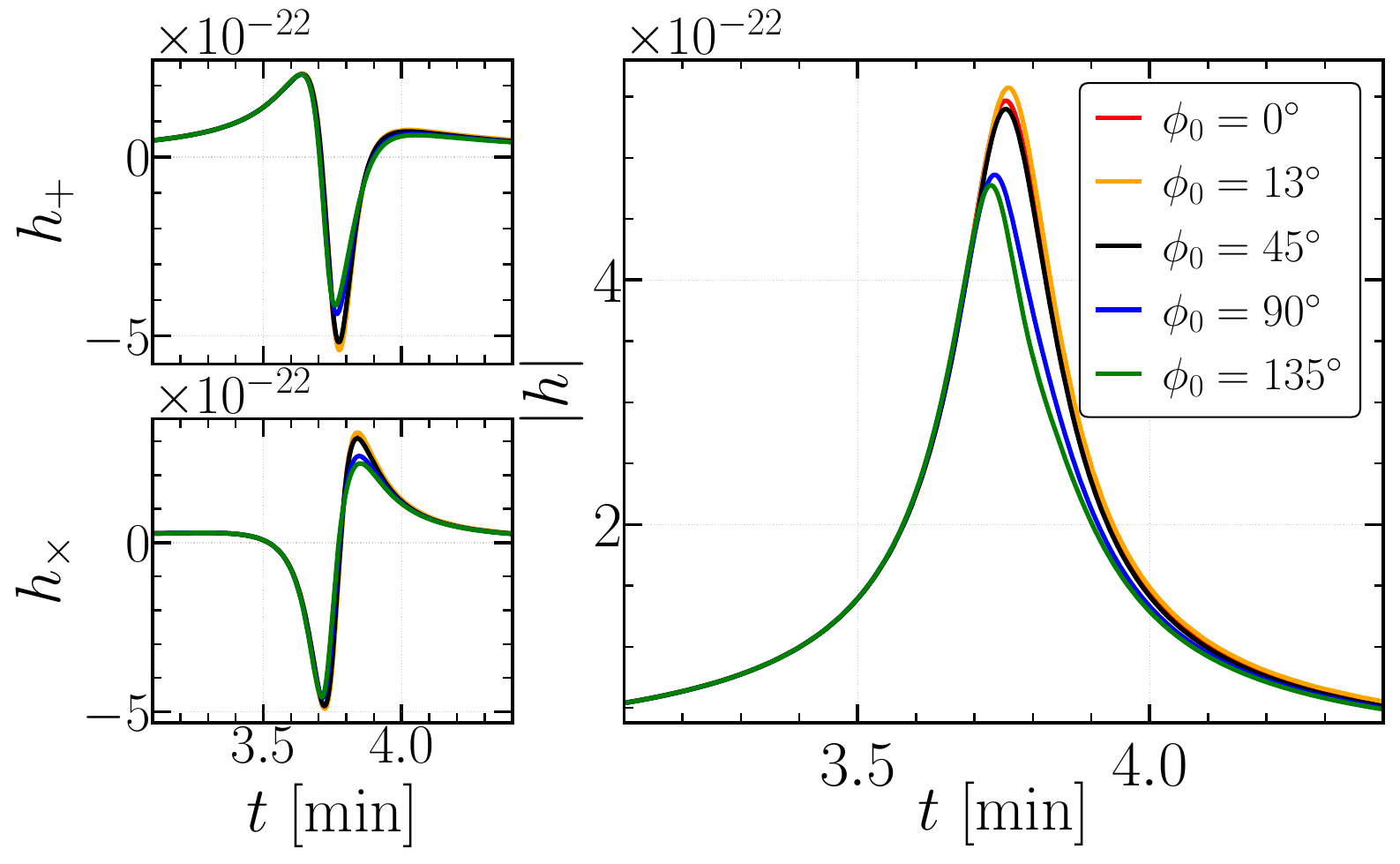}{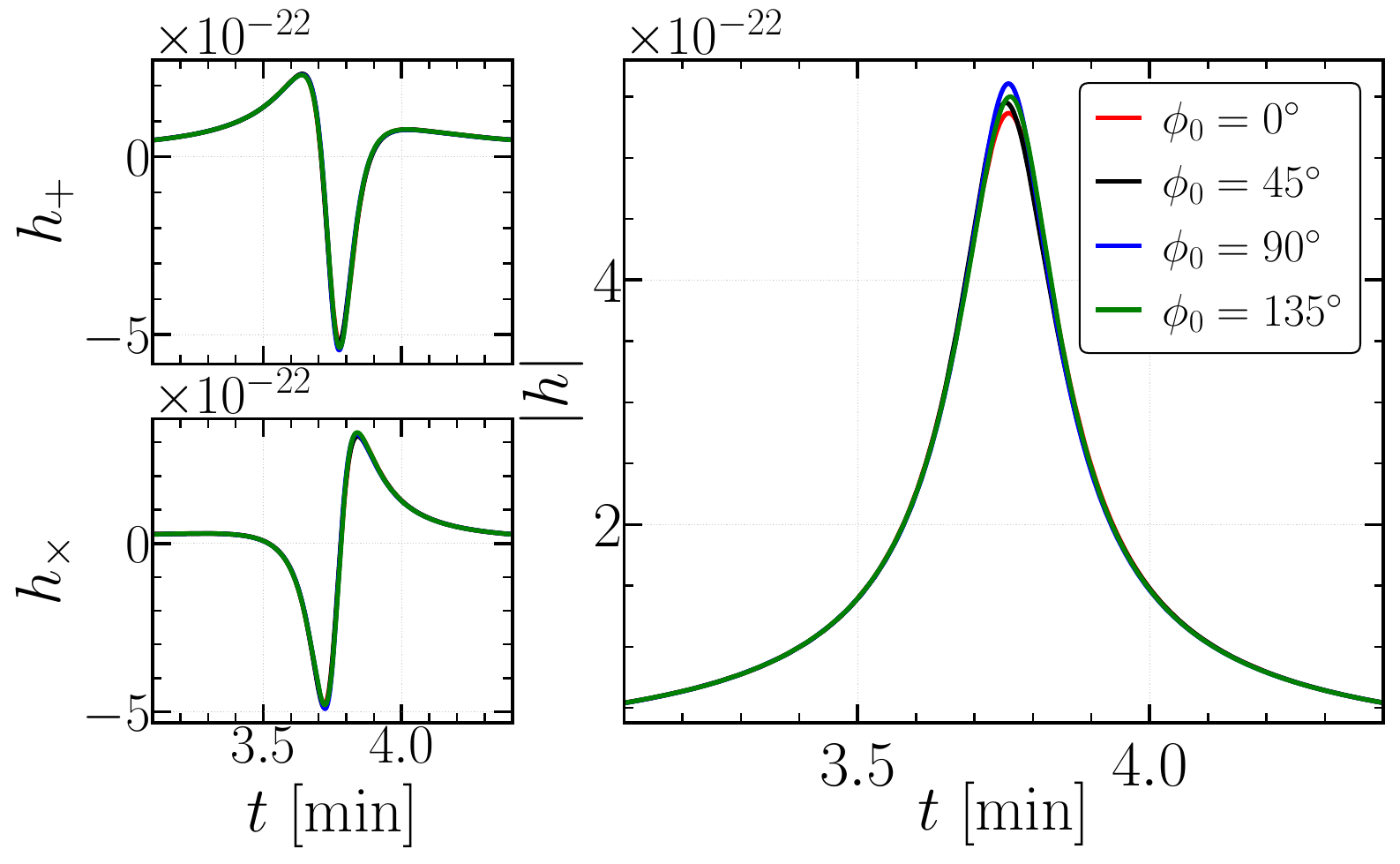}
\caption{{\small The GW amplitudes \( h_{+} \), \( h_{\times} \), and \( |h| \) are shown as functions of time \( t \) at a distance of $20~ \rm{Mpc}$. \textbf{Left Panel:} Identical Prograde binary. \textbf{Right Panel:} Identical Retrograde binary (The red, black, blue, green curves are almost visually overlapping).}}
\label{gw_IRNIR}
\end{figure}

We computed the GW emission using the quadrupole approximation, following the method outlined in \cite{1993ApJ...416..719C}. Figure \ref{gw_IRNIR} illustrates the GW polarization amplitudes, $h_+(t)$ and $h_\times(t)$, as well as the total amplitude, $|h(t)| = \sqrt{|h_+(t)|^2 + |h_\times(t)|^2}$ for the binary system, were evaluated over time for different configurations. 
\tb{The polarizations $h_+(t)$ and $h_\times(t)$ are computed for an observer located at a distant point perpendicular to the orbital plane of the event, i.e., along the $\hat{\mathbf{z}}$ direction, while the pericentre of the binary centre of mass orbit is aligned along $\hat{\mathbf{y}}$. Accordingly, we rotate the reference frame of the system and evaluate the GW amplitudes in the transformed frame following the conventions described in \citet{Maggiore}. For estimating them, we adopt a representative source-observer distance of $d \approx 20,\mathrm{Mpc}$ \citep{Toscani}, which is commonly used when discussing detectability or event rates.}
The GW amplitudes for WD binaries are higher than those produced in the single-star disruption scenario.

%%%%%%%%%%%%%%%%%%%%%%%%%%%%%%%%%%%%%%%%%%%%%%%%
%%%%%%%%%%%%%%%%%%%%%%%%%%%%%%%%%%%%%%%%%%%%%%%%

In the IP configuration (see Left Panel of Figure \ref{gw_IRNIR}), both $\phi_0 = 0^{\circ}$ and $\phi_0 = 45^{\circ}$ cases yield comparable GW amplitudes. This is because, in each case, one WD undergoes partial disruption while the other experiences a stronger disruption, leading to similar overall GW emission. However, $\phi_0 = 0^{\circ}$ produces a slightly higher amplitude. A similar trend is observed for $\phi_0 = 90^{\circ}$ and $\phi_0 = 135^{\circ}$, with $\phi_0 = 90^{\circ}$ again resulting in a marginally stronger signal. Overall, the percentage difference between the maximum(at the offset phase $\phi_0 =  13^{\circ}$ where both WD merge at pericentre) and minimum GW amplitudes among these cases is approximately $14.5\%$. 
The IR configuration (see right panel of Figure~\ref{gw_IRNIR}) exhibits behavior that is notably different from that of the IP configuration. Here, both WDs experience nearly uniform disruption for the various initial phases, leading to GW amplitudes that differ only marginally from each other. There is an approximate $4.6\%$ difference between the maximum and minimum amplitudes.

%%%%%%%%%%%%%%%%%%%%%%%%%%%%%%%%%%%%%%%%%%%%%%%%
\begin{figure}[H]
	\centering
	\includegraphics[height=5.4cm]{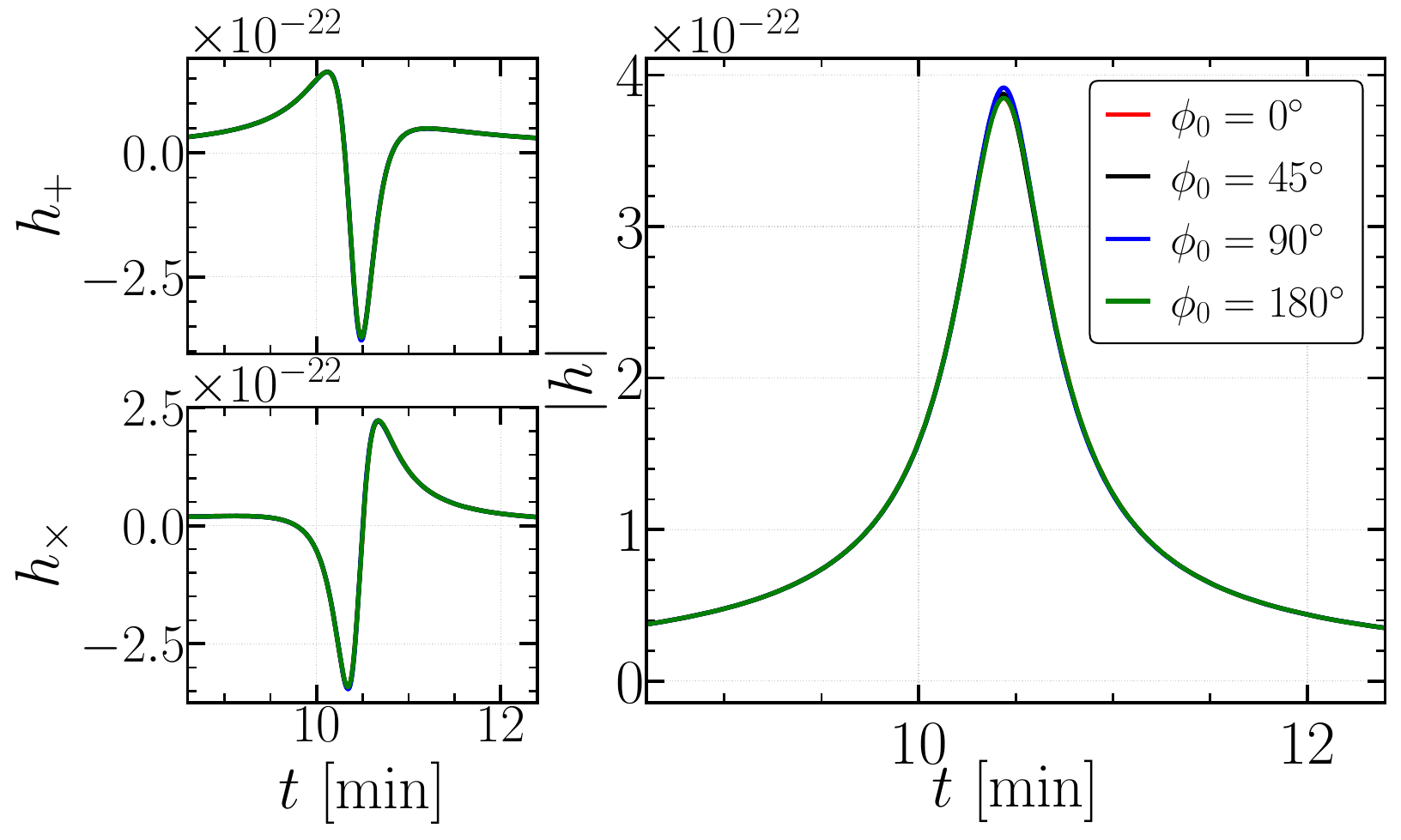}
	\includegraphics[height=5cm]{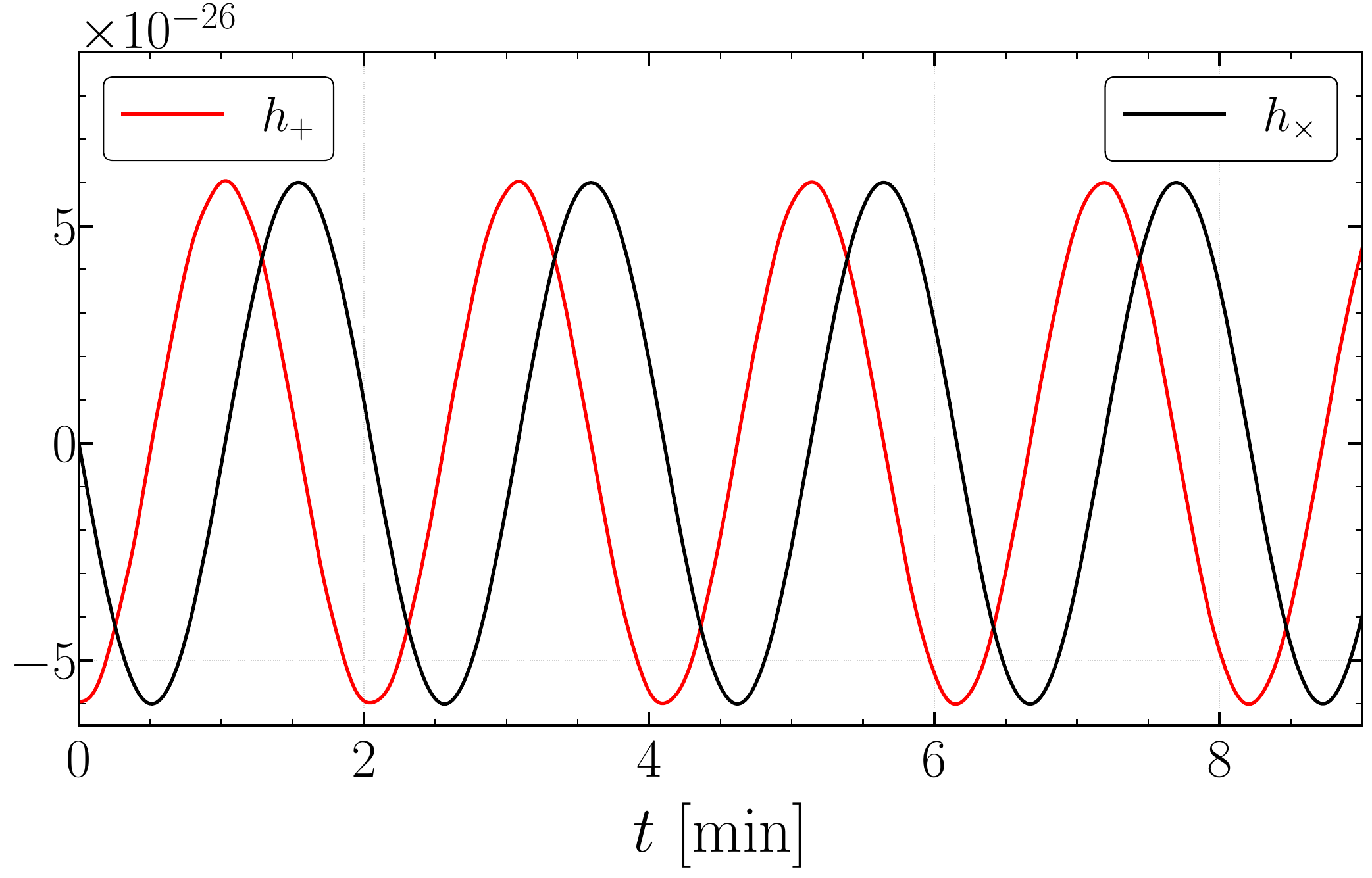}
	\caption{{\small  \textbf{Left Panel:}  The GW amplitudes \(h_+\), \(h_\times\), and \(|h|\) are shown as functions of time \(t\) for the Non Identical retrograde binary at a distance of $20~ \rm{Mpc}$ (The red, black, blue, green curves are almost visually overlapping).  \textbf{Right Panel:} \tb{The polarizations \(h_+\) and \(h_\times\) calculated only due to the DWD binary motion in the $x-y$ plane without BH background at a distance of $20~ \rm{Mpc}$.}}}
	\label{gwNIR}
\end{figure}
%%%%%%%%%%%%%%%%%%%%%%%%%%%%%%%%%%%%%%%%%%%%%%%%

%%%%%%%%%%%%%%%%%%%%%%%%%%%%%%%%%%%%%%%%%%%%%%%%

\begin{figure}[H]
	\epsscale{0.6}
	\plotone{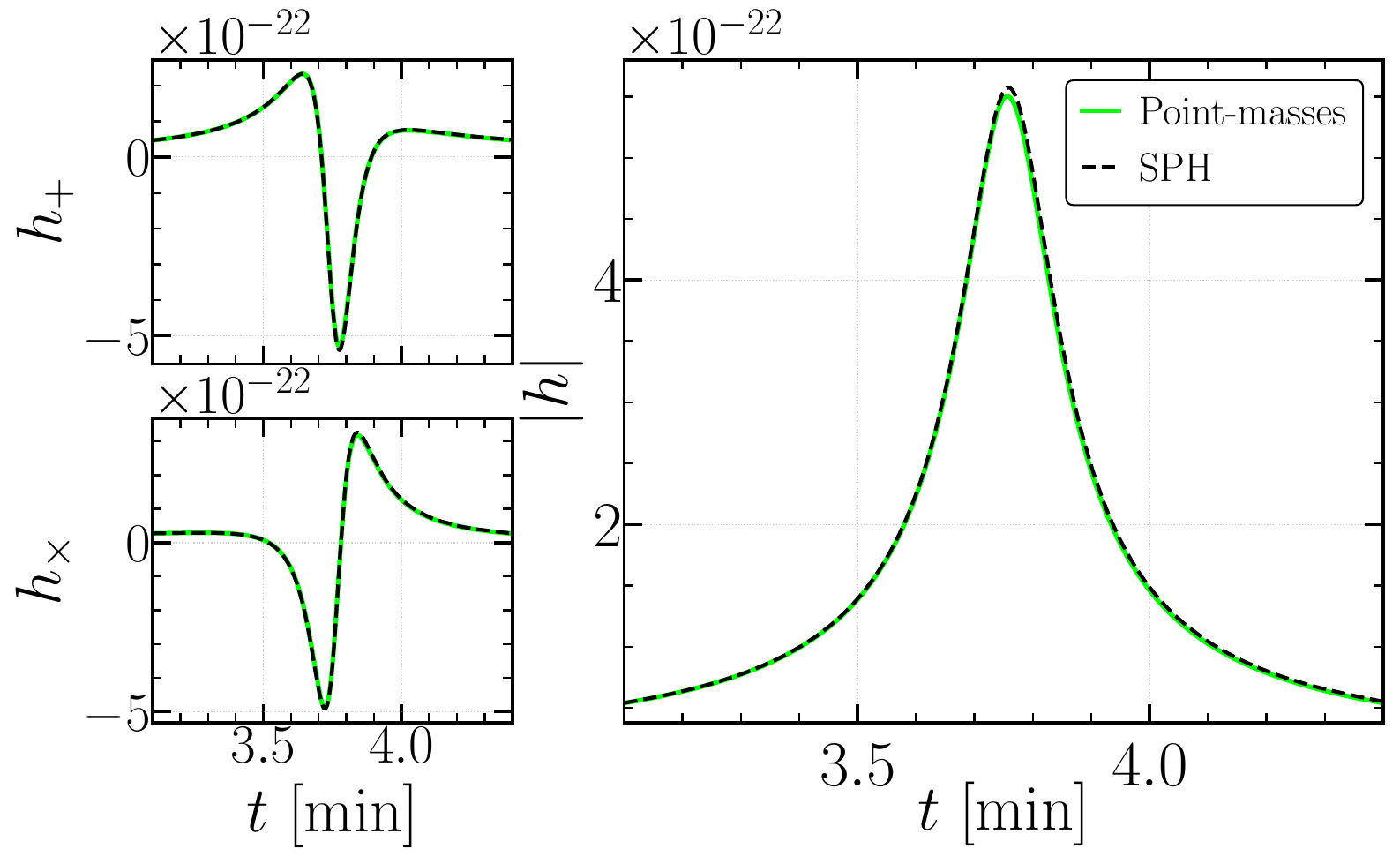}
	\caption{\tb{{\small The GW amplitudes $h_+$, $h_\times$, and $|h|$ are shown as
						functions of time $t$ for the phase $\phi_0=45^\circ$ of Identical Prograde binary at a distance of $20~ \rm{Mpc}$. The SPH waveform is compared with that obtained from the point-particle motion of the binary centre of mass around the IMBH. The two results agree very well in both phase and amplitude level, indicating that the GW signal is primarily driven by the orbital motion, while hydrodynamic effects introduce only minor corrections. The light green and black curves are almost visually overlapping.}}}
	\label{gw_cm}
\end{figure}

The NIR configuration (see right panel of Figure~\ref{gwNIR}) exhibits interesting behavior. In this case, the signal duration is approximately $1$ minute, nearly twice that observed in the IR and IP configurations, where it is about $0.5$ minutes. This extended duration results from the combined effects of mass asymmetry and positional orientation. In NIR binaries, the lighter WD undergoes gradual stripping over an extended period, while the heavier WD is undisrupted. Moreover, the retrograde orientation reduces the efficiency with which orbital angular momentum is transferred to the disrupted stellar material, leading to slower debris dispersal and a more prolonged disruption process. 
The overall GW amplitudes for different initial phases, however, show only minimal variation. 
%Notably, in the case of $\phi_0 = 45^\circ$, a nearly complete mass transfer from one WD to the other induces a small fluctuation of the order of $10^{-24}$ (which may not be measurable) in the GW amplitude, as illustrated in the inset of the right panel of Figure~\ref{gwNIR}.

\tb{We make an important remark on the GW amplitudes presented here. The overall behaviour and magnitude of the GW signal are primarily governed by the orbital motion of the DWD centre of mass around the IMBH. The intrinsic evolution of the compact binary system alone (i.e., without the IMBH background) produces a much weaker GW signal, with amplitudes of the order of $\sim 10^{-26}$ (see the right panel of Figure~\ref{gwNIR}). In Figure~\ref{gw_cm}, we compare the GW signals obtained from the full SPH simulation with those computed using a point-particle approximation for the motion of the binary centre of mass around the IMBH. The two waveforms agree very well in their overall phase and profile. The difference in the peak amplitude is very small, for the IP binary with $\phi_0 = 45^\circ$, the relative difference is only $\sim 1.85\%$. This comparison indicates that the dominant contribution to the GW emission arises from the orbital motion of the DWD around the IMBH, while hydrodynamic effects introduce only minor corrections. A similar comparison of the GW emission has been discussed previously in the context of TDEs (e.g., \citealt{Kobayashi2004}).}

% In contrast, the $\phi_0 = 90^\circ$ case avoids such interaction, allowing one WD to undergo a typical disruption while the other remains largely intact. This leads to a higher total GW amplitude compared to $\phi_0 = 0^\circ$. Notably, the $\phi_0 = 180^\circ$ configuration produces a GW amplitude similar to that of $\phi_0 = 0^\circ$. Percentage change between maximum and minimum amplitudes is less than $2\%$. In this case, the GW amplitude for $\phi_0 = 45^\circ$ lies between those of $\phi_0 = 0^\circ$ and $\phi_0 = 90^\circ$. Although the bound core analysis (Figure~\ref{mcorenIR}) shows that $\phi_0 = 0^\circ$ results in more disruption than $\phi_0 = 90^\circ$, the GW pattern is reversed. This is because, in the $\phi_0 = 0^\circ$ case,  one WD collides with the tidal tail of its companion, interrupting the disruption process. As a result, the nearly intact WD gains some mass, which reduces the overall GW amplitude.

\section{Discussions and conclusions}
\label{sec4}
Our aim in this paper was to quantify the effects of stellar hydrodynamics on the physics of TDEs involving a close non-accreting WD binary 
and an IMBH. As we have discussed in the introduction, the scenario is important and interesting as there is growing observational 
evidence for such binaries as well as BHs with intermediate masses. 
In this work, we restricted ourselves to a close binary, with
the binary components on the equatorial plane. 
Our study here compliments well known results of three-body 
scattering simulations of these scenarios that are very well studied in the literature. As we saw in section \ref{sec2}, 
Figures \ref{sep_nbh}, \ref{sep_b0p2}, \ref{sep_b0p8}, \ref{trajlz} indicate that inclusion of relativistic effects due to the Schwarzschild background
on the stellar trajectory are crucial to the analysis, as  a Newtonian formalism is prone to substantial error in the regime of our interest. 
Further, our results show the crucial dependence of observables such as kick velocities, fallback rates and GW emissions on 
initial conditions, i.e., if the initial motion of the binary is pro or retro-grade, and also on whether the binary components are
of same or unequal masses. 

Note that in this work, we have taken the pericenter distance to be $r_p=25r_g$, so that the stars of the binary 
are far enough from the BH to warrant a Newtonian treatment of their self gravity sufficiently robust \citep{Stone2019}. 
However, the gravitational effects of the remnant core in case of partial distributions is non-trivial. This is reflected in the nature of 
the masses of the remnant core in Figures \ref{mcore_IRIP} and \ref{mcorenIR}. In particular,
for prograde binaries, the binaries detach and individual cores largely follow the pattern of single star disruptions, 
the behaviour in retrograde binaries are very different, as the system is more stable (i.e., the components do not separate), 
and the core masses are 
influenced by the interactions between disrupted stellar matter. The patterns shown in the Right Panel of Figure  \ref{mcore_IRIP} and
in Figure \ref{mcorenIR} cannot arise from single-star disruptions. Next, the Left Panel of Figure \ref{clump} indicates 
that the clump formation in our cases can be as small as a fraction of an hour, whereas for TDEs involving SMBHs, it is
typically of the order of years. Further, Figure \ref{vIRIP} indicates that IMBHs can impart high kick velocities on 
the ejected WD, of the order of a few $\times 10^4~{\rm Km/sec}$, thus resulting in detectable HVSs. 
Figure \ref{dmdtIPIR} highlights the differences in the fallback rate of stellar debris for binary stellar systems, as compared
to solitary star TDEs. Whereas the Left Panel of this figure indicates that TDEs involving prograde binaries 
always have a typical ``double-hump'' structure (since the stars separate and then disrupt individually), this feature may
or may not be present in TDEs of retrograde binaries, as seen from the Right Panel of this figure. 
Finally, Figures \ref{gw_IRNIR} and \ref{gwNIR} detail the GW amplitudes for the identical and non-identical binary 
systems, respectively and our results indicate that the GW amplitudes are typically higher in magnitude for 
binary systems compared to solitary star disruptions. 

\begin{center}{\bf Acknowledgements}\end{center}
%\addcontentsline{toc}{section}{Acknowledgements}

We acknowledge the support and resources provided by PARAM Sanganak under the National Supercomputing Mission, Government of India, 
at the Indian Institute of Technology Kanpur. The work of AM is supported by Prime Minister’s Research Fellowship by the Ministry of 
Education, Govt. of India. PB acknowledges financial support from Anusandhan National Research Foundation, Government of India, 
File Number PDF/2022/000332.
The work of TS is supported in part by the USV Chair Professor position at IIT Kanpur, India. 

\begin{center}{\bf Data Availability Statement}\end{center}

The data underlying this article will be shared upon reasonable request to the corresponding author.

\appendix 
\section{Resolution Dependence}\label{appendix}
\begin{figure}[H]
	\epsscale{1.15}
	\plottwo{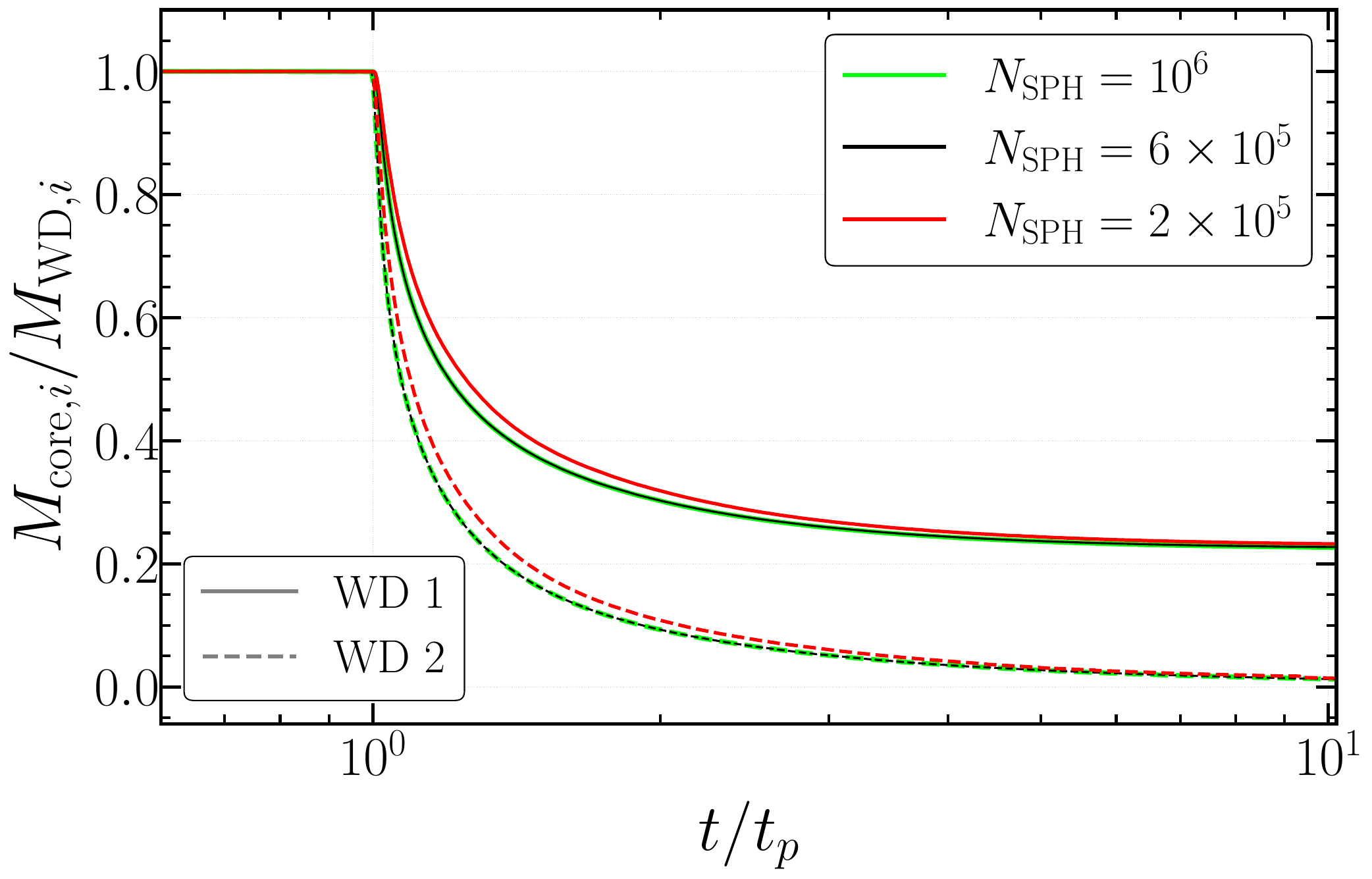}{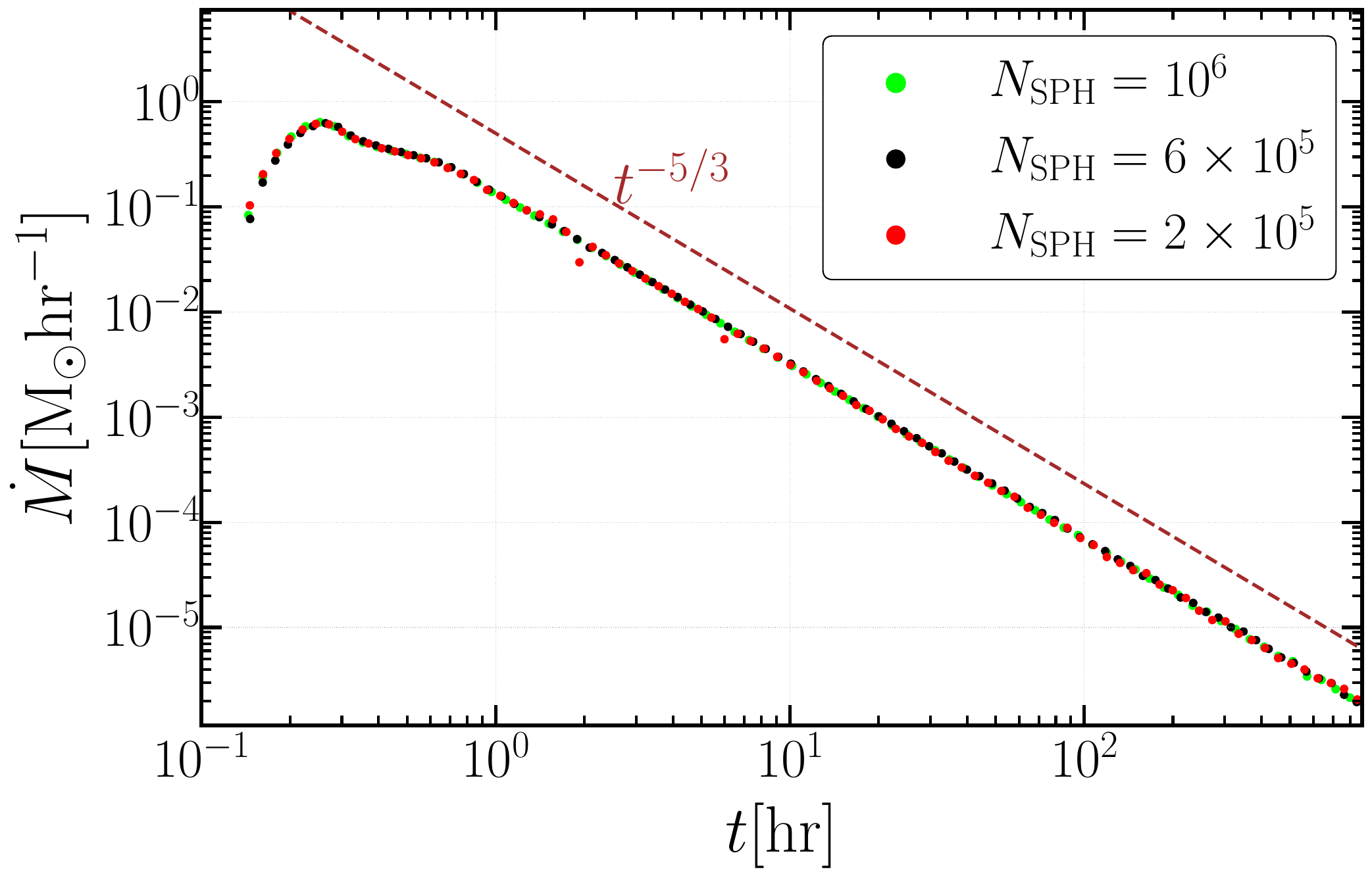}
	\caption{{\small \tb{Resolution dependence.  \textbf{Left Panel:}  core mass fraction $M_{{\rm core},i}/M_{{\rm WD},i}$ \textbf{Right Panel:} fallback rate $\dot{M}$, The light green, black, and red curves correspond to $ N_{\rm SPH} = 10^6$, $6\times10^5$, and $2\times10^5$, respectively, and are visually overlapping, indicating good numerical convergence.}}}
	\label{res_1}
\end{figure}

\begin{figure}[H]
	\epsscale{0.6}
	\plotone{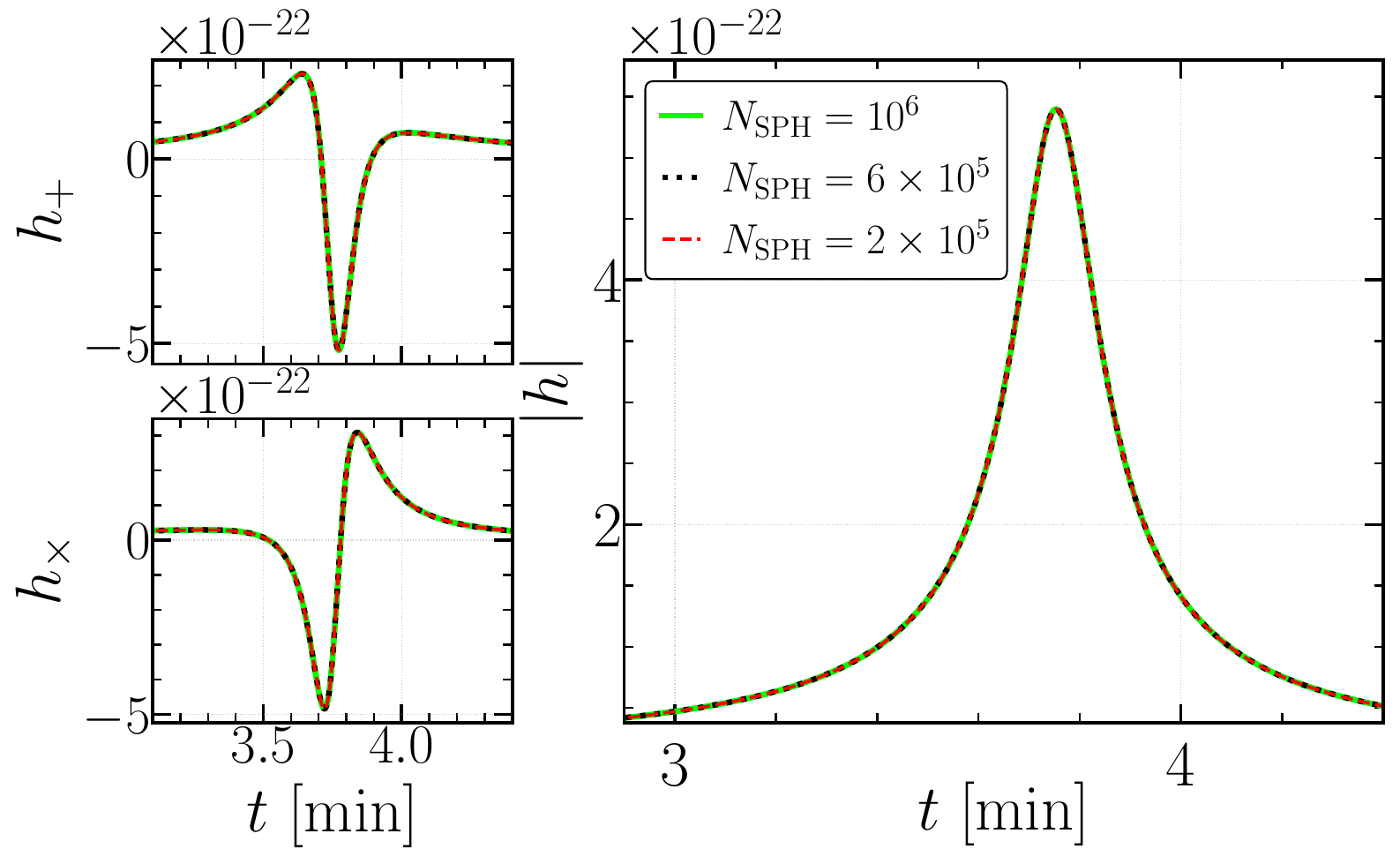}
	\caption{{\small \tb{Resolution dependence for the gravitational wave amplitude. The light green, black, and red curves correspond to $N_{\rm SPH} = 10^6$, $6\times10^5$, and $2\times10^5$, respectively, and are visually overlapping, indicating convergence of the gravitational wave signal.}}}
	\label{res_2}
\end{figure}

\tb{We verify the numerical consistency of our SPH results by performing convergence tests at both lower and higher resolutions, using $2 \times 10^5$ and $10^6$ SPH particles, respectively. In Figures~\ref{res_1} and \ref{res_2}, we show the convergence of key quantities, including the fallback rate and gravitational wave amplitude, for the identical prograde configuration with $\phi_0 = 45^\circ$. For the core mass fraction $M_{{\rm core},i}/M_{{\rm WD},i}$, we observe a small initial deviation at the lowest resolution ($N_{\rm SPH} = 2 \times 10^5$) compared to the higher-resolution runs ($N_{\rm SPH} = 6 \times 10^5$ and $10^6$). However, the saturated value of $M_{{\rm core},i}/M_{{\rm WD},i}$ is well converged, with relative deviations of $\lesssim 2.4 \%$ between $N_{\rm SPH} = 2 \times 10^5$ and $10^6$, and $\lesssim 0.25\%$ between $N_{\rm SPH} = 6 \times 10^5$ and $10^6$. Similar relative differences also exist in the peak values of fallback rate and GW amplitude. Particularly, the SPH results obtained with $N_{\rm SPH} = 6 \times 10^5$ and $N_{\rm SPH} = 10^6$ are in good agreement, demonstrating full convergence. The results obtained with $N_{\rm SPH} = 2 \times 10^5$ are also nearly converged. Therefore, the results presented in this work are insensitive to further increases in SPH resolution.}

%%%%%%%%%%%%%%%%%%%%%%%%%%%%%%%%%%%%%%%%%%%%%%%%
%%%%%%%%%%%%%%%%%%%%%%%%%%%%%%%%%%%%%%%%%%%%%%%%

%%%%%%%%%%%%%%%%%%%%%%%%%%%%%%%%%%%%%%%%%%%%%%%%
%%%%%%%%%%%%%%%%%%%%%%%%%%%%%%%%%%%%%%%%%%%%%%%%


\begin{thebibliography}{999}


\bibitem[Abbott et al.(2020)]{IMBH3} Abbott, R., Abbott, T.~D., Abraham, S., et al.\ 2020, \prl, 125, 101102. doi:10.1103/PhysRevLett.125.101102

\bibitem[Amaral et. al. (2024)]{Amaral} Amaral, L. A., et. al. 2024, \aap,  685, A9. doi:10.1051/0004-6361/202348564

\bibitem[Antonini et. al. (2011)]{Antonini} Antonini, F., Lombardi, Jr., J. C., \& Merritt, D., 2011, ApJ, 731, 128. doi: 10.1088/0004-637X/731/2/128

\bibitem[Balsara(1995)]{Balsara1995} Balsara, D.~S.\ 1995, Journal of Computational Physics, 121, 357. doi:10.1016/S0021-9991(95)90221-X

\bibitem[Banerjee et al.(2023)]{Banerjee2023} Banerjee, P., Garain, D., Chowdhury, S., et al.\ 2023, \mnras, 522, 4332. doi:10.1093/mnras/stad1284

\bibitem[Baumgardt et al.(2004)]{Baumgardt2004} Baumgardt, H., Makino, J., \& Ebisuzaki, T.\ 2004, \apj, 613, 2, 1143. doi:10.1086/423299

\bibitem[Bonnerot et al.(2016)]{Bonnerot2016} Bonnerot, C., Rossi, E.~M., Lodato, G., et al.\ 2016, \mnras, 455, 2253. doi:10.1093/mnras/stv2411

%\bibitem[Brown et. at. (2016)]{Brown} Brown, W. R. et. al. 2016 \apj, 818 155. doi:10.3847/0004-637X/818/2/155

\bibitem[Brown et al.(2015)]{Brown} Brown, G.~C., Levan, A.~J., Stanway, E.~R., et al.\ 2015, \mnras, 452, 4297. doi:10.1093/mnras/stv1520

\bibitem[Brown et al.(2016)]{Brown2016} Brown, W.~R., Gianninas, A., Kilic, M., et al.\ 2016, \apj, 818, 2, 155. doi:10.3847/0004-637X/818/2/155

\bibitem[Carter \& Luminet(1982)]{CarterLumineta} Carter, B. \& Luminet, J.~P.\ 1982, \nat, 296, 211. doi:10.1038/296211a0

\bibitem[Carter \& Luminet(1983)]{CarterLuminetb} Carter, B. \& Luminet, J.-P.\ 1983, \aap, 121, 97


\bibitem[Centrella \& McMillan(1993)]{1993ApJ...416..719C} Centrella, J.~M. \& McMillan, S.~L.~W.\ 1993, \apj, 416, 719. doi:10.1086/173272

\bibitem[Chen \& Shen(2018)]{Chen2018} Chen, J.-H. \& Shen, R.-F.\ 2018, \apj, 867, 20. doi:10.3847/1538-4357/aadfda

%\bibitem[Chen et al.(2023)]{Chen2023} Chen, J.-H. Shen, R.-F. \& Liu S-F., \apj, 947, 32. doi:10.3847/1538-4357/acbfb6

\bibitem[Chilingarian et al.(2018)]{IMBH1} Chilingarian, I.~V., Katkov, I.~Y., Zolotukhin, I.~Y., et al.\ 2018, \apj, 863, 1. doi:10.3847/1538-4357/aad184

\bibitem[Clerici \& Gomboc(2020)]{Clerici2020} Clerici, A. \& Gomboc, A.\ 2020, \aap, 642, A111. doi:10.1051/0004-6361/202037641

\bibitem[Coleman Miller \& Colbert(2004)]{Coleman} Coleman Miller, M. and Colbert, E. J. M., 2004, I.J.M.P D13, 1. doi:10.1142/S0218271804004426

\bibitem[Coughlin \& Nixon(2015)]{Coughlin2015} Coughlin, E.~R. \& Nixon, C.\ 2015, \apjl, 808, L11. doi:10.1088/2041-8205/808/1/L11.

%\bibitem[Coughlin et al.(2016a)]{Coughlin2016a} Coughlin, E.~R., Nixon, C., Begelman, M.~C., et al.\ 2016a, \mnras, 455, 3612. doi:10.1093/mnras/stv2511
%
%\bibitem[Coughlin et al.(2016b)]{Coughlin2016b} Coughlin, E.~R., Nixon, C., Begelman, M.~C., et al.\ 2016b, \mnras, 459, 3089. doi:10.1093/mnras/stw770

\bibitem[Coughlin \& Nixon(2019)]{CoughlinNixon} Coughlin, E.~R. \& Nixon, C.~J.\ 2019, \apjl, 883, L17. doi:10.3847/2041-8213/ab412d

\bibitem[Cufari et al.(2022)]{Cufari2022} Cufari, M., Coughlin, E.~R., \& Nixon, C.~J.\ 2022, \apj, 924, 34. doi:10.3847/1538-4357/ac32be

\bibitem[Darbha et al.(2019)]{Darbha2019} Darbha, S., Coughlin, E.~R., Kasen, D., et al.\ 2019, \mnras, 488, 5267. doi:10.1093/mnras/stz1923

\bibitem[Evans \& Kochanek(1989)]{Evans1989} Evans, C.~R. \& Kochanek, C.~S.\ 1989, \apjl, 346, L13. doi:10.1086/185567

\bibitem[Frank \& Rees(1976)]{FrankRees} Frank, J. \& Rees, M.~J.\ 1976, \mnras, 176, 633. doi:10.1093/mnras/176.3.633

\bibitem[Fuller \& Lai(2012)]{Fuller} Fuller, J. \& Lai, D.\ 2012, \apjl, 756, 1, L17. doi:10.1088/2041-8205/756/1/L17


\bibitem[Gafton et al.(2015)]{Gafton2015} Gafton, E., Tejeda, E., Guillochon, J., et al.\ 2015, \mnras, 449, 771. doi:10.1093/mnras/stv350

\bibitem[Gafton \& Rosswog(2019)]{Gafton2019} Gafton, E. \& Rosswog, S.\ 2019, \mnras, 487, 4790. doi:10.1093/mnras/stz1530

\bibitem[Garain \& Sarkar (2024 a)]{Garain1} Garain, D., \& Sarkar, T., 2024 \apj, 967, 167. doi:10.3847/1538-4357/ad3dfa

\bibitem[Garain \& Sarkar (2024 b)]{Garain2} Garain, D., \& Sarkar, T., 2024 ArXiv: 2401.17031 [astro-ph.HE]

\bibitem[Garain et. al. (2023)]{Garain2023a} Garain D., Banerjee P., Chowdhury S., Sarkar T., 2023, JCAP, 2023, 062. doi:10.1088/1475-7516/2023/11/062

\bibitem[Garain et al.(2023)]{Garain2023} Garain, D., Banerjee, P., Chowdhury, S., et al.\ 2023, arXiv:2307.03142. doi:10.48550/arXiv.2307.03142

\bibitem[Gebhardt et al.(2002)]{2002ApJ...578L..41G} Gebhardt, K., Rich, R.~M., \& Ho, L.~C.\ 2002, \apjl, 578, 1, L41. doi:10.1086/342980

\bibitem[Golightly et al.(2019a)]{Golightly2019a} Golightly, E.~C.~A., Nixon, C.~J., \& Coughlin, E.~R.\ 2019a, \apjl, 882, L26. doi:10.3847/2041-8213/ab380d

\bibitem[Golightly et al.(2019b)]{Golightly2019b} Golightly, E.~C.~A., Coughlin, E.~R., \& Nixon, C.~J.\ 2019b, \apj, 872, 163. doi:10.3847/1538-4357/aafd2f

\bibitem[Greene et al.(2020)]{Green} Greene, J.~E., Strader, J., \& Ho, L.~C.\ 2020, \araa, 58, 257. doi:10.1146/annurev-astro-032620-021835

\bibitem[Guillochon \& Ramirez-Ruiz(2013)]{Guillochon2013} Guillochon, J. \& Ramirez-Ruiz, E.\ 2013, \apj, 767, 25. doi:10.1088/0004-637X/767/1/25

\bibitem[Guillochon et al.(2014)]{Guillochon2014} Guillochon, J., Manukian, H., \& Ramirez-Ruiz, E.\ 2014, \apj, 783, 23. doi:10.1088/0004-637X/783/1/23

\bibitem[Hayasaki et al.(2013)]{Hayasaki2013} Hayasaki, K., Stone, N., \& Loeb, A.\ 2013, \mnras, 434, 909. doi:10.1093/mnras/stt871

\bibitem[Hayasaki et al.(2016)]{Hayasaki2016} Hayasaki, K., Stone, N., \& Loeb, A.\ 2016, \mnras, 461, 3760. doi:10.1093/mnras/stw1387

\bibitem[Heggie(1975)]{Heggie} Heggie, D.~C.\ 1975, \mnras, 173, 729. doi:10.1093/mnras/173.3.729

\bibitem[Hermes et. al. (2012)]{Hermes} Hermes, J. J., et. al., 2012, \apjl, 757 L21. doi:10.1088/2041-8205/757/2/L21

\bibitem[Hills(1975)]{Hills} Hills, J.~G.\ 1975, \nat, 254, 295. doi:10.1038/254295a0

\bibitem[Hills(1988)]{Hills1} Hills, J.~G.\ 1988, \nat, 331, 687. doi: 10.1038/331687a0

\bibitem[Holoien et al.(2019)]{Holoien} Holoien, T.~W.-S., Vallely, P.~J., Auchettl, K., et al.\ 2019, \apj, 883, 111. doi:10.3847/1538-4357/ab3c66

\bibitem[Hut(1981)]{Hut} Hut, P.\ 1981, \aap, 99, 126. 

\bibitem[Ivanova et al.(2005)]{Ivanova} Ivanova, N., Belczynski, K., Fregeau, J.~M., et al.\ 2005, \mnras, 358, 2, 572. doi:10.1111/j.1365-2966.2005.08804.x

\bibitem[Ivanova et al.(2006)]{Ivanova2} Ivanova, N., Heinke, C.~O., Rasio, F.~A., et al.\ 2006, \mnras, 372, 3, 1043. doi:10.1111/j.1365-2966.2006.10876.x

\bibitem[Jonker et al.(2022)] {Jonker} Jonker, P.~G, Arcavi, I., Phinney, E.~S, et al.\ 2022, The Tidal Disruption of Stars by Massive Black Holes (Springer) 

\bibitem[Kagaya et al.(2019)]{Kagaya2019} Kagaya, K., Yoshida, S., \& Tanikawa, A.\ 2019, arXiv:1901.05644. doi:10.48550/arXiv.1901.05644

\bibitem[Kilic et. al. (2007)]{Kilic1} Kilic, M., et. al., 2007, \apj 664, 1088. doi:10.1086/518735

\bibitem[Kilic et. al. (2014)]{Kilic2} Kilic, M., et. al., 2014, \mnras, 444, L1. doi:https://doi.org/10.1093/mnrasl/slu093

\bibitem[Kirk et al.(2016)]{Kirk} Kirk, B., Conroy, K., Pr{\v{s}}a, A., et al.\ 2016, \aj, 151, 3, 68. doi:10.3847/0004-6256/151/3/68

\bibitem[K{\i}z{\i}ltan et al.(2017)]{IMBH0} K{\i}z{\i}ltan, B., Baumgardt, H., \& Loeb, A.\ 2017, \nat, 542, 203. doi:10.1038/nature21361

\bibitem[Kobayashi et. al. (2012)]{Sari2} Kobayashi, S., Hainick, Y., Sari, R., et. al. 2012, \apj, 748, 105. doi:10.1088/0004-637X/748/2/105

\bibitem[Kobayashi et al.(2004)]{Kobayashi2004} Kobayashi, S., Laguna, P., Phinney, E.~S., et al.\ 2004, \apj, 615, 2, 855. doi:10.1086/424684

\bibitem[Kochanek(1994)]{Kochanek1994} Kochanek, C.~S.\ 1994, \apj, 422, 508. doi:10.1086/173745

\bibitem[Korol et. al. (2017)]{Korol} Korol, V., et. al., 2017, \mnras 470, 1894. doi:10.1093/mnras/stx1285

\bibitem[Lacy et al.(1982)]{Lacy1} Lacy, J.~H., Townes, C.~H., \& Hollenbach, D.~J.\ 1982, \apj, 262, 120. doi:10.1086/160402

\bibitem[Law-Smith et al.(2017)]{Law2017} Law-Smith, J., MacLeod, M., Guillochon, J., et al.\ 2017, \apj, 841, 132. doi:10.3847/1538-4357/aa6ffb

\bibitem[Law-Smith et al.(2019)]{Law2019} Law-Smith, J., Guillochon, J., \& Ramirez-Ruiz, E.\ 2019, \apjl, 882, L25. doi:10.3847/2041-8213/ab379a

%\bibitem[Law-Smith et al.(2020)]{Law2020} Law-Smith, J.~A.~P., Coulter, D.~A., Guillochon, J., et al.\ 2020, \apj, 905, 141. doi:10.3847/1538-4357/abc489

\bibitem[Li et al.(2021)]{2021MNRAS.501.1621L} Li, J., Lai, D., Anderson, K.~R., et al.\ 2021, \mnras, 501, 2, 1621. doi:10.1093/mnras/staa3779

\bibitem[Lin et al.(2020)]{IMBH4} Lin, D., Strader, J., Romanowsky, A.~J., et al.\ 2020, \apjl, 892, L25. doi:10.3847/2041-8213/ab745b

\bibitem[Liptai et al.(2019)]{Liptai2019} Liptai, D., Price, D.~J., Mandel, I., et al.\ 2019, arXiv:1910.10154. doi:10.48550/arXiv.1910.10154

\bibitem[Lodato et al.(2009)]{Lodato} Lodato, G., King, A.~R., \& Pringle, J.~E.\ 2009, \mnras, 392, 332. doi:10.1111/j.1365-2966.2008.14049.x

\bibitem[MacLeod et al.(2013)]{MacLeod2013} MacLeod, M., Ramirez-Ruiz, E., Grady, S., et al.\ 2013, \apj, 777, 133. doi:10.1088/0004-637X/777/2/133

\bibitem[Maeda et. al. (2023 a)]{Maeda1} Maeda, K., Gupta, P.,\& Okawa, H., 2023, \prd, 107, 12, 124039. doi:10.1103/PhysRevD.107.124039

\bibitem[Maeda et. al. (2023 b)]{Maeda2} Maeda, K., Gupta, P.,\& Okawa, H., 2023, \prd, 108, 12, 123041. doi:10.1103/PhysRevD.108.123041

\bibitem[Maeda \& Okawa (2024)]{Maeda3} Maeda, K., \& Okawa, H., 2024 ArXiv: 2504.18934 [gr-qc]

\bibitem[Maggiore(2007)]{Maggiore} Maggiore, M.\ 2007, . doi:10.1093/acprof:oso/9780198570745.001.0001


\bibitem[Mahapatra et. al. (2024)]{Garain3} Mahapatra, A., Pandey, A., Garain, D. et. al. 2024 ArXiv: 2410.12727 [gr-qc]

%\bibitem[Mainetti et al.(2017)]{Mainetti2017} Mainetti, D., Lupi, A., Campana, S., et al.\ 2017, \aap, 600, A124. doi:10.1051/0004-6361/201630092
\bibitem[Mahapatra et. al. (2026)]{MahapatraLatest} Mahapatra, A., Pandey, A., Banerjee, P. et. al., 2026, ArXiv:2601.04870 [astro-ph.HE]

\bibitem[Manukian et al.(2013)]{Manukian2013} Manukian, H., Guillochon, J., Ramirez-Ruiz, E., et al.\ 2013, \apjl, 771, L28. doi:10.1088/2041-8205/771/2/L28

\bibitem[Manzaneda et al.(2024)]{Manzaneda} Manzaneda, L.~A., Navarrete, C.~O., \& Tejeda, E.\ 2024, , arXiv:2404.16270. doi:10.48550/arXiv.2404.16270

\bibitem[Mandel \& Levin(2015)]{2015ApJ...805L...4M} Mandel, I. \& Levin, Y.\ 2015, \apjl, 805, 1, L4. doi:10.1088/2041-8205/805/1/L4

\bibitem[Meibom \& Mathieu(2005)]{Meibom} Meibom, S. \& Mathieu, R.~D.\ 2005, \apj, 620, 2, 970. doi:10.1086/427082

\bibitem[Miles et al.(2020)]{Milesetal} Miles, P.~R., Coughlin, E.~R., \& Nixon, C.~J.\ 2020, \apj, 899, 36. doi:10.3847/1538-4357/ab9c9f

\bibitem[Misner et al.(1973)]{Misner1973}
Misner, C. W., Thorne, K. S., \& Wheeler, J. A. 1973, 
\textit{Gravitation} (San Francisco: W. H. Freeman)

\bibitem[Moe \& Di Stefano(2017)]{Moe} Moe, M. \& Di Stefano, R.\ 2017, \apjs, 230, 2, 15. doi:10.3847/1538-4365/aa6fb6

\bibitem[Munday et. al. (2023)]{Munday} Munday, J., et. al., 2023, \mnras, 518, 5123. doi:10.1093/mnras/stac3385

\bibitem[Musielak \& Quarles (2014)]{MQ} Musielak, Z. E., \& Quarles, B., 2014, Rep. Prog. Phys. 77, 6,  065901. doi:10.1088/0034-4885/77/6/065901

\bibitem[Naoz (2016)]{Naoz} Naoz, S., 2016, \araa, 54, 441. doi:10.1146/annurev-astro-081915-023315

\bibitem[Noyola et al.(2008)]{2008ApJ...676.1008N} Noyola, E., Gebhardt, K., \& Bergmann, M.\ 2008, \apj, 676, 2, 1008. doi:10.1086/529002

\bibitem[Nixon et al.(2021)]{Nixon2021} Nixon, C.~J., Coughlin, E.~R., \& Miles, P.~R.\ 2021, \apj, 922, 168. doi:10.3847/1538-4357/ac1bb8

%\bibitem[Open TDE Catalog (2023)]{OpenTDE} 
%The Open TDE Catalog, 2023, available at https://tde.space/

\bibitem[Park \& Hayasaki(2020)]{Park2020} Park, G. \& Hayasaki, K.\ 2020, \apj, 900, 3. doi:10.3847/1538-4357/ab9ebb

\bibitem[Peters(1964)]{Peters} Peters, P.~C.\ 1964, Phys.\ Rev., 136, B1224

\bibitem[Phinney(1989)]{Phinney} Phinney, E.~S.\ 1989, The Center of the Galaxy, 136, 543

%\bibitem[Price(2007)]{Price2007} Price, D.~J.\ 2007, \pasa, 24, 159. doi:10.1071/AS07022

%\bibitem[Price et al.(2018)]{Price2018} Price D.~J., Wurster J., Tricco T.~S., Nixon C., Toupin S., Pettitt A., Chan C., et al., 2018, 
%\pasa, 35, e031. doi:10.1017/pasa.2018.25
\bibitem[Price(2011)]{splash} Price, D.~J.\ 2011, Astrophysics Source Code Library. ascl:1103.004

\bibitem[Price et al.(2018)]{2018PASA} Price, D.~J., Wurster, J., Tricco, T.~S., et al.\ 2018, \pasa, 35, e031. doi:10.1017/pasa.2018.25

\bibitem[Price-Whelan et al.(2020)]{Price-Whelan} Price-Whelan, A.~M., Hogg, D.~W., Rix, H.-W., et al.\ 2020, \apj, 895, 1, 2. doi:10.3847/1538-4357/ab8acc


\bibitem[Raghavan \& McAlister(2009)]{Raghavan} Raghavan, D. \& McAlister, H.~A.\ 2009, \aas, 213, 330.04. 

\bibitem[Rasio \& Shapiro (1994)]{RasioShapiro1} Rasio, F. A., \& Shapiro, S. L.,  1994, \apj, 432, 242. doi:10.1086/174566

\bibitem[Rasio \& Shapiro (1995)]{RasioShapiro2} Rasio, F. A., \& Shapiro, S. L.,  1995, \apj, 438, 887. doi:10.1086/175130

\bibitem[Roelofs et. al. (2010)]{Roelofs} Roelofs, G. H. A. et. al., 2010 \apjl, 7, 11 L138. doi:10.1088/2041-8205/711/2/L138

\bibitem[Rosswog et al.(2009)]{Rosswog2009} Rosswog, S., Ramirez-Ruiz, E., \& Hix, W.~R.\ 2009, \apj, 695, 1, 404. doi:10.1088/0004-637X/695/1/404

\bibitem[Rosswog et al.(2008)]{Rosswog2008} Rosswog, S., Ramirez-Ruiz, E., \& Hix, W.~R.\ 2008, \apj, 679, 2, 1385. doi:10.1086/528738

\bibitem[Rees(1988)]{Rees} Rees, M.~J.\ 1988, \nat, 333, 523. doi:10.1038/333523a0

\bibitem[Ren et. al. (2023)]{Ren} Ren, L., et. al., 2023 \apjs, 264 39. doi:10.3847/1538-4365/aca09e

%\bibitem[Rosswog et al.(2009)]{2009ApJ...695..404R} Rosswog, S., Ramirez-Ruiz, E., \& Hix, W.~R.\ 2009, \apj, 695, 404. doi:10.1088/0004-637X/695/1/404

\bibitem[Ryu et al.(2020a)]{Ryu2020a} Ryu, T., Krolik, J., Piran, T., et al.\ 2020a, \apj, 904, 98. doi:10.3847/1538-4357/abb3cf

\bibitem[Ryu et al.(2020b)]{Ryu2020b} Ryu, T., Krolik, J., Piran, T., et al.\ 2020b, \apj, 904, 100. doi:10.3847/1538-4357/abb3ce

%\bibitem[Ryu et al.(2020c)]{Ryu2020c} Ryu, T., Krolik, J., Piran, T., et al.\ 2020c, \apj, 904, 100. doi:10.3847/1538-4357/abb3ce

\bibitem[Sacchi \& Lodato(2019)]{Sacchi2019} Sacchi, A. \& Lodato, G.\ 2019, \mnras, 486, 1833. doi:10.1093/mnras/stz981

\bibitem[Sacchi et al.(2020)]{Sacchi 2020} Sacchi, A., Lodato, G., Toci, C., et al.\ 2020, \mnras, 495, 1, 1227. doi:10.1093/mnras/staa1299

\bibitem[Sari et al.(2010)]{Sari} Sari, R., Kobayashi, S., \& Rossi, E.~M.\ 2010, \apj, 708, 1, 605. doi:10.1088/0004-637X/708/1/605

\bibitem[Sethi et al.(2026)]{Sethi} Sethi, R., Martin, D.~V., Barker, A., et al.\ 2026, arXiv:2603.04554. doi:10.48550/arXiv.2603.04554

%\bibitem[Kobayashi et. al. (2012)]{Sari2} Kobayashi, S., Hainick, Y., Sari, R., et. al. 2012, \apj, 748, 105. doi:10.1088/0004-637X/748/2/105

\bibitem[Shara \& Hurley(2002)]{shara} Shara, M.~M. \& Hurley, J.~R.\ 2002, \apj, 571, 2, 830. doi:10.1086/340062

\bibitem[Stone et. al. (2019)]{Stone2019} Stone, N. C., Kesden, M., Cheng, R. M., et al. 2019, General Relativity and Gravitation, 51, 30. doi:10.1007/s10714-019-2510-9

\bibitem[Takekawa et al.(2019)]{IMBH2} Takekawa, S., Oka, T., Iwata, Y., et al.\ 2019, \apjl, 871, L1. doi:10.3847/2041-8213/aafb07

\bibitem[Tejeda \& Rosswog(2013)]{Rosswog_TR} Tejeda, E. \& Rosswog, S.\ 2013, \mnras, 433, 1930. doi:10.1093/mnras/stt853

\bibitem[Tejeda et al.(2017)]{Tejeda2017} Tejeda, E., Gafton, E., Rosswog, S., et al.\ 2017, \mnras, 469, 4483. doi:10.1093/mnras/stx1089

\bibitem[Toonen et. al. (2017)]{Toonen} Toonen, S., M., Hollands, M., Gansicke, B. T., et al. 2017 \aap,\  602, A16. doi:10.1051/0004-6361/201629978

\bibitem[Tokovinin(2021)]{Tokovinin} Tokovinin, A.\ 2021, Universe, 7, 9, 352. doi:10.3390/universe7090352

\bibitem[Toscani et al.(2022)]{Toscani} Toscani, M., Lodato, G., Price, D.~J., et al.\ 2022, \mnras, 510, 1, 992. doi:10.1093/mnras/stab3384

%\bibitem[Ulmer(1999)]{Ulmer1999} Ulmer, A.\ 1999, \apj, 514, 180. doi:10.1086/306909.

\bibitem[Vick et al.(2017)]{Vick2017} Vick, M., Lai, D., \& Fuller, J.\ 2017, \mnras, 468, 2, 2296. doi:10.1093/mnras/stx539

\bibitem[Volonteri(2012)]{Volonteri} Volonteri, M.\ 2012, Science, 337, 544. doi:10.1126/science.1220843

\bibitem[Voltonen \& Karttunen (2005)]{VK} Voltonen, M., \& Karttunen, H., {\it The three body problem} Cambridge University Press, 2005.

\bibitem[Vynatheya et al.(2026)]{Vynatheya} Vynatheya, P., Dessart, L., Ryu, T., et al.\ 2026, arXiv:2601.20133. doi:10.48550/arXiv.2601.20133

\bibitem[Wang et al.(2021)]{Wang2021} Wang, Y.-H., Perna, R., \& Armitage, P.~J.\ 2021, \mnras, 503, 6005. doi:10.1093/mnras/stab802

\bibitem[Woosley et al.(2004)]{Woosely2004} Woosley, S.~E., Wunsch, S., \& Kuhlen, M.\ 2004, \apj, 607, 2, 921. doi:10.1086/383530

\bibitem[Will(2017)]{Will2017} Will, C.~M.\ 2017, \prd, 96, 2, 023017. doi:10.1103/PhysRevD.96.023017

\bibitem[Xuan et al.(2025)]{Xuan} Xuan, Z., Shariat, C., \& Naoz, S.\ 2025, \apj, 995, 1, 27. doi:10.3847/1538-4357/ae160e

\bibitem[Yu \& Lai (2024)]{YuLai1} Yu, F. \& Lai, D., 2024, \apj,  977, 268. doi:10.3847/1538-4357/ad93a6

\bibitem[Yu \& Lai (2025)]{YuLai2} Yu, F. \& Lai, D., 2025, ArXiv: 2504.14146 [astro-ph.HE]

\bibitem[Zahn(1977)]{Zahn} Zahn, J.-P.\ 1977, \aap, 57, 383. 



%%%%%%%%%%%%%%%%%%%%%%%%%%%%%%%%%%%%%%%%%%%%%%%%%%%

%\bibitem[Open TDE Catalog (2023)]{OpenTDE} 
%The Open TDE Catalog (2023) available at https://tde.space/


\end{thebibliography}
\end{document}